\documentclass[a4paper,12pt]{article}

\usepackage{cmap}					
\usepackage{mathtext} 				
\usepackage[T2A]{fontenc}			
\usepackage[utf8]{inputenc}			
\usepackage[english,main=russian]{babel}	

\usepackage{etoolbox,pdftexcmds}
\let\englishtableofcontents\tableofcontents
\patchcmd\englishtableofcontents{{toc}}{{tec}}{}{}
\preto\englishtableofcontents{\begin{otherlanguage}{english}}
\appto\englishtableofcontents{\end{otherlanguage}}
\addto\captionsenglish{
  \renewcommand{\contentsname}%
    {Content}%
}
\let\englishlistoffigures\listoffigures
\patchcmd\englishlistoffigures{{lof}}{{lef}}{}{}
\preto\englishlistoffigures{\begin{otherlanguage}{english}}
\appto\englishlistoffigures{\end{otherlanguage}}
\let\englishlistoftables\listoftables
\patchcmd\englishlistoftables{{lot}}{{let}}{}{}
\preto\englishlistoftables{\begin{otherlanguage}{english}}
\appto\englishlistoftables{\end{otherlanguage}}

\newcommand{\addetoc}[2]{%
  \addcontentsline{tec}{#1}{\protect\numberline{\csname the#1\endcsname}#2}%
}
\makeatletter
\newcommand{\englishcaption}[1]{%
  \ifnum\pdf@strcmp{\@captype}{figure}=\z@
    \addcontentsline{lef}{figure}{\protect\numberline{\thefigure}#1}%
  \else
    \addcontentsline{let}{table}{\protect\numberline{\thetable}#1}%
  \fi
}
\makeatother

\usepackage{amsmath,amsfonts,amssymb,amsthm,mathtools} 
\usepackage{bm}
\usepackage{icomma} 
\usepackage[short]{optidef}
\renewcommand{\theequation}{\arabic{section}.\arabic{equation}}

\usepackage{graphicx}  
\graphicspath{{./images/}}  
\usepackage{wrapfig} 
\usepackage{caption}
\usepackage{float}
\usepackage{floatflt}
\usepackage{listings}

\usepackage{array,tabularx,tabulary,booktabs} 
\usepackage{longtable}  
\usepackage{multirow} 

\usepackage{etoolbox} 

\usepackage{geometry} 
\usepackage{indentfirst}
\usepackage{fancyhdr} 
\usepackage{setspace} 
\usepackage{lastpage} 

\usepackage{soulutf8} 

\usepackage{multicol}

\usepackage{csquotes}
\usepackage
[backend=biber,
bibencoding=utf8,
language=english,
sorting=none
]
{biblatex}
\usepackage{tikz} 
\usepackage{pstool}
\usepackage{pgfplots}
\pgfplotsset{compat=1.9}
\usepackage{pgfplotstable}
\usepackage{color}
\usepackage{ifthen}
\usepackage{animate}

\renewcommand{\epsilon}{\ensuremath{\varepsilon}}
\renewcommand{\phi}{\ensuremath{\varphi}}
\renewcommand{\kappa}{\ensuremath{\varkappa}}
\renewcommand{\le}{\ensuremath{\leqslant}}

\renewcommand{\ge}{\ensuremath{\geqslant}}

\renewcommand{\vec}{\ensuremath\bm}
\DeclareMathOperator*{\argmin}{arg\,min}

\renewcommand{\eqref}[1]{(\ref{#1})}

\usepackage{adjustbox}

\usepackage{tikz}
\usetikzlibrary{shapes.geometric, arrows}

\tikzstyle{cdft} = [rectangle, rounded corners, minimum width=1cm, minimum height=1cm,text centered,text width=3.5cm, draw=black]

\tikzstyle{vfdft} = [rectangle, rounded corners, minimum width=1cm, minimum height=1cm,text centered,text width=2.6cm, draw=red]

\tikzstyle{empty} = [minimum width=1cm, minimum height=1cm,text centered,text width=1cm]

\tikzstyle{arrow} = [thick,->,>=stealth]

\usepackage{nomencl}
\makenomenclature    

\renewcommand{\author}[1]{\def\mtauthortext{#1}}
\renewcommand{\title}[1]{\def\mttitletext{#1}}
\newcommand{\supervisor}[1]{\def\mtsupervisortext{#1}}

\usepackage[justification=centerlast]{caption}
\usepackage{subcaption}

\usepackage[section]{placeins}

\renewcommand{\titlepage}{%
    \thispagestyle{empty}
    \begin{center}
        {\bf Moscow Institute of Physics and Technology (State University) \\ }
        \vspace{1cm}

        Phystech School of Radio Engineering and Computer Technology \\
        Department of Modeling and Technologies for Oil Field Development \\

        \vspace{3em}

        Master's thesis on a speciality 03.04.01\\
         "Applied Mathematics and Physics" 
    \end{center}

    \begin{center}
        \vspace{\fill}
        \singlespacing

        \LARGE \mttitletext{}

        \vspace{\fill}
    \end{center}

    \begin{flushright}
        \mtauthortext{} \\
        yuriy.kanygin@phystech.edu\\
        \vspace{2.5em}
        Scientific Supervisor \\
        \mtsupervisortext{} \\
        \vspace{2.5em}
        Co-Supervisor \\
        Pavel Lomovitskiy \\
    \end{flushright}

    \vspace{6em}

    \begin{center}
        Moscow \\
        \the\year{}
    \end{center}
\newpage
}
\usetikzlibrary{external}

\title{Development of fast numerical density functional theory methods for studying the structures of nanoporous materials}

\author{Yuriy Kanygin}
\supervisor{Aleksey Khlyupin}

\begin{document} 

\selectlanguage{english}
\titlepage

\begin{abstract}
Density functional theory (DFT) has been actively used and developed recently. DFT is an efficient instrument for describing a wide range of nanoscale phenomena: wetting transition, capillary condensation, adsorption, and others. In this work, we suggest a method for obtaining the equilibrium fluid density in a pore using DFT without calculating the free energy variation ---~
Variation Free Density Functional Theory (VF-DFT). This technique is applicable to explore fluids with complex interactions and speed up calculations for simple fluids. In VF-DFT the fluid density is represented as a decomposition over a limited set of basis functions and decomposition coefficients. To construct basis functions, we applied principal component analysis (PCA). PCA is used to extract the main patterns of the fluid densities in the nanopore. Decomposition coefficients are sought with stochastic gradient-free optimization methods, such as genetic algorithm (GA), particle swarm optimization (PSO) to minimize the free energy of the system. We also suggest the Hybrid Density Functional Theory (H-DFT) approach based on stochastic optimization methods and the classical Piccard iteration method to find the equilibrium fluid density in the pore. Combining these methods helps to significantly speed up the calculations of equilibrium density in the system without losing quality. We considered two fluids: nitrogen at the temperature of $T=77.4$~K and argon $T=87.3$~K, at the pore width of $3.6$~nm. VF-DFT, H-DFT with different optimization algorithms were compared with each other and with classical Piccard iteration technique.  Furthermore, the problem of calculation pore size distribution for nanoporous materials is discussed. The Tikhonov regularization method was applied to reconstruct of pore size distributions from experimental data on low-temperature adsorption. This method is proved to be very sensitive to the quality of adsorption data.
\end{abstract}
\newpage

\englishtableofcontents

\newpage
\selectlanguage{russian}
\begin{abstract}
    Теория функционала плотности (DFT) активно используется и развивается в последнее время. DFT является важным инструментом для описания широкого круга явления на наномасштабах: смачивание, капиллярная конденсация, адсорбция и многое другое. В данной работе рассматривается подход, позволяющий использовать теорию функционала плотности для расчета термодинамических свойств флюида в нанопорах без вычисления вариации свободной энергии (Variation Free Density Functional Theory --- VF-DFT). Такой подход позволяет исследовать флюиды со сложным типом взаимодействия и ускорить вычисление для простых флюидов. В безвариационном методе равновесная плотность флюида в поре представляется в виде разложения по ограниченному набору базисных функций. Коэффициенты разложения функции плотности ищутся с помощью стохастических методов оптимизации (генетический алгоритм --- GA, метод роя частиц --- PSO) так, чтобы минимизировать свободную энергию системы. В работе рассматриваются два флюида: азот при температуре $77.4$~K и аргон при температуре $87.3$~K в поре $3.6$~нм и проводится сравнение алгоритмов оптимизации. Также рассматривается гибридный подход (Hybrid Density Functional Theory --- H-DFT) на основе стохастических методов оптимизации и метода простой итерации для поиска равновесной плотности флюида в поре. Такое объединение методов помогает значительно сократить время поиска решения без потери качества. Также рассматривается задача восстановления распределения пор по размерам на основе экспериментальных данных по низкотемпературной адсорбции методами теории функционала плотности. В работе приводятся результаты по решению модельных задач восстановления распределения пор по размерам, рассматриваются случаи унимодальных и бимодальных гауссовских распределения с различными значениями параметров дисперсии и математического ожидания.
\end{abstract}
\newpage
\tableofcontents
\newpage

\section{Введение}
\addetoc{section}{Intoduction}
Задача исследования флюидов на масштабах от микро- до нанометров становится все более актуальной как для нефтегазовой индустрии (сланцевые углеводороды), так и для других технологических отраслей, в которых необходимо производить сепарацию флюидов, улавливание и хранение углекислого газа, исследовать различного рода течения флюидов в природных или искусственных наноразмерных структурах \cite{RezaeiGomari2019NewReservoirs,Sumida2012CarbonFrameworks,Wang2018AtomisticCapillary,Wu2006DensityMaterials}. В частности, для нефтегазовой индустрии, в нетрадиционных коллекторах большая часть порового пространства представляет собой структуры размера нескольких десятков нанометров. Однако, на таких масштабах классические подходы имеют ограничения, становится необходимым более точно учитывать силы межмолекулярного взаимодействия, чтобы корректно описывать и учитывать такие физические явления, как: адсорбция \cite{Neimark1998PoreAdsorption, Ravikovitch2001DensityNanopores}, образование двойного электрического слоя \cite{Yu2004Density-functionalSolutions,Wu2006DensityMaterials}, капиллярная конденсация \cite{Wu2006DensityMaterials}, смачивание \cite{Berim2008NanodropConsiderations, Wu2006DensityMaterials}, исследовать гетерогенность стенок коллектора \cite{Aslyamov2017DensitySurfaces, Neimark2009QuenchedCarbons, Jagiello20132D-NLDFTCorrugation, Yang2011SolvationSurfaces}. Корректное описание этих явлений позволяет изучать структуру материалов на наномасштабе, поведение флюидов и их взаимодействие как между собой, так и с твердой средой.\\

С открытием высоко упорядоченный мезопористых материалов начали развиваться методы описания поведения флюида в нанопорах и характеризации порового пространства. Традиционные методы, такие как BET \cite{brunauer1938adsorption}, BJH \cite{barrett1951determination} не подошли для таких задач так, как они не способны учесть различную морфологию порового пространства, учесть влияние микропористости и предсказать распределение пор по размерам, которые могли бы быть независимо определены с помощью рентгеновской дифракции (XRD) и просвечивающей электронной микроскопии (TEM) с точностью, недоступной ранее. Новые нанопористые материалы и новые экспериментальные подходы высокого разрешения требуют новых адекватных теоретических методов для анализа экспериментальных данных.\\

Одним из наиболее распространенных инструментов для описания флюидов и их взаимодействий с поверхностями на таких масштабах является Теория Функционала Плотности (DFT). DFT предоставляет собой компромисс между классическими полуэмпирическими методами и молекулярным моделированием. С одной стороны, DFT способна учитывать микроскопическую структуру макроскопической системы при относительно низких вычислительных затратах, а с другой, DFT --- более строгая теория, чем классические феноменологические подходы. Несмотря на то, что DFT, как инструмент исследования термодинамических свойств флюидов в классических системах, был успешно применен в 1976 году \cite{ebner1976density}, теория функционала плотности только в последнее время стала активно использоваться и развивается для изучение равновесия и кинетики фазовых переходов, свойств полимерных материалов, тонких пленок, биологических систем.\\

В 1989 году, теория функционала плотности была впервые применена для расчета распределения пор по размерам из данных по низкотемпературной изотерме адсорбции \cite{seaton1989new}, и вскоре было признано, что DFT обеспечивает более обоснованный и универсальный подход к расчету параметров структуры пор по сравнению с традиционными методами, основанными на уравнении Кельвина \cite{cohan1938sorption}.\\

С математической точки зрения, молекулярная теория функционала плотности напоминает DFT из квантовой механики (оба имеют один и тот же акроним) за исключением того, что в первом случае функционал плотности относится к строению атомов или элементам полимерной молекулы, тогда как во втором случае идет речь об электронах. Молекулярная теория функционала плотности опирается на теорему о том, что система с определенной температурой $T$, объемом $V$, химическим потенциалом каждого из компонент флюида $\mu_i$, внешним потенциалом для каждой компоненты $V_{ext}\left(\vec{r}\right)$ однозначно определяется равновесной плотностью распределения молекул $\rho\left(\vec{r}\right)$ \cite{Kohn1964}. Следствием этой теоремы является то, что для энергия Гельмгольца для системы может быть записана как функционал плотности $F\left[\rho\left(\vec{r}\right)\right]$ и в равновесии свободная энерия минимальна.  Математическую основу для DFT в квантово-механических терминах сформулировали Хоэнберг и Кон \cite{Kohn1964}.\\

Таким образом, для исследования сложных систем методами теории функционала плотности необходимо две вещи. Во-первых, корректно задать энергию Гельмгольца для системы, именно это определяет физику системы. Во-вторых, найти распределение плотности флюида, которое бы минимизировало свободную энергию системы. На основе минимальности свободной энергии делается вывод о том, что вариация свободной энергии по плотности должна равняться нулю. На этом утверждении строится вся дальнейшая теория вычисления равновесной плотности \cite{Roth2010FundamentalReview,Sears2003AFluids}. К сожалению, для реальных флюидов свободная энергия Гельмгольца имеет сложную структуру, из-за чего вычисление ее вариации по плотности является трудоемкой задачей. Поэтому приходится строить различного рода приближения, которые бы позволяли с определенной точностью описывать физику системы. В последнее время разрабатываются подходы, позволяющие учитывать гетерогенность поверхности пор \cite{Aslyamov2017DensitySurfaces, Neimark2009QuenchedCarbons, Jagiello20132D-NLDFTCorrugation,khlyupin2016effects,aslyamov2019theoretical}, кулоновскую природу частиц флюида. В работе \cite{Aslyamov2017DensitySurfaces} наглядно продемонстрировано, что усложнение физической модели и учет гетерогенности стенок поры через потенциал взаимодействия молекул флюида с поверхностью может значительно изменить вид свободной энергии и ее вариации, наиболее значительное изменение претерпело слагаемое с вариацией от диполь-дипольного взаимодействия. Кроме изменения физики системы на вид вариации влияет геометрия задачи. В зависимости от того, какой материал исследуется, вариацию свободной энергии приходится считать в разных системах координат \cite{Roth2010FundamentalReview}, потому что для одних материалов поры можно рассматривать как цилиндрические трубки, а для других материалов стенки поры представляются как две параллельные стенки, и задача в таком случае имеет одно выделенное направление.  \\  

В данной работе представлен новый подход вычисления равновесной плотности, который не требует вычисления вариации свободной энергии Гельмгольца (Variation Free Density Functional Theory). На начальном этапе один раз для всех последующих задач рассчитывается датасет, состоящий из функций, поведение которых максимально близко к поведению исследуемой функции плотности флюида. Конкретно в данной работе, в качестве функций для датасета были взяты равновесные плотности азота при различных относительных давлениях в поре 3.6 нм (эти функции были рассчитаны с помощью классического подхода с методом простой итерации \cite{Roth2010FundamentalReview}). Также были добавлены плотности флюидов с параметрами взаимодействия азота, но с отличными радиусами молекул. Добавление таких <<искусственных>> функций позволяет внести разнообразие в данные и применять алгоритм для флюида, отличного от азота.\\

На следующем этапе с помощью метода главных компонент (principle component analysis --- PCA) из датасета выделяются наиболее значимые паттерны, а неизвестная функция плотности флюида представляется в виде разложения по рассчитанным с помощью PCA функциям, которые содержат эти наиболее значимые паттерны. Таким образом, задача поиска равновесной плотности сводится к задаче оптимизации коэффициентов разложения по базису. Благодаря тому, что используется не весь базис, а только его наиболее значимая часть, удается значительно понизить размерность оптимизационной задачи без заметной потери качества решения. На основе энергетического критерия удалось сократить количество базисных функций со 140 до 10. Полученные 10 векторов содержат 95\% информации о исходном базисе.\\

В безвариационном алгоритме поиск равновесной плотности осуществлялся с помощью стохастических методов оптимизации, в отличие от классических вариационных подходов, на основе метода простой итерации или ньютоновского метода \cite{Roth2010FundamentalReview,Wu2017VariationalModeling,Edelmann2016ATheory,Sears2003AFluids}. Среди широкого спектра стохастических оптимизационных алгоритмов были выбраны и сравнены два наиболее распространенных и популярных генетический алгоритм (genetic algorithm --- GA), метод роя частиц (particle swarm optimization --- PSO). Оптимизационные алгоритмы ищут коэффициенты разложения плотности флюида по базису так, чтобы минимизировать свободную энергию системы. \\

Также был разработан гибридный подход (Hybrid Density Functional Theory), который основывается на безвариационном методе и методе простой итерации. Работа гибридного алгоритма H-DFT отличается от VF-DFT только тем, что решение, которое выдает стохастический алгоритм оптимизации подается как начальное приближение для метода простой итерации. Такая комбинация позволяет сократить время поиска равновесной плотности без потери качества решения.\\

Важно отметить, что несмотря на то, что базис был построен на основе флюидов с параметрами азота, удается вычислить равновесную плотность флюида, информации о котором в базисе не было, VF-DFT хорошо справился с задачей вычисления равновесной плотности аргона. Из преимуществ VF-DFT то, что для его работы не нужно вычислять вариацию свободной энергии, а время поиска решения значительно меньше, чем у DFT с методом простой итерации, но качество решения получается незначительно хуже того, что дает классический DFT с методом простой итерации и зависит от выбора базисных функций. Чтобы сохранить преимущество в скорости и улучшить качество получаемого решения, в работе рассматривается гибридный алгоритм H-DFT на основе стохастических методов оптимизации и метода простой итерации. Комбинирование подходов позволило ускорить в несколько раз вычисления равновесной плотности без потери качества.\\

Разработанные подходы были применены к задаче нахождения равновесной плотности флюида в поре 3.6 нм. Рассматривались флюиды азот и аргон, стенки поры ---~углерод, геометрия задачи ---~планарная, то есть стенки поры представляются как две параллельные пластины. В качестве потенциала взаимодействия флюида со стенкой был взят потенциал Стилла 10-4-3. Для описания флюид-флюид взаимодействия использовался подход фундаментальной теории меры (Fundamental Measure Theory --- FMT) для короткодействующего отталкивания, разработанный Розенфельдом \cite{Roth2010FundamentalReview,Rosenfeld1989Free-energyFreezing}, и схема WCA \cite{Weeks1971RoleLiquids} для учета диполь-дипольного взаимодействия. Исследуемый флюид и поверхность, азот, аргон и карбон, достаточно хорошо изучены как экспериментально, так и с помощью классического DFT. Выбор именно такого флюида и поверхности обусловлен простотой их моделирования и наличием большого количества данных для валидации результатов разработанных методов.\\

В разделе \ref{sec:DFT} будет приведена справка по теории функционала плотности и методе простой итерации, в разделе \ref{sec:VF-DFT} будет приведено описание метода анализа главных компонент, методов оптимизации, которые использовались в данной работе, и подробнее рассмотрены преимущества подходов VF-DFT и H-DFT. Затем, в разделе \ref{sec:results_vfdft} будут рассмотрены задачи поиска равновесной плотности азота и аргона и валидация полученных решений с классическим подходом DFT с методом простой итерации. В разделе \ref{sec:psd} будет рассмотрен один из классических подходов для восстановления распределения пор по размерам, а в разделе \ref{sec:results_psd} буду приведены результаты работы этого подхода для решения нескольких модельных задач. \\


\section{Теория функционала плотности}\label{sec:DFT}
\renewcommand{\theequation}{\arabic{section}.\arabic{equation}}
\addetoc{section}{Density Functional Theory}

Будем рассматривать систему с заданными температурой $T$, объемом $V$, химическим потенциалом $\mu$. Согласно второму началу термодинамики, для системы с внешним потенциалом $V_{ext}\left(\vec{r}\right)$
термодинамический потенциал большого канонического ансамбля $\Omega$ минимален в равновесном состоянии. Так как $\Omega$ потенциал связан со свободной
энергией Гельмгольца $F\left[\rho\left(\vec{r}\right)\right]$, то $\Omega$ потенциал также является функционалом плотности \cite{Aslyamov2017DensitySurfaces,Roth2010FundamentalReview,Wu2006DensityMaterials}:

\begin{equation}\label{eq:Omega}
    \Omega\left[\rho\left(\vec{r}\right)\right]=F\left[\rho\left(\vec{r}\right)\right]+\int d^3r \rho\left(\vec{r}\right)\left(V_{ext}\left(\vec{r}\right)-\mu\right),
\end{equation}
где $F\left[\rho\left(\vec{r}\right)\right]$ --- энергия Гельмгольца, $\rho\left(\vec{r}\right)$ --- плотность флюида в точке $\vec{r}$, $V_{ext}$ --- внешний потенциал (потенциал стенки поры в нашем случае), $\mu$ --- химический потенциал флюида; температура, объем системы и химический потенциал считаются фиксированными.\\

Из условия экстремальности функционала следует, что его вариация должна обращаться в ноль в равновесии \cite{Wu2006DensityMaterials}

\begin{equation}\label{eq:variation}
    \dfrac{\delta\Omega\left[\rho\left(\vec{r}\right)\right]}{\delta \rho\left(\vec{r}\right)} = 
    \dfrac{\delta F\left[\rho\left(\vec{r}\right)\right]}{\delta \rho\left(\vec{r}\right)} + V_{ext}\left(\vec{r}\right) - \mu = 0. 
\end{equation}

Решением уравнения \eqref{eq:variation} является равновесная плотность, чтобы решить это уравнение необходимо знать вид энергии Гельмгольца. Несмотря на то, что формализм свободной энергии является точным, свободная энергия неизвестна для большинства систем. Задача вычисления свободной энергии Гельмгольца равносильна задаче вычисления статистической сумм системы, поэтому для решения прикладных задач часто разбивают свободную энергию на идеальную часть и поправки к ней, которые учитывают различного рода взаимодействия \cite{Wu2006DensityMaterials}:
\begin{align}
    F\left[\rho\right] &=  F^{id}\left[\rho\right]+F^{ex}\left[\rho\right]\label{eq:sum_ener}\\
    F^{id}\left[\rho\right] &= k_B T\int d\vec{r}\,\rho\left(\vec{r}\right)\left(\ln{{(\Lambda}^3 \rho\left(\vec{r}\right))}-1\right)\label{eq:id_ener}.
\end{align}

Выражение \eqref{eq:id_ener} учитывает свободную энергию флюида, как если бы он был идеальным газом,  где $k_B$ --- константа Больцмана, $T$ --- температура, $\Lambda = \frac{h}{\sqrt{2\pi mT}}$ --- тепловая длина волны де-Бройля, $h$ --- постоянная Планка, $m$ --- масса молекулы газа.

Слагаемое $F^{ex}$ содержит в себе информацию о взаимодействии молекул и именно оно определяет физику молекул флюида. В общем виде его можно представить через функции Майера через разложение по плотности следующим образом:

\begin{equation}\label{eq:excees}
    F^{ex}\left[\rho\left(\vec{r}\right)\right] = -\dfrac{k_{B}T}{2}\int d\vec{r}_1 \int d\vec{r}_2\, \rho\left(\vec{r}_1\right)\rho\left(\vec{r}_2 \right) f\left(r_{12}\right) + \mathcal{O}\left(\rho^3\right),
\end{equation}
здесь $r_{12} = \vert\vec{r}_1 - \vec{r}_2 \vert$, $f\left(r\right)$ --- функция Майера, которая выражается через потенциал парного межмолекулярного взаимодействия $U\left(\vec{r}_1, \vec{r}_2\right)$
\begin{equation}
    f\left(r_{12}\right) = \exp{\left(-\frac{U\left(r_{12}\right)}{k_{B}T}\right)} - 1.
\end{equation}

Парным потенциалом взаимодействия простых молекул является потенциал Леннарда--Джонса, но так как этот потенциал обладает особенностью, его проблематично напрямую учесть в выражении для свободной энергии. Поэтому в работе в дальнейшем будет идти разделение добавочной части свободной энергии на два слагаемых, сумма которых обеспечит корректный учет взаимодействия простых молекул через потенциал Леннарда--Джонса:
\begin{equation}
    F^{ex} = F^{HS} + F^{att}.
\end{equation}
Первое слагаемое отвечает за короткодействующее отталкивание, а второе за дальнодействующее притяжение. Каждое из слагаемых будет подробнее рассмотрено ниже.
\subsection{Короткодействующее отталкивание}
\addetoc{subsection}{Hard sphere repulsion}

Для учета взаимодействия двух атомов на расстояниях порядка радиуса в теории функционала плотности атомы представляют как абсолютно жесткие сферы, таким образом каждая частица имеет объем, недоступный для других атомов \cite{chandler1983van}. Корректное описание короткодействующего взаимодействия позволяет исследовать такое сложное явление, как фазовые переходы.

В теории функционала плотности для описания короткодействующего отталкивания популярен подход, основанный на теории меры (Fundamental measure theory --- FMT), который разработал Розенфельд \cite{Rosenfeld1989Free-energyFreezing}. Подход Розенфельда обладает преимущетвом в универсальности модели для описания различных явлений и позволяет рассматривать систему твердых сфер: в свободном объеме (3D); флюид жестких дисков, зажатый между двумя плоскостями (2D); флюид жестких стержней в цилиндрической поре (1D), и в кавернах (0D). Эта замкнутая, математически строгая теория применима как к однокомпонентному флюиду, так и к полидисперсной смеси флюидов в разных фазах, состоящей из несферических частиц. В данной работе будет рассматриваться модель жестких сфер для однокомпонентного флюида. В данном случае функция Майера имеет наглядную геометрическую интерпретацию ($R$ --- здесь и в дальнейшем радиус частицы, если это не оговорено):
\begin{equation}\label{eq:fmauer}
    U\left(r_{12}\right) = \left\{
    \begin{aligned}
        &\infty & &r_{12} \le 2R\\
        &0 & &\text{иначе}, 
    \end{aligned}
    \right.
    \qquad\Longrightarrow\qquad
    f\left(r_{12}\right) = \left\{
    \begin{aligned}
        &-1 & &r_{12} \le 2R\\
        &0 & &\text{иначе}, 
    \end{aligned}
    \right.
\end{equation}

Таким образом, функции Майера позволяет учитывать объем, недоступный для центров различных частиц. Так как выражение \eqref{eq:excees} с учетом вида функции Майера для жестких сфер \eqref{eq:fmauer} является сложным для прямого аналитического решения, функцию Майера представляют в виде разложение на более простые слагаемые.

Объем двух тел $i$ и $j$ можно представить следующим образом:
\begin{equation}\label{eq:measures}
    V_{i+j} = V_i + R_i S_j + R_j S_i + V_j ,
\end{equation}
где $V_k ,\, S_k ,\, R_k$ --- объем, площадь поверхности и радиус кривизны соответственно. С учетом этого, Розенфельд \cite{Rosenfeld1989Free-energyFreezing} отметил, что функцию Майера можно представить в виде следующей суммы:
\begin{equation}\label{eq:f_deconvolution}
    -f\left(r_{ij}\right) = \omega_3^i \otimes\omega_0^j + \omega_0^i \otimes\omega_3^j + \omega_2^i \otimes\omega_1^j + \omega_1^i \otimes\omega_2^j - \bm{\omega}_2^i \otimes\bm{\omega}_1^j - \bm{\omega}_1^i \otimes\bm{\omega}_2^j
\end{equation}
с весовыми функциями
\begin{align}
    \omega_3 \left(\bm{r}\right) &= \Theta\left(R - r\right),\\
    \omega_2 \left(\bm{r}\right) &= \delta\left(R - r\right),\\
    \bm{\omega}_2 \left(\bm{r}\right) &= \dfrac{\bm{r}}{r} \delta\left(R - r\right),
\end{align}
и $\omega_1 \left(\bm{r}\right) = \omega_2 \left(\bm{r}\right)/\left(4\pi R\right)$, $\omega_0 \left(\bm{r}\right) = \omega_2 \left(\bm{r}\right)/\left(4\pi R^2\right)$, $\bm{\omega}_1 \left(\bm{r}\right) = \bm{\omega}_2 \left(\bm{r}\right)/\left(4\pi R\right)$. Здесь $\Theta\left(r\right)$ --- функция Хэвисайда, а $\delta\left(r\right)$ --- функция Дирака, символом $\otimes$ обозначена свертка функций ($\alpha,\,\beta = 1,2,3$)
\begin{equation}
    \omega^i_\alpha \otimes\omega^j_\beta \left(\bm{r} = \bm{r}_i - \bm{r}_j \right) = \int d\bm{r}'\, \omega^i_\alpha \left( \bm{r}' - \bm{r}_i\right) \omega^j_\alpha \left( \bm{r}' - \bm{r}_j\right).
\end{equation}

Связь между разложением функции Майера \eqref{eq:f_deconvolution} и выражением для объема двух тел \eqref{eq:measures} становится очевидной, если проинтегрировать весовые функции $\omega^i_\alpha \left(\bm{r}\right)$, для $\alpha = 3$ получается объем частицы $V_i$, $S_i$ для $\alpha = 2$, $R_i$ для $\alpha = 1$ и $1$ для $\alpha = 0$, что является основными геометрическими мерами для сферической частицы в трехмерном пространстве (Fundamental Measures) \cite{Rosenfeld1989Free-energyFreezing,Roth2010FundamentalReview}.

С помощью весовых функций $\omega_\alpha$ можно перейти к взвешенным плотностям $n_\alpha$, через которые в дальнейшем будет выражена свободная энергия Гельмгольца
\begin{equation}\label{eq:weighted_dens}
    n_\alpha \left(\vec{r}\right)=\int d^3r^\prime \rho\left(\vec{r}^\prime\right)\omega_\alpha\left(\vec{r}-\vec{r}^\prime\right).
\end{equation}

Если рассмотреть предел в выражении для свободной энергии для короткодействующего отталкивания \eqref{eq:excees} при плотности стремящейся к нулю, то с учетом \eqref{eq:weighted_dens} можно получить точное выражение, в дальнейшем это будет учтено для явной записи свободной энергии взаимодействия жестких сфер
\begin{equation}\label{eq:lim_hardsphere}
    \lim\limits_{\rho\rightarrow 0} F^{HS}\left[\rho\left(\bm{r}\right)\right] = k_B T \int d\bm{r} \, \{ n_0 \left(\bm{r}\right) n_3 \left(\bm{r}\right) + n_1 \left(\bm{r}\right) n_2 \left(\bm{r}\right) - \bm{n}_1 \left(\bm{r}\right) \cdot\bm{n}_2 \left(\bm{r}\right)\}
\end{equation}

Получив предельное выражение \eqref{eq:lim_hardsphere} Розенфельд предположил, что и для высоких плотностей вид свободной энергии должен зависеть только от взвешенных плотностей \cite{Rosenfeld1989Free-energyFreezing,Roth2010FundamentalReview}
\begin{equation}
    F^{HS}\left[\rho\left(\bm{r}\right)\right] = k_B T \int d\bm{r}\, \Phi^{RF}\left[n_\alpha \left(\bm{r}\right)\right]
\end{equation}

Из анализа размерности подинтегральной функции можно предположить, что ее вид должен быть следующим:
\begin{equation}\label{eq:ansatz}
    \Phi^{RF} = g_1 \left(n_3\right) n_0 + g_2 \left(n_3\right) n_1 n_2 + g_3 \left(n_3\right) \bm{n}_1 \cdot \bm{n}_2 + g_4 \left(n_3\right) n_2^3 + g_5 \left(n_3\right) n_2 \bm{n}_2 \cdot \bm{n}_2
\end{equation}

Каждое слагаемое в выражении \eqref{eq:ansatz} имеет размерность (длина$^{-3}$), функции $g_1, \dots ,g_5$ являются неизвестными. Однако с учетом выражения \eqref{eq:lim_hardsphere} можно сделать вывод о поведении функций $g_1, \dots ,g_5$ при низки плотностях $g_1 = n_3 + n_3^2 /2 + \mathcal{O}\left(n_3^3\right)$, $g_2 = 1 + n_3 + \mathcal{O}\left(n_3^2 \right)$, $g_3 = -1 - n_3 + \mathcal{O}\left(n_3^2 \right)$, $g_4 = 1/24\pi  + \mathcal{O}\left(n_3 \right)$, $g_5 = -3/24\pi  + \mathcal{O}\left(n_3 \right)$. Отсюда можно сделать вывод о взаимосвязи функций:
\begin{align}
    g_5 &= -3g_4,\nonumber\\
    g_2 &= -g_3. \nonumber
\end{align}

С учетом этой взаимосвязи выражение для плотности свободной энергии будет иметь вид
\begin{equation}
    \Phi^{RF} = g_1 \left(n_3\right) n_0 + g_2\left( \left(n_3\right) n_1 n_2 -  \left(n_3\right) \bm{n}_1 \cdot \bm{n}_2\right) + g_4\left( \left(n_3\right) n_2^3 - 3 \left(n_3\right) n_2 \bm{n}_2 \cdot \bm{n}_2\right).
\end{equation}

Функции $g_1, g_2, g_4$ могут быть определены из требований выполнения термодинамических соотношений для функционала $\Phi^{RF}$. В работе \cite{Rosenfeld1989Free-energyFreezing} Розенфельд использовал соотношение из теории масштабированных частиц (Scaled Particle Theory --- SPT):
\begin{equation}\label{eq:lim_chempot}
    \lim\limits_{R\rightarrow\infty} \dfrac{\mu^{HS}}{k_B T V} = \dfrac{p}{k_B T}
\end{equation}
где $V$ --- объем сферической частицы с радиусом $R$, $p$ --- давление, $\mu^{HS}$ --- химический потенциал от взаимодействия жестких сфер. Химический потенциал от взаимодействия жестких сфер может быть найден варьированием функционала $\Phi^{RF}$ по плотности (из определения химического потенциала по свободной энергии):
\begin{equation}
    \dfrac{\mu^{HS}}{k_B T} = \dfrac{\delta F^{ex}\left[\rho\left(\bm{r}\right)\right]}{\delta \rho\left(\bm{r}\right)} = \sum\limits_\alpha \dfrac{\partial \Phi^{RF}}{\partial n_\alpha}\dfrac{\partial n_\alpha}{\partial \rho}.
\end{equation}
Учитывая геометрический смысл весовых функций $\partial n_3/\partial \rho = \int d\bm{r}\,\omega_3\left(\bm{r}\right) \equiv V$, $\partial n_2/\partial \rho \equiv S$, $\partial n_1/\partial \rho \equiv R$, $\partial n_0/\partial \rho \equiv 1$ (производные с векторными взвешенными плотностями равны нулю), предельное выражение \eqref{eq:lim_chempot} получается следующим:
\begin{equation}
    \lim\limits_{R\rightarrow\infty} \dfrac{\mu^{HS}}{k_B T V} =\dfrac{\partial \Phi^{RF}}{\partial\rho} = \dfrac{p}{k_B T}.
\end{equation}

С другой стороны можно воспользоваться тождеством для большого канонического ансамбля для флюида в свободном объеме (bulk) $\Omega_{bulk} = -pV$, а плотность свободной энергии большого канонического ансамбля $\Omega_{bulk}/V = \Phi + f_{id} + \mu\rho_{bulk}$. Таким образом, получается система уравнений:
\begin{equation}
    \dfrac{\partial \Phi^{RF}}{\partial n_3} = -\Phi^{RF} + \sum\limits_\alpha \dfrac{\partial\Phi^{RF}}{\partial n_\alpha} n_\alpha + n_0.
\end{equation}
Собирая все члены, содержащие $n_0$ получается дифференциальное уравнение для $g_1$ ('~обозначает производную по $n_3$)
\begin{equation}
    g'_1 \left(n_3\right)\left(1-n_3\right) = 1\qquad\Longrightarrow\qquad g_1 \left(n_3\right) = const_1 - \ln{\left(1 - n_3\right)} 
\end{equation}
Аналогично и для других неизвестных функций $g_2 ,\, g_4$
\begin{equation}
    g'_2 \left(n_3\right)\left(1-n_3\right) = g_2 \left(n_3\right)\qquad\Longrightarrow\qquad g_2 \left(n_3\right) = \dfrac{const_2}{1- n_3} 
\end{equation}
\begin{equation}
    g'_4 \left(n_3\right)\left(1-n_3\right) = 2g_4 \left(n_3\right)\qquad\Longrightarrow\qquad g_4 \left(n_3\right) = \dfrac{const_3}{\left(1-n_3\right)} 
\end{equation}
Константы интегрирования можно подобрать, если подставить полученные выражения в выражение для предела свободной энергии от взаимодействия жестких сфер при плотности, стремящейся к нулю \eqref{eq:lim_hardsphere}. Таким образом, функционал Розенфельда \cite{Rosenfeld1989Free-energyFreezing,Roth2010FundamentalReview} окончательно можо записать в следующем виде:
\begin{equation}
    \Phi^{RF} = -n_0 \ln{\left(1-n_3\right)} + \frac{n_1 n_2- \vec{n_1}\cdot\vec{n_2}}{1-n_3} + \frac{n_2^3-3n_2 \vec{n_2}\cdot\vec{n_2}}{24\pi\left(1-n_3\right)^2}.
\end{equation}

Важно отметить, что функционал Розенфельда согласуется с уравнением состояние Перкус–Йевика и прямой корреляционной функцией Перкус–Йевика.
\begin{figure}[htb!]
    \centering
    \includegraphics[page=1]{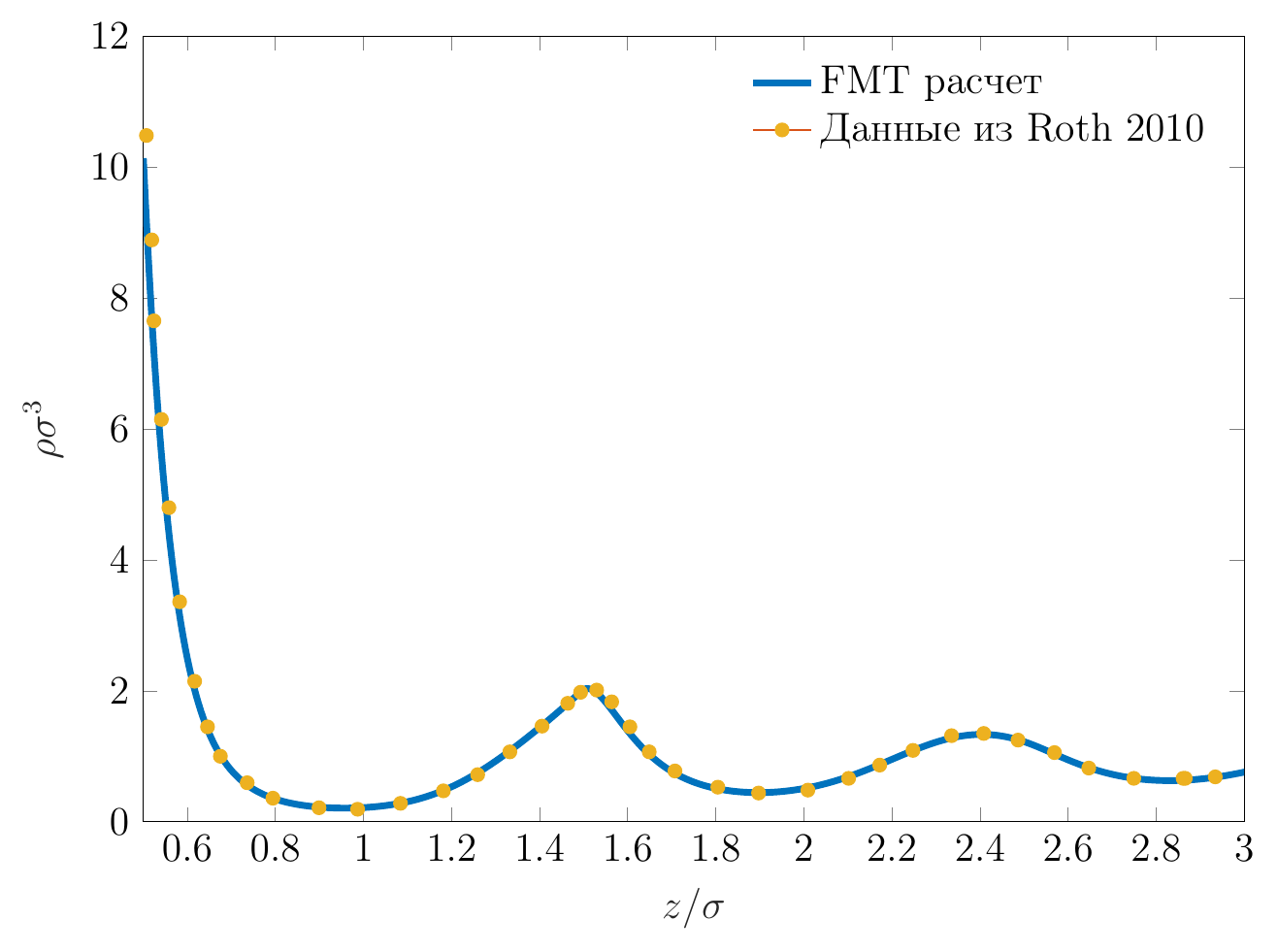}
    \caption{Распределения профиля плотности жестких сфер в поре из абсолютно жестких стенок при $\eta = 0.4783$ \cite{Roth2010FundamentalReview}}
    \label{fig:hard_sphere}
\end{figure} 
На рис.~\ref{fig:hard_sphere} приведен профиль плотности флюида жестких сфер в поре с абсолютно жесткими стенками. Результаты расчета свалидированы с результатами из \cite{Roth2010FundamentalReview} при значении плотности упаковки флюида $\eta = 4\pi R^3 \rho_{bulk} /3 = 0.4783$. Внешний потенциал в такой поре задается следующим образом:
\begin{equation}
    V_{ext}\left(r\right) = \left\{
    \begin{aligned}
        &\infty & &r \ge H,\,r \le 0\\
        &0 & &\text{иначе}, 
    \end{aligned}
    \right.
\end{equation}

Таким образом, в данной работе учитывается короткодействующее отталкивание между молекулами флюида в свободной энергии Гельмгольца через потенциал Розенфельда. 
\FloatBarrier
\subsection{Диполь-дипольное взаимодействие}
\begin{figure}[htb!]
    \centering
    \includegraphics[scale=1]{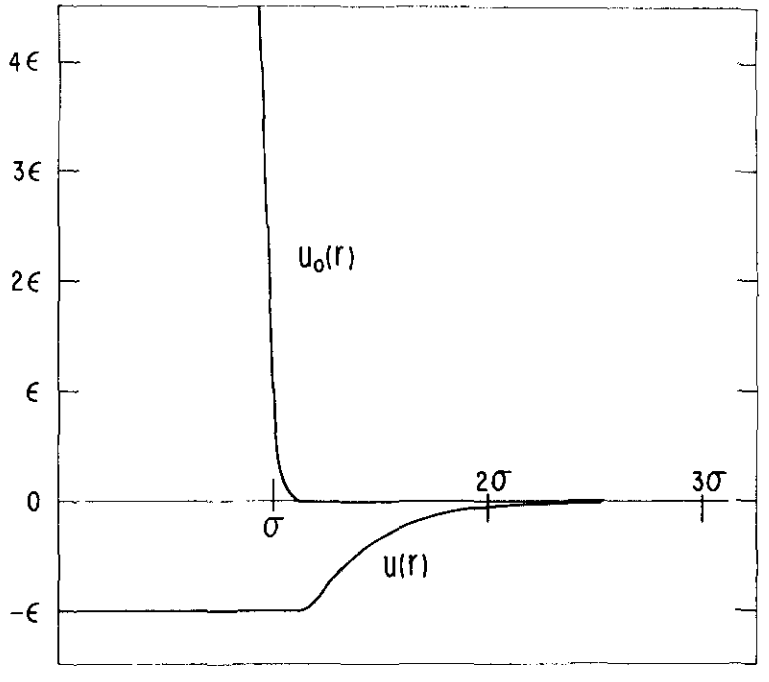}
    \caption{Схематичная иллюстрация модифицированного потенциала Леннарда--Джонса по схеме WCA из \cite{Weeks1971RoleLiquids} (на рисунке $U \left(\bm{r}\right) = U_{att} \left(\bm{r}\right)$)}
    \label{fig:wca}
\end{figure}

\addetoc{subsection}{Attraction interaction}
Большинство версий теорий функционала плотности учитывают диполь-дипольное взаимодействие между атомами и молекулами через приближение среднего поля. Несмотря на то, что использование приближения среднего поля накладывает ряд ограничений на исследуемую систему и обладает меньшей общностью, существуют модификации функционала Гельмгольца, которые помогают обойти это препятствие. В данной работе диполь-дипольное взаимодействие рассматривается как возмущение над потенциалом короткодействующего отталкивания по схеме, которую разработали Weeks, Chandler, Anderson (WCA) \cite{Weeks1971RoleLiquids}

\begin{equation}
    U\left(\bm{r}\right) = U_0 \left(\bm{r}\right) + U_{att} \left(\bm{r}\right)
\end{equation}
где $U_0 \left(\bm{r}\right)$ --- потенциал взаимодействия жестких сфер, $U_{att} \left(\bm{r}\right)$ --- возмущение. Из работы \cite{Weeks1971RoleLiquids} потенциал возмущения представляется следующим образом: 
\begin{equation}\label{eq:U_att}
    U_{att}\left(r\right)= \left\{
    \begin{matrix}
        -\varepsilon_{ff}&r<\lambda\\
        U_{LJ}&\lambda<r<r_{cut}\\
        0&r>r_{cut}\\
    \end{matrix}\right.
\end{equation}

\begin{equation}
    U_{LJ}=4\varepsilon_{ff}\left(\left(\frac{\sigma_{ff}}{r}\right)^{12}- \left(\frac{\sigma_{ff}}{r}\right)^6\right)
\end{equation}
$\lambda = 2^{1/6} \sigma_{ff}$ --- положение минимума потенциала Леннарда--Джонса, $\sigma_{ff}$ --- диаметр частицы флюида, $\varepsilon_{ff}$ --- значение потенциала Леннарда--Джонса в минимуме, $r_{cut}$ --- расстояние действия потенциала. На рис.~\ref{fig:wca} схематично проиллюстрирован потенциал взаимодействия двух молекул \cite{Weeks1971RoleLiquids}.

\begin{figure}[htb!]
    \centering
    \includegraphics[page=1]{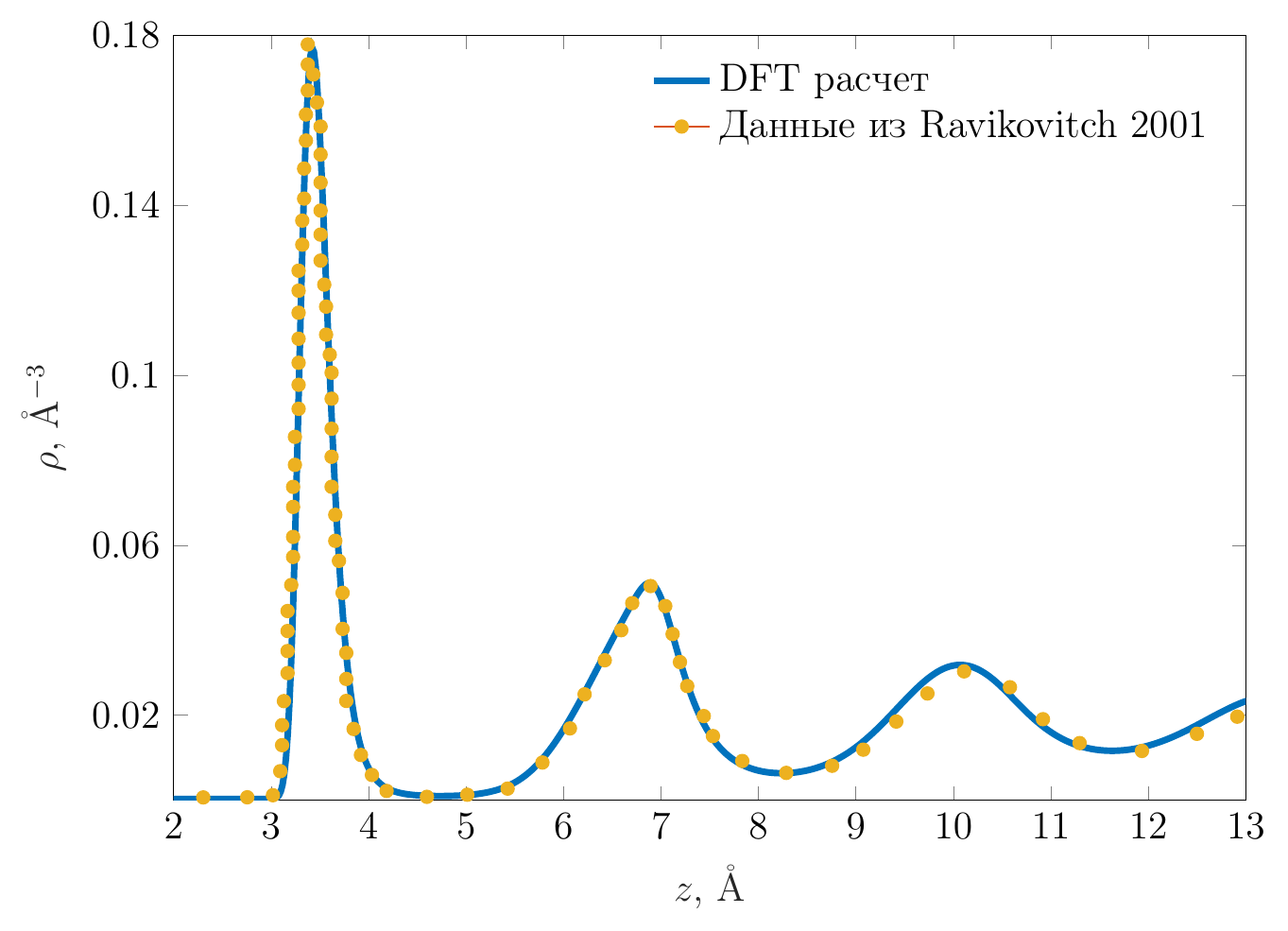}
    \caption{Распределение профиля плотности азота в поре, шириной $H_{cc} = 10\sigma_{ff}$ при относительном давлении в свободном объеме $P/P_0 = 0.7$ \cite{Ravikovitch2001DensityNanopores}}
    \label{fig:attraction}
\end{figure}

В итоге, в приближении среднего поля свободная энергия диполь-дипольного взаимодействия записывается следующим образом:
\begin{equation}
    F^{att}=\frac{k_B T}{2}\iint d^3 r_1d^3 r_2 \,\rho\left(\bm{r}_1\right)\rho\left(\bm{r}_2\right)U_{att}\left(r\right).
\end{equation}
На рис.~\ref{fig:attraction} приведен график распределения плотности азота в поре, который был посчитан с помощью приведенной в данной работе модели DFT и свалидирован с результатами из статьи \cite{Ravikovitch2001DensityNanopores}.

\FloatBarrier
\subsection{Метод простой итерации (МПИ)}
\addetoc{subsection}{Piccard iteration}

Чтобы найти равновесную плотность флюида, которая бы минимизировала свобоную энергию большого канонического ансамбля \eqref{eq:Omega}, методом просто итерации, необходимо знать вариацию омега-потенциала \cite{Roth2010FundamentalReview}. С учетом выражения \eqref{eq:sum_ener} и \eqref{eq:id_ener} равенство нулю вариации свободной энергии большого канонического ансамбля запишется в виде:

\begin{equation} \label{eq:omega_variation}
    k_B T\ln{\Lambda^3\rho\left(\vec{r}\right)} + \dfrac{\delta F^{HS}\left[\rho\left(\vec{r}\right)\right]}{\delta \rho\left(\vec{r}\right)} + \dfrac{\delta F^{att}\left[\rho\left(\vec{r}\right)\right]}{\delta \rho\left(\vec{r}\right)} + V_{ext}\left(\vec{r}\right) - \mu = 0.
\end{equation}
Аналогично энергии Гельмгольца, химический потенциал флюида можно также условно представить как сумму трех слагаемых
\begin{align}
    \mu&= \mu^{id}+\mu^{HS}+\mu^{att},\\
    \mu^{id}\left(\rho\right)&=k_B T \left(\ln{\Lambda^3 \rho^{bulk}}\right)=const,\\
    \mu^{HS}\left(\rho\right) &=k_B T\left(\sum{\frac{\partial\Phi}{\partial n_\alpha}\frac{\partial n_\alpha }{\partial\rho}}\right)=const,\\
    \mu^{att}\left(\rho\right)&=\rho^{bulk}\int{d^3r\, U_{att}\left(r\right)}=const. \label{eq:chem_att}
\end{align}
Под $\rho^{bulk} =\Lambda^{-3}\exp{\left(\mu_{id}/k_B T\right)} $ подразумевается плотность флюида в свободном объеме. В уравнении \eqref{eq:omega_variation} плотность входит явно в первом слагаемом и ее можно выразить
\begin{equation}\label{eq:density}
    \hat{\rho}\left(\vec{r}\right) = \rho^{bulk}\exp{\left\lbrace \frac{1}{k_B T}\left( -\dfrac{\delta F^{HS}\left[\rho\left(\vec{r}\right)\right]}{\delta \rho\left(\vec{r}\right)} - \dfrac{\delta F^{att}\left[\rho\left(\vec{r}\right)\right]}{\delta \rho\left(\vec{r}\right)} - V_{ext}\left(\vec{r}\right) + \mu^{HS} + \mu^{att}\right)\right\rbrace}.
\end{equation}

Таким образом, получается выражение для равновесной плотности флюида, однако плотность неявно входит в правую часть \eqref{eq:density}, через вариацию энергии Гельмгольца. Для того, чтобы решить это нелинейное уравнение обычно используют метод простой итерации с релаксацией. В качестве начального приближения выбирается плотность флюида в свободном объеме $\rho^{j=0} = \rho^{bulk}$ (от выбора начального приближения зависит устойчивость и скорость сходимости метода). Решение на следующем шаге итерации определяется как сумма решений на предыдущей итерации и на текущей
\begin{equation}
    \rho^{j+1} = \left(1-\gamma\right)\rho^{j} + \gamma\hat{\rho}^j
\end{equation}

Параметр $\gamma\in\left[0,1\right]$ определяет степень доверия новому решению на последующей итерации, чем он больше, тем быстрее сходится метод, но при слишком больших значениях решение будет неустойчиво и сходимости не будет. Метод простой итерации ---~весьма надежен и с его помощью, рано или поздно решение будет найдено, вот только чаще всего работает он достаточно долго. Кроме того, для работы этого метода необходимо вычислять вариацию свободной энергии Гельмгольца и это накладывает некоторые ограничения на вид энергии Гельмгольца. Усложнение модели, например учет гетерогенности поверхности \cite{Aslyamov2017DensitySurfaces, Jagiello20132D-NLDFTCorrugation, Neimark2009QuenchedCarbons}, или учет кулоновского взаимодействия, приводит к изменению вида энергии Гельмгольца, что влечет изменение соответствующих вариаций. Становится необходимым изменение практически каждого слагаемого в \eqref{eq:density} и добавление новых. В работе \cite{Aslyamov2017DensitySurfaces} наглядно видно то, как учет неоднородности поверхности может сильно повлиять на выражения для вариаций, и какие сложности могут при этом возникнуть в вычислении равновесной плотности, несмотря на то, что рассматриваемый флюид является простым для моделирования. 

\subsection{Геометрия поры}
\addetoc{subsection}{The geometry of the pores}

В работе рассматривается планарная геометрия поры (стенки поры ---~параллельные пластины). В такой постановке, задача становится симметричной по двум направлениям ($x,\ y$ условно) и плотность будет зависеть только от одной координаты $\rho\left(\vec{r}\right) = \rho\left(z\right)$. С учетом такой геометрии можно проинтегрировать по переменным $x,\, y$ выражения для взвешенных плотностей:
\begin{equation}
    n_\alpha\left(z\right)=\int d^3 z^\prime\ \rho\left(z^\prime\right)\omega_\alpha\left(z-z^\prime\right),  
\end{equation}
где к весовым функциям после интегрирования: $\omega_3 = \pi\left(R^2 - z^2 \right) \Theta\left(R - \vert z\vert\right)$, $\omega_2 = 2\pi R \Theta\left(R - \vert z\vert\right)$, $\bm{\omega}_2 = 2\pi z\Theta\left(R - \vert z\vert\right)$, $\omega_1 = \omega_2 / 4\pi R$, $\bm{\omega}_1 = \bm{\omega}_2 / 4\pi R$, $\omega_0 = \omega_2/4\pi R^2$. Также можно выписать вариацию для свободной энергии взаимодействия жестких сфер
\begin{equation}
    \dfrac{\delta F^{HS}\left[\rho\left(z\right)\right]}{\delta\rho\left(z\right)} = k_B T \sum\limits_\alpha \int dz'\, \dfrac{\partial \Phi^{RF}\{n_\alpha \left(z'\right)\}}{\partial n_\alpha \left(z'\right)}\dfrac{\delta n_\alpha \left(z'\right)}{\delta\rho\left(z' \right)},
\end{equation}
\begin{equation}
    \dfrac{\delta n_\alpha \left(z'\right)}{\delta\rho\left(z' \right)} = \frac{\delta}{\delta\rho\left(z\right)}\int{dz^{\prime\prime}\rho\left(z^{\prime\prime}\right)\omega_\alpha\left(z^\prime-\ z^{\prime\prime}\right)}=\omega_\alpha\left(z^\prime-z\right).
\end{equation}

Также можно упростить выражения для вариации энергии от диполь-дипольного взаимодействия
\begin{multline}
    F^{att}=\frac{k_B T}{2}\int dz_1 dz_2\,\rho\left(z_1\right)\rho\left(z_2\right)\int dx_1 dx_2 dy_1 dy_2 \,U_{att}\left(r\right) =\\= \frac{k_B T}{2}\iint dz_1 dz_2\,\rho\left(z_1\right)\rho\left(z_2\right)G\left(z_1,z_2\right), 
\end{multline}
где с учетом \eqref{eq:U_att} при $\vert \Delta z\vert \le \lambda$
\begin{multline}
    G\left(z_1,z_2\right) = 2\pi\left\{\int\limits_{0}^{\sqrt{\lambda^2-\Delta z^2}}{-\varepsilon_{ff}rdr+\ \int\limits_{\sqrt{\lambda^2-\Delta z^2}}^{\sqrt{r_{cut}^2-\Delta z^2}}{dr\ 4\varepsilon_{ff}r\left[\frac{\sigma_{ff}^{12}}{\left(\Delta z^2+r^2\right)^6}-\ \frac{\sigma_{ff}^6}{\left(\Delta z^2+r^2\right)^3}\right]}}\right\} =\\=
    -\pi\varepsilon_{ff}\left(\lambda^2-\Delta z^2\right)+\frac{4}{5}\pi\varepsilon_{ff}\sigma_{ff}^{12}\left(\frac{1}{\lambda^{10}}-\ \frac{1}{r_{cut}^{10}}\right)-2\pi\varepsilon_{ff}\sigma_{ff}^6\left(\frac{1}{\lambda^4}-\ \frac{1}{r_{cut}^4}\right),
\end{multline}
при $\vert \Delta z\vert > \lambda$ и $\vert \Delta z\vert \le r_{cut}$
\begin{multline}
    G\left(z_1,z_2\right) = 2\pi\int\limits_{0}^{\sqrt{r_{cut}^2-\Delta z^2}}{dr\ 4\varepsilon_{ff}r\left[\frac{\sigma_{ff}^{12}}{\left(\Delta z^2+r^2\right)^6}-\ \frac{\sigma_{ff}^6}{\left(\Delta z^2+r^2\right)^3}\right]} =\\=
    \frac{4}{5}\pi\varepsilon_{ff}\sigma_{ff}^{12}\left(\frac{1}{{\Delta z}^{10}}-\ \frac{1}{r_{cut}^{10}}\right)-2\pi\varepsilon_{ff}\sigma_{ff}^6\left(\frac{1}{{\Delta z}^4}-\ \frac{1}{r_{cut}^4}\right),
\end{multline}
а при $\vert \Delta z\vert > r_{cut}$, $G\left(z_1,z_2\right) = 0$. Таким образом, вариация энергии Гельмгольца от диполь-дипольного взаимодействия
\begin{equation}
    \dfrac{\delta F^{att}\left[\rho\left(z\right)\right]}{\delta \rho\left(z\right)} = \dfrac{k_B T}{2} \int dz'\, \rho\left(z'\right)G\left(z - z'\right)
\end{equation}

Стенки поры в работе рассматриваются углеродные и потенциал взаимодействия флюида со стенкой описывается потенциалом Стилла 10-4-3 \cite{steele1974interaction}
\begin{equation}
    V_{sf}\left(z\right)=2\pi\varepsilon_{sf}\rho_V\sigma_{sf}^3\Delta\left(\frac{2}{5}\left(\dfrac{\sigma_{sf}}{z}\right)^{10}-\left(\dfrac{\sigma_{sf}}{z}\right)^4-\ \dfrac{\sigma_{sf}^4}{3\Delta\left(0.61\Delta+z\right)^3}\right).
\end{equation}
Полный потенциал в поре от двух стенок:
\begin{equation}
    V_{ext}\left(z\right)= V_{sf}\left(z\right)+ V_{sf}\left(H_{cc}-z\right), 
\end{equation}
$\rho_V$ ---~плотность материала стенки $0.114 \text{ \AA}^{-3}$ в нашем случае для углерода, $\Delta$ ---~расстояние между слоями атомов углерода $3.35 \text{ \AA}$, $H_{cc}$ ---~диаметр поры рассчитанный между центрами атомов углерода противоположных стенок, ширина поры доступная для флюида $H=H_{cc}-2z_0+\sigma_{ff}$, где $z_0$ ---~глубина проникновения молекулы флюида в поверхность (значение $z$, при котором потенциал стенки обращается в ноль), в нашей работе мы использовали приближение $z_0 = 0.9\sigma_{sf}$.
\section{Безвариационный подход в теории функционала плотности (VF-DFT)}\label{sec:VF-DFT}
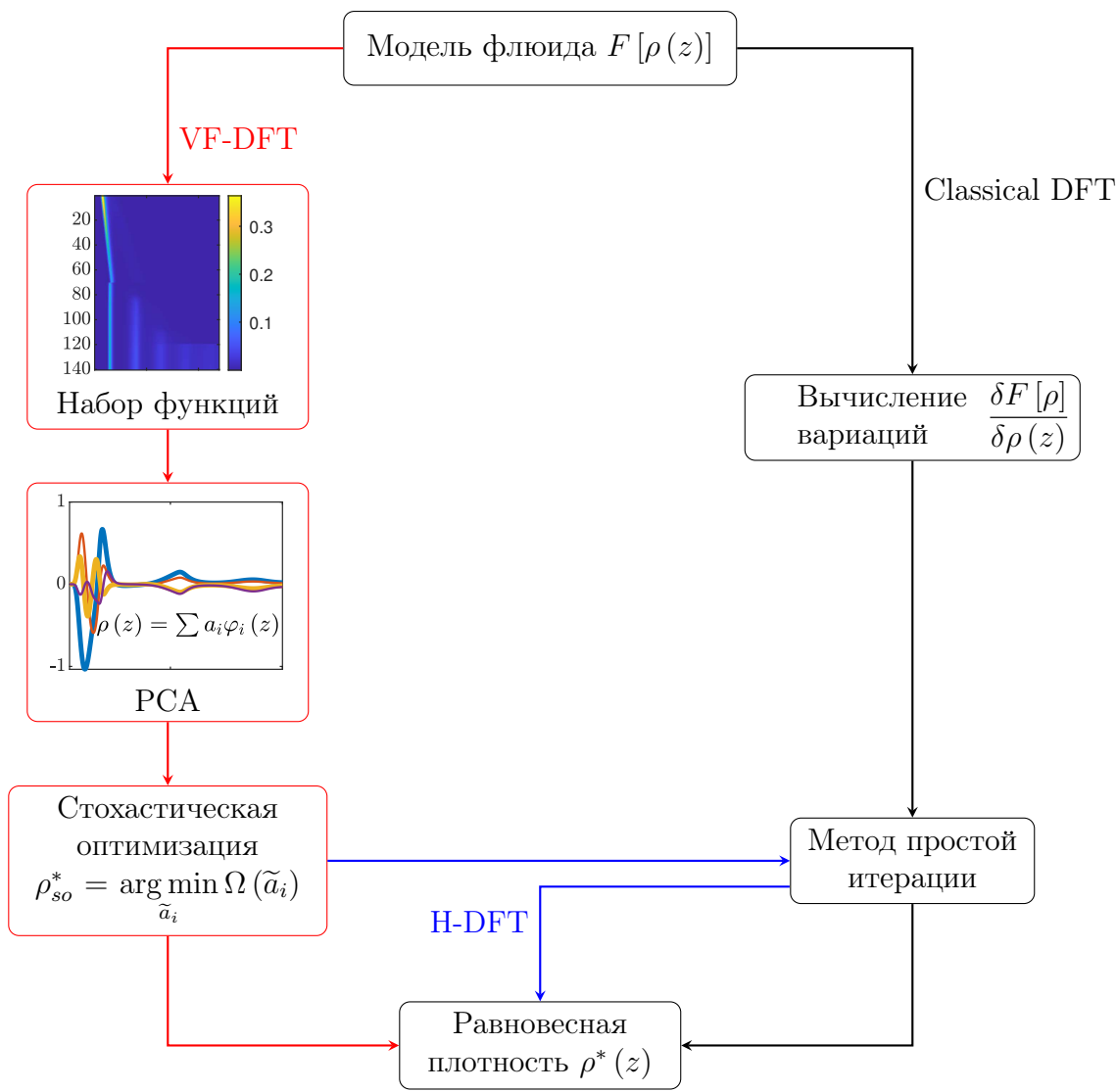
\begin{figure}[htb!]
\centering
\begin{tikzpicture}[node distance=3.5cm]

\node (start) [cdft,text width=5cm] {Модель флюида $F\left[\rho\left(z\right)\right]$};

\node (varcalc) [cdft,below of=start,xshift=5cm,minimum width=4.5cm,text width=3.1cm,align=left,yshift=-1.5cm] {Вычисление вариаций};
\node [empty,below of=start,xshift=6.5cm,yshift=-1.5cm] {$\dfrac{\delta F\left[\rho\right]}{\delta\rho\left(z\right)}$};
\node (picard) [cdft,below of=varcalc,yshift=-2.5cm,text width=3cm] {Метод простой итерации};

\node (basisfunc) [vfdft,below of=start,xshift=-5cm, text width=3.5cm] {
\includegraphics[scale=0.9]{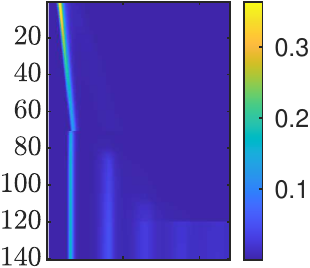}
Набор функций};
\node (PCA) [vfdft,below of=basisfunc,text width=3.5cm,yshift=-0.5cm] {
\includegraphics[scale=0.9]{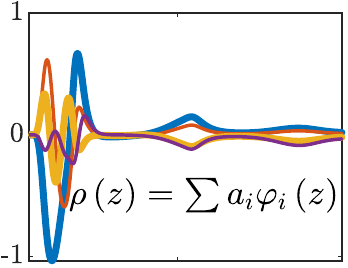}
PCA};
\node (stochopt) [vfdft,below of=PCA,text width=4cm] {Стохастическая оптимизация\\ $\rho^*_{so}=\argmin\limits_{\widetilde{a}_i}\Omega\left(\widetilde{a}_i\right)$};

\node (dens) [cdft,below of=start,yshift=-10cm] {Равновесная плотность $\rho^{*}\left(z\right)$};

\draw [arrow,red] (start) --(-5,0) -- node[anchor=north west, red] {VF-DFT} (basisfunc);
\draw [arrow] (start) --(5,0) -- node[anchor=south west] {Classical DFT} (varcalc);
\draw [arrow,red] (basisfunc) -- (PCA);
\draw [arrow,red] (PCA) -- (stochopt);
\draw [arrow] (varcalc) -- (picard);
\draw [arrow] (picard) |- (dens);
\draw [arrow,red] (stochopt) |- (dens);
\draw [arrow,blue] (stochopt) -- (picard);
\draw [arrow,blue] (picard)+(-1.65, -0.35)-- (0, -11.35) -- node[anchor=south east, blue] {H-DFT} (dens);
\end{tikzpicture}
\caption{Схематическое описания принципа работы классического подхода в DFT с методом простой итерации (черная линия), подхода Variation Free DFT (красная линия), гибридного подхода Hybrid DFT (красные и синии линии)}
\label{fig:TOC}
\end{figure}

\setcounter{equation}{0}
\addetoc{section}{Variation Free Density Functional Theory (VF-DFT)}
В отличие от классического DFT, который использует метод простой итерации для поиска равновесного решения, безвариационный подход в теории функционала плотности (Variation Free Density Functional Theory VF-DFT) использует стохастические методы оптимизации. На рис.~\ref{fig:TOC} схематично представлен принцип работы разработанных алгоритмов в сравнении с классическим методом. Применение безградиентных стохастических методов оптимизации, таких как генетический алгоритм (Genetic Algorithm --- GA) и метод роя частиц (Particle Swarm Optimization --- PSO), позволяет найти равновесную плотность $\rho^*\left(z\right)$ без вычисления вариаций и сократить время поиска решения. 
Отсутствие необходимости считать вариацию свободной энергии является главным преимуществом VF-DFT. Для разных систем вид свободной энергии Гельмгольца различен, следовательно, отличается и ее вариация. Для некоторых систем вычисление вариации не представляется возможным из-за сложной структуры свободной энергии системы. Например, в работе \cite{Aslyamov2017DensitySurfaces} от стандартной модели адсорбции газа на идеально гладкой поверхности перешли к рассмотрению адсорбции на гетерогенной поверхности, это повлекло сильные изменения в вариации энергии Гельмгольца. Для того, чтобы учитывать более сложную структуру молекул флюида, требуется усложнять и менять вид свободной энергии Гельмгольца. Для описания сложных молекул активно развивается в настоящее время статистическая теория ассоциативных флюидов (Statistical Associating Fluid Theory --- SAFT) \cite{Papaioannou2016ApplicationIndustry,Lafitte2013AccurateSegments,aslyamov2019random}, из-за учета цепочечных типов связей в молекуле в выражении для свободной энергии появляется дополнительное слагаемое. Помимо того, что необходимо выбирать физическую модель и соответствующий ей вид свободной энергии, исследуемые задачи часто наделяют определенными геометрическими свойствами, из-за которых диктуется тип системы координат и упрощается вычисление вариаций \cite{Roth2010FundamentalReview}, переход от одной системы координат к другой может также вызывать трудности с точки зрения пересчета вариаций. Для некоторых физических моделей, из-за сложной структуры энергии Гельмгольца не всегда возможно посчитать вариацию по плотности не прибегая к каким-либо ухищрениям и упрощениям модели, что является проблемой для исследования этих систем. Использование VF-DFT, дает возможность обойти эти преграды и получить достаточно точное решение, зная лишь вид свободной энергии Гельмгольца.

\subsection{Построение базисных функций}
\addetoc{subsection}{Basic pattern analysis}
Задача оптимизации свободной энергии системы для поиска равновесной плотности флюида обладает слишком большим пространством поиска. Чтобы уменьшить размерность оптимизационной задачи можно использовать информацию о характерном поведении искомой функции, вычленить основные закономерности и искать плотность флюида в виде линейной комбинации функций, которые содержат в себе эту информацию. В таком представлении искомой функции, задача поиска равновесной плотности сводится к поиску оптимальных коэффициентов разложения $a_i$, которые бы минимизировали $\Omega$~потенциал $\Omega\left[\rho\left(z\right)\right] = \Omega\left[a_i\right]$. На качество и скорость поиска решения будет влиять количество и вид базисных функций. 

\begin{figure}[htb]
    \centering
    \includegraphics[page=1]{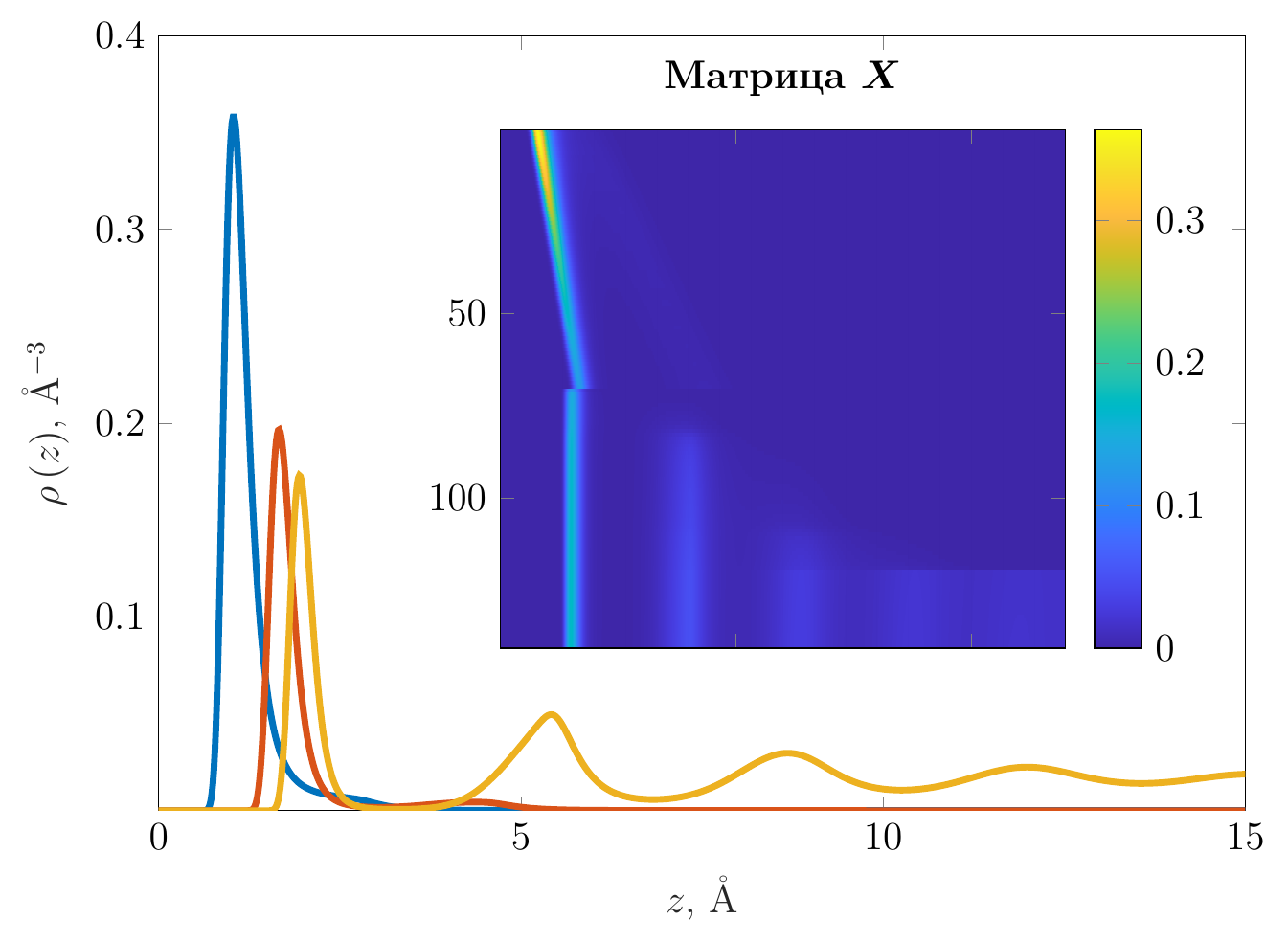}
    \caption{Три различных элемента $\rho\left(z\right)$ из матрицы $\vec{X}$. На вставленном изображении представлено изображение датасета $\bm{X}$, где цветом отмечено значение соответствующей функции при заданной координате}
    \label{fig:basis_matrix}
\end{figure}

Для того, чтобы базисные функции наиболее точно описывали поведение плотности произвольного флюида в поре, было решено составить датасет $\vec{X}$ из функций плотности азота при температуре $77.4$~К и разных относительных давлениях (от $10^{-6}$ до $0.99$). Эти функции плотности были посчитаны классическим DFT с методом простой итерации один раз для всех рассматриваемых кейсов в разделе~\ref{sec:results_vfdft}. Также, для того, чтобы добавить в базис как можно больше информации о поведении искомой функции, были взяты равновесные плотности флюидов с радиусами молекул от $0.85$~\AA,\ до $2$~\AA\ при температуре  $77.4$~K, остальные параметры взаимодействия соответствуют параметрам азота и были зафиксированы. Такой выбор базисных функций никак не ограничивает область применимости алгоритма, в разделе \ref{sec:results_vfdft} будет продемонстрировано, что VF-DFT применим для флюида, отличного от азота, при других термодинамических условиях. В итоге, получился датасет $\vec{X}$ из 140 векторов рис.~\ref{fig:basis_matrix}.
\begin{figure}[htb]
    \centering
    \includegraphics[page=1]{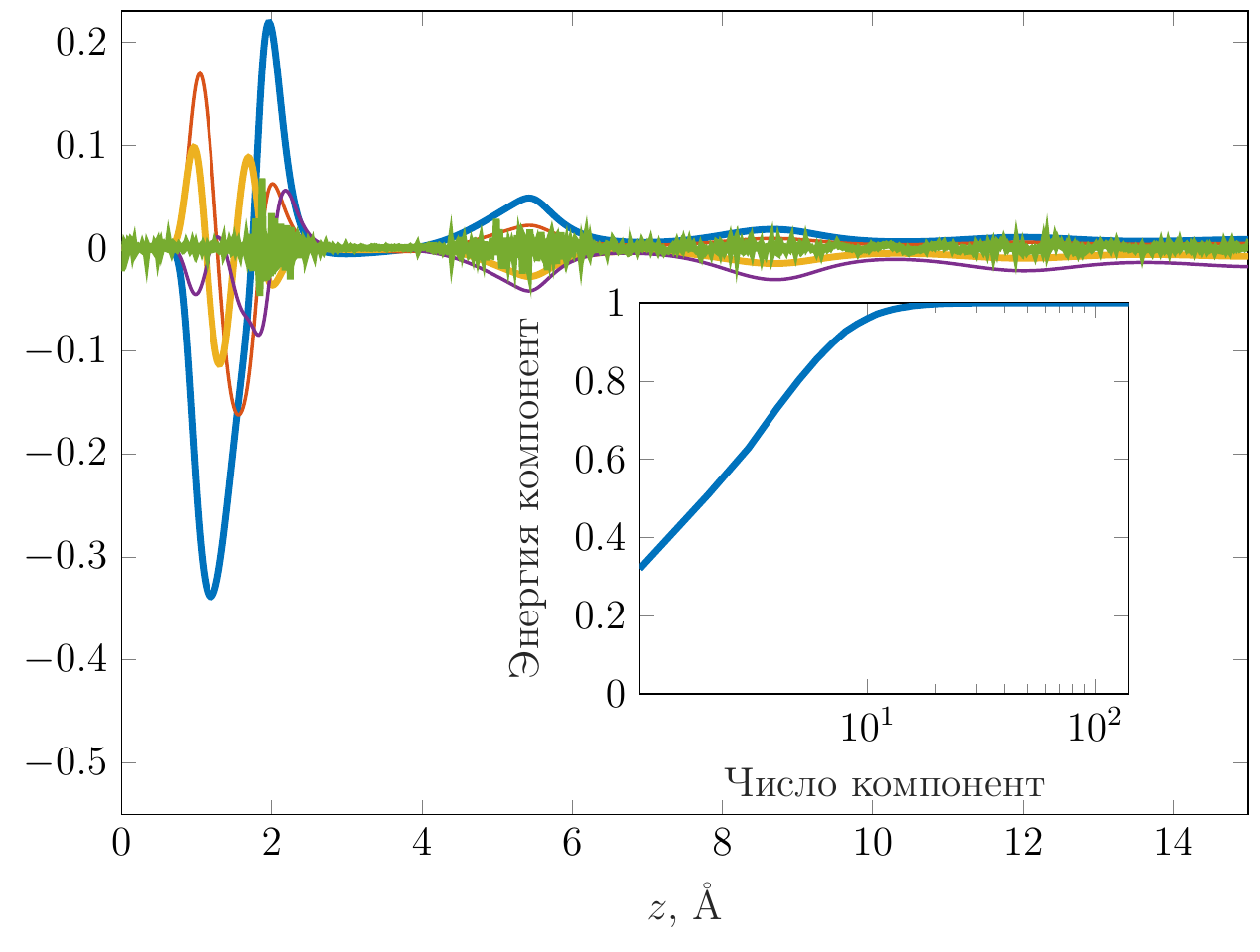}
    \caption{Пять первых главных компонент матрицы~$\vec{X}$. Относительная ошибка реконструкции полного датасета составляет 4.84\%, график распределения ошибки показан темно-зеленой линией}
    \label{fig:PCA_basis}
\end{figure}
Чтобы выделить характерные паттерны для функции плотности и наиболее точно описать поведение плотности произвольного флюида в поре, можно применить метод главных компонент \cite{elizarev2020objective} (Principle Component Analysis --- PCA) к датасету $\vec{X}$ с разнообразными решениями. Пусть вектор-столбец $\vec{x}$ ---~одно из распределений плотности (в нашей работе их 140). Отнормируем матрицу $\vec{X}(Nx \times Ns)$, $Nx$ ---~размерность по $\rho$, $Ns$ ---~число функций в датасете $\bm{X}$, следующим образом:

\begin{equation}
    \vec{X}\left(i\right) = \vec{x} - \vec{x}_{mean},
\end{equation}
где $\vec{x}_{mean}(Nx \times 1)$ ---~среднее распределение плотности по датасету $\vec{X}$.

Согласно \cite{Abdi2010PrincipalStatistics}, разложим центрированную матрицу $\vec{X}$ с помощью сингулярного разложения (Singular Value Decomposition --- SVD):

\begin{equation}
    \vec{X} = \vec{U}\cdot\vec{\Sigma}\cdot \vec{V}^{\text{T}},
\end{equation}
где $\vec{U}\left(Nx \times Ns\right)$ ---~матрица левых сингулярных векторов (ортонормированный базис), $\vec{\Sigma}\left(Ns \times Ns\right)$ ---~квадратная диагональная матрица, в которой содержатся сингулярные числа, отсортированные по убыванию, $\vec{V}\left(Ns \times Ns\right)$ ---~матрица правых сингулярных векторов (тоже ортонормированный базис). Как будет показано далее, левые сингулярные вектора как раз и несут информацию об искомых паттернах, а соответствующие им сингулярные числа показывают, насколько часто встречается данный паттерн в нашем датасете. Для облегчения вычислений в РСА и оптимизации применим усеченный (truncated) SVD, выбрав число компонент $K$ из энергетического критерия на $\vec{\Sigma}$. Квадрат сингулярного числа равен дисперсии по соответствующему направлению в пространстве признаков, проекция на такое направление будет минимальна. Под энергией элемента базиса понимается квадрат соответствующего сингулярного числа, нормированный на сумму квадратов всех сингулярных чисел, соответственно чем больше энергии у выбранных направлений, тем меньше будет проекция у признаков и тем лучше выбранные компоненты будут описывать датасет \cite{Sun2017AProblems}
\begin{equation}
    I = \sigma^2/\sum\sigma^2.
\end{equation}

Таким образом, $\vec{X}\approx \widetilde{\vec{X}}= \widetilde{\vec{U}}\cdot \widetilde{\vec{\Sigma}}\cdot\widetilde{\vec{V}}^{\text{T}}$. Для датасета $\widetilde{\vec{X}}$ можно вычислить сэмпловую (оценочную) ковариационную матрицу $\vec{Q}=\widetilde{\vec{X}}\cdot \widetilde{\vec{X}}^{\text{T}}/{(Ns-1)}$. Обратим внимание, что $\widetilde{\vec{X}}\cdot \widetilde{\vec{X}}^{\text{T}}=\widetilde{\vec{U}}\cdot\widetilde{\vec{\Sigma}}\cdot \widetilde{\vec{V}}^{\text{T}}\cdot \widetilde{\vec{V}}\cdot\widetilde{\vec{\Sigma}}^{\text{T}}\cdot \widetilde{\vec{U}}^{\text{T}}=\widetilde{\vec{U}}\cdot\widetilde{\vec{\Sigma}}^2 \cdot \widetilde{\vec{U}}^{\text{T}}$, то есть квадрат сингулярных чисел матрицы $\widetilde{\vec{X}}$ равен собственным числам $\vec{S}$ матрицы $\vec{Q}$ с точностью до поправки Бесселя $1/(Ns-1)$, а левые сингулярные вектора $\widetilde{\vec{U}}$ равны собственным векторам ковариационной матрицы $\vec{Q}$. Зная $\widetilde{\vec{U}}$ и $\vec{S}$, можно воспользоваться подходом из \cite{Sarma2006EfficientUpdating}, в котором новая реализация генерируется с использованием ковариационной матрицы следующим образом:

\begin{equation}\label{eq:new_dens}
    \vec{x}_{new} = \widetilde{\vec{U}}\cdot\vec{S}\cdot\vec{a} + \vec{x}_{mean}.
\end{equation}
Здесь $\vec{a}$ ---~вектор $K\times 1$ из распределения Гаусса $\mathcal{N}\left(0,1\right)$, его можно трактовать как вектор оптимизируемых параметров (компоненты этого вектора по сути являются коэффициентами разложения искомой плотности по базисным функциям, которые содержат информацию о характерном поведении плотности флюида вблизи стенок), изменяя который мы будем подбирать $\vec{x}_{new}$, минимизирующий Омега-потенциал.

В результате, после анализа главных компонент на основе матрицы $\vec{X}$ был вычислен новый базис размером 10 векторов (вид первых пяти главных компонент базиса на рис.~\ref{fig:PCA_basis}), которые охватывают 95\% информации по энергетическому критерию. Это значит, что вместо того, чтобы искать 140 неизвестных чисел $\vec{a}$, придется искать всего 10, причем качество такого подхода практически не будет отличаться от решения с поиском всех 140 коэффициентов.
\subsection{Стохастическая оптимизация}
\addetoc{subsection}{Stochastic optimization}
Как уже было показано, рассматриваемая задача оптимизации характеризуется тем, что вычислить вариацию свободной энергии достаточно сложно аналитически, а изменение вида потенциала Гельмгольца ведет к необходимости пересчета каждого слагаемого в \eqref{eq:density}. В работе была применена стохастическая оптимизация, в которой исследуемая функция используется как черный ящик, без вычисления вариаций. Искомая функция представляется в виде \eqref{eq:new_dens}, а алгоритмы оптимизации подбирают элементы вектора $\vec{a}$ так, чтобы минимизировать функционал свободной энергии $\Omega\left[\vec{a}\right]$. Размерность оптимизационной задачи равняется размерности вектора $\vec{a}$, в данной работе эта размернасть равна $10\times 1$. Были рассмотрены два итерационных эвристических подхода ---~ГА и МРЧ, широко применяющихся в науке и технике и зарекомендовавшие себя во многих отраслях \cite{Eberhart2001ParticleResources,Montes2001TheOptimization,NejadEbrahimi2013GeneticImages,Onwunalu2009DevelopmentDevelopment}. В этом разделе будут представлены краткие описания принципов работы алгоритмов, а также рекомендации по настройке их параметров.
\subsubsection{Генетический алгоритм (ГА)}
\addetoc{subsubsection}{Genetic Algorithm (GA)}

Создание генетический алгоритм было вдохновлено идеей Дарвиновской теории эволюции. В основе алгоритма лежит идея, что наиболее приспособленные особи, которые эволюционируют, наследуют признаки у своих родителей, мутируют, именно они и выживают (<<выживает сильнейший>>).
ГА был успешно применен для решения самых разнообразных задач из различных дисциплин: отбор признаков в машинном обучении, обработки изображений \cite{Huang2007AInformation,Ouellette2004GeneticDetection,Sharma2010GeneticExcitation}. Описать все модификации и методы разработанные на основе классического генетического алгоритма не представляется возможным, наиболее полное изложение можно найти в статьях \cite{NejadEbrahimi2013GeneticImages,Montes2001TheOptimization}.

\textit{Инициализация}: будем называть вектор $\vec{a}$ хромосомой, набор хромосом ---~поколение. Рекомендуем брать $Nc$ хромосом равным число оптимизируемых параметров. Сначала они все случайные в заданных границах, для каждой вычисляется целевая функция (фитнес функция) ---~в задаче минимизации чем она меньше, тем более хромосома <<приспособлена>>. В данном случае роль целевой функции играет омега потенциал $\Omega\left[\vec{a}\right]$.
В генетическом алгоритме оптимизации каждая итерация представляет собой процесс формирования дочернего поколения из родительского. Этот процесс происходит с помощью генетических операторов. В работе использовались самые базовые операторы, распишем их подробнее.

\textit{Селекция}: одна хромосома с лучшим значением целевой функции переносится в дочернее поколение из родительского без изменений. Это позволяет как минимум не потерять лучшее найденное к данному моменту решение, что обеспечивает немонотонную сходимость алгоритма.

\textit{Скрещивание}: каждая новая хромосома ($Nc-1$ штук) получается путем скрещивания пары родительских. Вероятность выбора хромосомы из родительского поколения для скрещивания может быть как заданной, так и зависеть от фитнес-функции (roulette wheel \cite{NejadEbrahimi2013GeneticImages}). Использование того или иного подхода изменяет свойства алгоритма, в данной работе используется комбинированный подход. Наиболее распространенные типы скрещивания ---~через точку, когда у родительских хромосом выбирается точка $c$ и все гены с индексами меньше чем $c$ у родительских хромосом меняются между друг другом, или через индексы, когда выбирается несколько точек в родительских хромосомах и обмен происходит блоками генов. В данной работе был использован подход скрещивания через индексы.

\textit{Мутация}: новая хромосома, полученная на предыдущем этапе из пары родительских, с заданной вероятностью подвергается мутации, т.е. случайному изменению одной или нескольких ее компонент. Высокий уровень мутации замедляет сходимость алгоритма, но позволяет более успешно находить глобальный оптимум. Уровень мутации в работе был определен на уровне 0.5 --- это значит, что мутации подвержено половина хромосом популяции.

После формирования дочернего поколения для всех хромосом вычисляется фитнес-функция (если она не была известна). Алгоритм останавливается по достижении отведенного числа вызовов фитнес-функции.
\subsubsection{Метод роя частиц (МРЧ)}
\addetoc{subsubsection}{Particle Swarm Optimization (PSO)}
Генетические алгоритмы на сегодняшний день являются одними из самых популярных инструментов решения задач оптимизации. Однако их использование не лишено недостатков. Решение сходится к оптимуму лишь после большого числа вычислений, при этом его глобальность не гарантируется. Нет четкого критерия останова --~при недостаточной мутации решение может очень долго находиться в локальном минимуме, монотонная сходимость не гарантируется. Все это приводит к появлению новых, более развитых ГА, использование ГА в связке с другими алгоритмами, а также развитие новых стохастических подходов.

Одним из таких подходов является метод роя частиц (в англоязычной литературе PSO ---~Particle Swarm Optimization). Впервые он был разработан в 1995 году Кеннеди, Эберхартом и Ши \cite{Kennedy1995ParticleOptimization} для исследования поведения роя насекомых или косяка рыб, передвигающихся по пространству в поисках пищи. В дальнейшем он был упрощен и применен для решения различных задач оптимизации \cite{Banks2007ADevelopment}. Рассмотрим этот метод подробнее, пользуясь терминологией из \cite{Onwunalu2009DevelopmentDevelopment}.

\textit{Инициализация}: вектор $\vec{a}$ называем частицей, набор частиц ---~рой (по аналогии с хромосомой и поколением). Как и в ГА, каждая частица хранит свое текущее положение $\bm{x}_i\left(k\right)$ и соответствующее значение целевой функции $F^{obj}\left(\bm{x}\right)$, а также вектор скорости $\bm{v}_i\left(k\right)$ (того же размера, что и вектор положений). Помимо этого, для каждой частицы хранится лучшее значение целевого функционала, на какой-либо итерации достигнутого ей и соответствующее положение. Целевая функция ($\Omega$-потенциал) чем меньше, тем лучше. 

Число частиц в работе задается равным числу оптимизируемых коэффициентов, а начальная скорость равна нулю. Начальное положение частиц задается из равномерного распределения.

МРЧ ---~итерационный алгоритм, в котором новое положение частицы $i$ на итерации~$k+1$, $\vec{x}_i\left(k+1\right)$, определяется путем прибавления скорости $\vec{v}_i\left(k+1\right)$ к $\vec{x}_i\left(k\right):
\vec{x}_i\left(k+1\right)=\mathbf{x}_i\left(k\right)+\vec{v}_i\left(k+1\right)$. Компоненты $\vec{v}_i\left(k+1\right)$ вычисляются следующим образом:

\begin{equation}
    \vec{v}_{i}\left(k+1\right)=\omega \vec{v}_{i}\left(k\right)+c_1\vec{r}_{1}\left({\hat{\vec{y}}}_{i}\left(k\right) -\vec{x}_{i}\left(k\right)\right)  +c_2\vec{r}_{2}\left(\vec{y}^\ast\left(k\right)-\vec{x}_{i}\left(k\right)\right),\label{eq:pso_velosity}
\end{equation}
где $\omega$, $c_1$, $c_2$ ---~некоторые веса, обозначающие степень привлекательности того или иного направления, $\vec{r}$ --- случайные величины в интервале от 0 до 1.

Из формулы \eqref{eq:pso_velosity} видно, что скорость складывается из трех компонент, называемых инерциальной, когнитивной и социальной соответственно. Инерциальная компонента $\omega \vec{v}_{i}\left(k\right)$ отвечает за движение частицы в направлении, в котором она двигалась на предыдущей итерации $k$. Когнитивное слагаемое $c_1\vec{r}_{1}\left({\hat{\vec{y}}}_{i}\left(k\right) -\vec{x}_{i}\left(k\right)\right)$ отвечает за движение в зависимости от предыдущего лучшего положения для частицы $i$. Социальная компонента $c_2\vec{r}_{2}\left(\vec{y}^\ast\left(k\right)-\vec{x}_{i}\left(k\right)\right)$ включает в себя информацию о лучшем положении для всех частиц и отвечает за движение по направлению к нему. На каждой итерации каждая частица в рое перемещается на новое место в пространстве поиска, и для каждого положения вычисляется целевая функция, определяющая, насколько хорошим является данное решение.

Алгоритм останавливается по достижении отведенного числа вызовов целевой функции, а лучшее положение частицы за все итерации считается ответом.

В работе коэффициенты $\omega=0.7298$, $c_1=1.4962$, $c_2=1.4962$. Эти значения являются наиболее универсальными и подходят для многих задач, именно с них стоит начинать тестировать PSO, стоит отметить, что эти три коэффициента не могут быть произвольными и связаны между собой соотношением из \cite{Trelea2003TheSelection}. В ходе работы изменение этих параметров не дало лучшего решения. 
\subsection{Гибридный подход (H-DFT)}
\addetoc{subsection}{Hybrid Density Functional Theory}
Подход VF-DFT не требует вычисление вариации энергии Гельмгольца и позволяет быстрее получать решение, чем классический DFT с методом простой итерации. Однако, классический DFT дает более точное решение. Чтобы сохранить преимущества в скорости вычислений и качестве решения от обоих подходов было решено совместить их в один гибридный алгоритм Hybrid Density Functional Theory (H-DFT) рис.~\ref{fig:H-DFT}, в котором на начальном этапе приближенное решение ищется с помощью VF-DFT и стохастических методов оптимизации. Затем, то решение, которое выдает VF-DFT подается в качестве начального приближения классическому DFT с методом простой итерации и решение уточняется. Такая комбинация позволила получить значительный выигрыш в скорости поиска равновесной плотности в сравнении с классическим DFT (результаты по скорости и качеству приведены в разделе~\ref{sec:results_vfdft}), при этом качество решения осталось на том же уровне, что и у классического алгоритма. Из-за того, что решение VF-DFT близко к равновесному, для метода простой итерации можно устанавливать значение $\gamma$  достаточно большим. Это влияет на скорость метода простой итерации и на число необходимых итераций для алгоритма. 

\begin{figure}[htbp]
\centering
\begin{tikzpicture}[node distance=4.5cm]
\node (start) [cdft,text width=2.4cm] {Свойства флюида $F\left[\rho\left(\vec{r}\right)\right]$};
\node (VF-DFT) [cdft,right of=start,text width=3.2cm,xshift=-0.8cm] {VF-DFT \\+\\ Стохастическая оптимизация};
\node (DFT) [cdft,right of=VF-DFT,xshift=0.8cm] {Классический DFT\\ +\\ Метод простой итерации};
\node (dens) [cdft,right of=DFT,text width=2.4cm] {Равновесная плотность $\rho^{*}\left(z\right)$};

\draw [arrow] (start) -- (VF-DFT);
\draw [arrow]  (VF-DFT) --node[anchor= north] {$\rho^0 = \rho^*_{so}$} (DFT);
\draw [arrow] (DFT) -- (dens);

\end{tikzpicture}
\caption{Схема работы гибридного алгоритма для теории функционала плотности.}
\label{fig:H-DFT}
\end{figure}
\section{Результаты работы безвариационного алгоритма}\label{sec:results_vfdft}
\setcounter{equation}{0}
\addetoc{section}{Results of VF-DFT}

В этом разделе представлены результаты тестирования метода VF-DFT и гибридного метода H-DFT, которые используют ГА и МРЧ для поиска равновесной плотности флюида в поре, с классическим DFT, который использует метод простой итерации. Было проведено сравнение методов по времени поиска решения и значению $\Omega$~потенциала для найденного решения. Методы тестируются на задачах поиска равновесного распределения плотности азота и аргона в поре 3.6 нм. Все необходимые параметры взаимодействия флюид-флюид и поверхность-флюид приведены в таблице \ref{tab:fluid-param}. Параметры для взаимодействия поверхности с флюидом были посчитаны по правилу Лоренца-Бертло \eqref{eq:Lor-Bert}, где $\epsilon_{ss} = 28$~K, $\sigma_{ss} = 3.4$~\AA 

\begin{equation}\label{eq:Lor-Bert}
    \epsilon_{sf} = \sqrt{\epsilon_{ss}\epsilon_{ff}}, \quad \sigma_{sf} = \dfrac{1}{2}\left(\sigma_{ss} + \sigma_{ff}\right)
\end{equation}

\renewcommand{\arraystretch}{1.1} 
\renewcommand{\tabcolsep}{0.7cm} 
\begin{table}[htb!]
\centering
\caption{Параметры взаимодействия флюид-флюид и поверхность-флюид.  Для потенциала ЛД $r_{cut}$ равен $5\sigma_{ff}$. Плотность поверхности стенки поры $\rho_V = 0.114$ \AA$^{-3}$, расстояние между слоями атомов углерода $\Delta = 3.35$ \AA.}
\label{tab:fluid-param}
\begin{tabular}{@{}clcccc@{}}
\hline\hline
\multicolumn{2}{c}{}                                 & \multicolumn{2}{c}{Флюид-Флюид} & \multicolumn{2}{c}{Поверхность-Флюид} \\ \midrule
\multirow{2}{*}{Флюид} &
  \multirow{2}{*}{$T,$ K} &
  \multirow{2}{*}{$\epsilon_{ff}/k_B,$ K} &
  \multirow{2}{*}{$\sigma_{ff},$ \AA} &
  \multirow{2}{*}{$\epsilon_{sf}/k_B$, K} &
  \multirow{2}{*}{$\sigma_{sf},$ \AA} \\
                          &                          &                 &               &                &                \\
\multicolumn{1}{l}{$N_2$} & \multicolumn{1}{c}{77.4} & 94.45           & 3.575         & 51.43          & 3.487          \\
\multicolumn{1}{l}{$Ar$}  & \multicolumn{1}{c}{87.3} & 111.95          & 3.358         & 55.99          & 3.379          \\\hline\hline 
\end{tabular}
\end{table}


\subsection{Азот}
\addetoc{subsection}{Nitrogen}
В самом простом приближении стоит тестировать алгоритм с определения равновесной плотности для флюида, информации о котором больше содержится в базисе. Как уже говорилось, базис строился на основе равновесных плотностей азота в поре $3.6$~нм при фиксированной температуре $77.4$~К и разных относительных давлениях. Также в базис вошли равновесные плотности  <<искусственных>> флюидов с разными радиусами молекул (от $0.85$~\AA, до $2$~\AA) с фиксированными остальными параметрами взаимодействия. Во всех задачах ниже используется один и тот же базис. 


\subsubsection{Высокое давление}
\addetoc{subsubsection}{High pressure}

\begin{figure}[htb!]
    \centering
    \includegraphics[page=1]{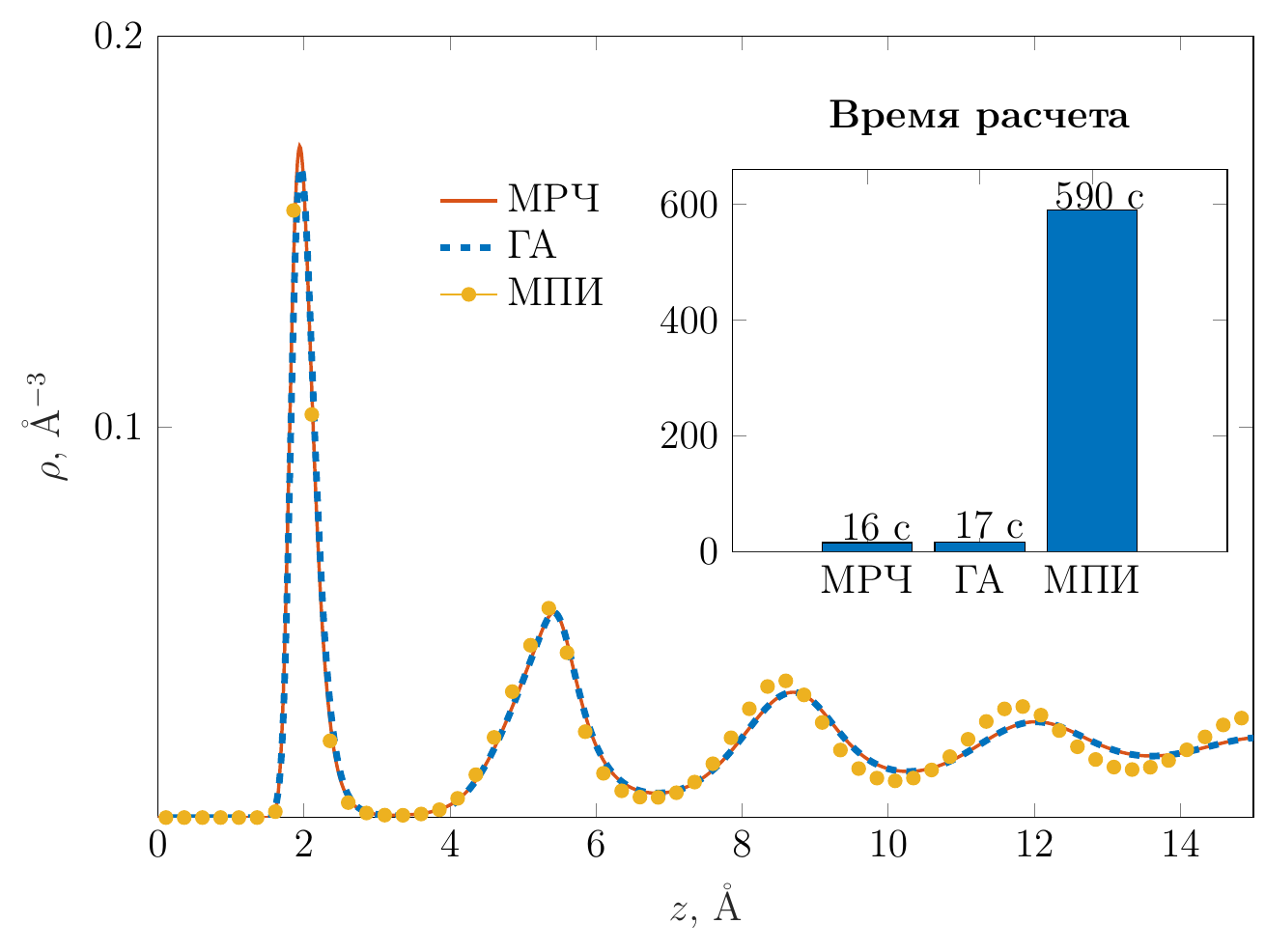}
    \caption{График равновесной плотности при $P/P_0 = 0.6924$ для азота. Красная сплошная линия --- VF-DFT с МРЧ $\Omega_\text{МРЧ} = -0.2630$, синяя прерывистая линия --- VF-DFT с ГА $\Omega_\text{ГА} = -0.2628$, желтыми кругами обозначен классический DFT с МПИ $\Omega_\text{МПИ} = -0.2634$}
    \label{fig:Density_hp_N}
\end{figure}

При расчете равновесной плотности на больших значениях относительного давления, как правило, классические алгоритмы долго ищут равновесное решение. Именно в таких задачах безвариационный подход дает наибольший выигрыш по времени расчета. Несмотря на то, что метод простой итерации весьма надежен, для обеспечения сходимости в таких задачах параметр $\gamma$ приходится выбирать очень маленьким, что сильно влияет на скорость сходимости. На рис.~\ref{fig:Density_hp_N}  показана плотность азота в поре 3.6 нм при температуре 77.4 K и относительном давлении для флюида в свободном объеме $P/P_0 = 0.6924$ ($P_0$ ---~давление насыщения). Важно отметить, что в датасет $\bm{X}$ не входит функция при тех относительных давлениях, при которых производятся расчеты в задачах, рассматриваемых ниже. Разница во времени расчета весьма значительна, при этом решение достаточно близко по значению $\Omega$~потенциала к классическому DFT.

\begin{figure}[htb!]
    \centering
    \includegraphics[page=1]{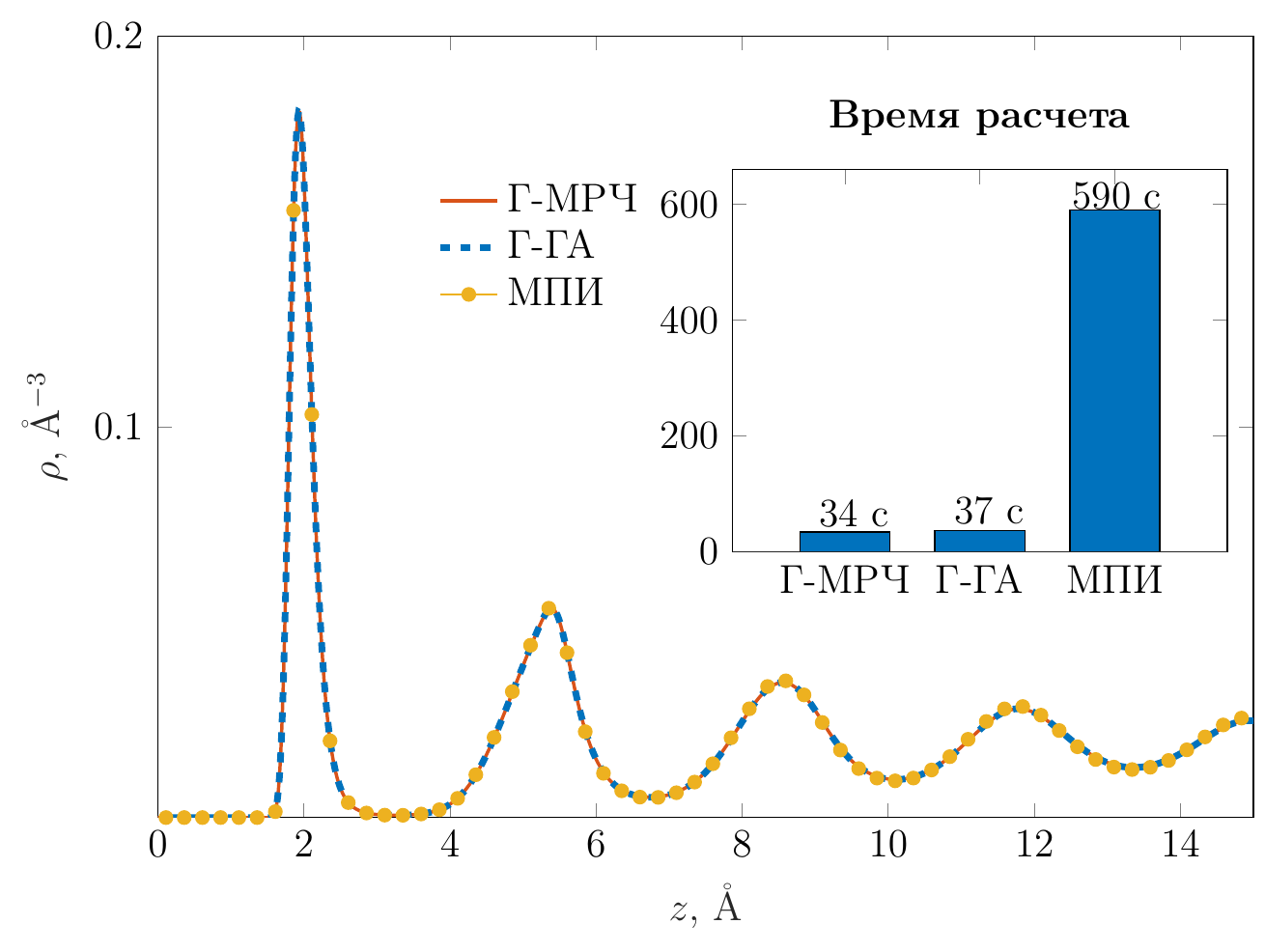}
    \caption{График равновесной плотности при $P/P_0 = 0.6924$ для азота. Красная сплошная линия --- H-DFT с МРЧ, синяя прерывистая линия --- H-DFT с ГА, желтыми кругами обозначен классический DFT с МПИ. Графики полностью совпадают, но время работы гибридных методов на порядок меньше $\Omega_\text{Г-МРЧ} =\Omega_\text{Г-ГА} =\Omega_\text{МПИ} =-0.2634$}
    \label{fig:Density_hp_N_C}
\end{figure}
\begin{figure}[htb!]
    \centering
    \includegraphics{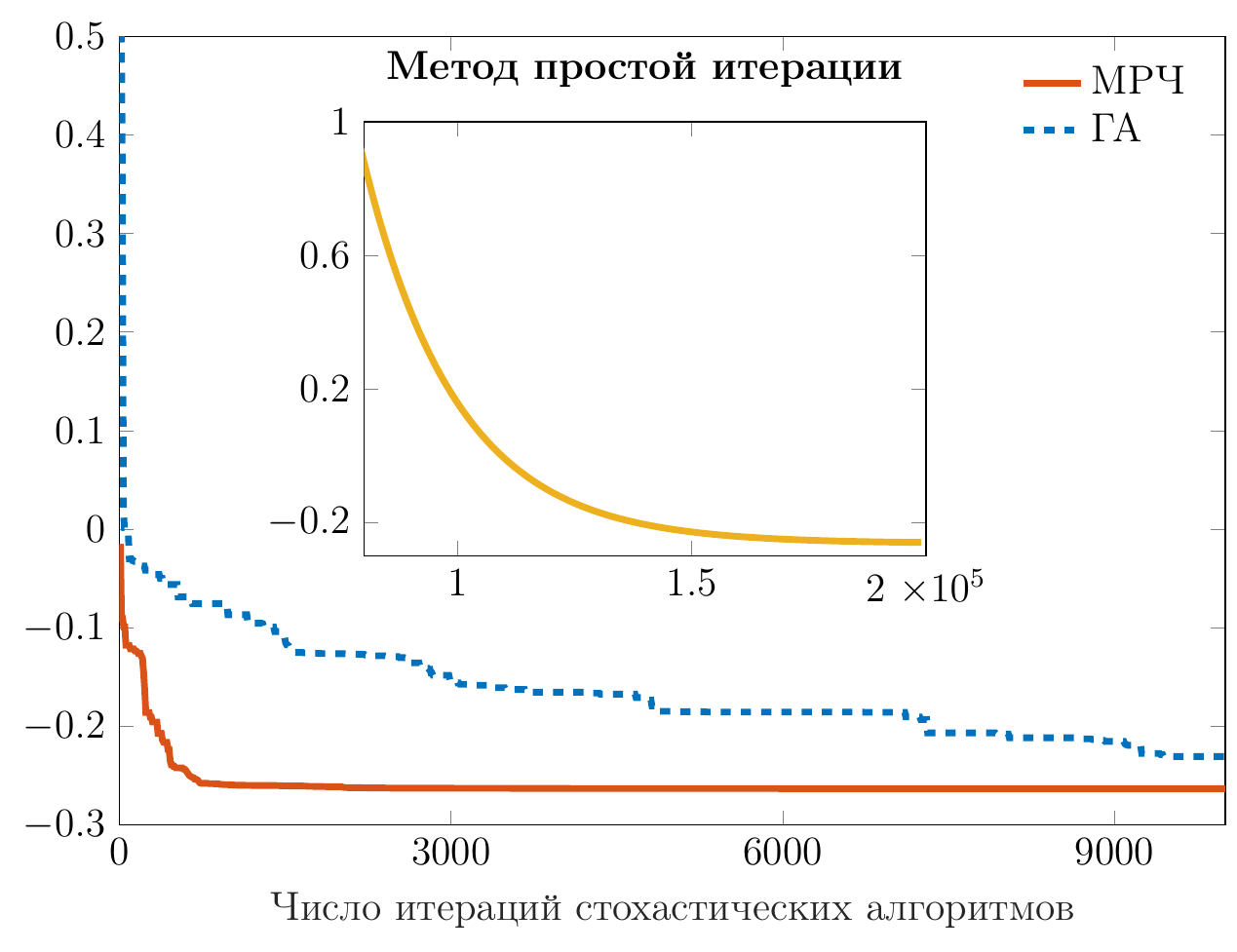}
    \caption{Значение $\Omega$~потенциала 
    на итерациях для классического DFT с методом простой итерации (желтая сплошная линия), значение $\Omega$~потенциала для VF-DFT с методом роя частиц (красная сплошная линия) и VF-DFT с генетическим алгоритмом (синяя пунктирная линия)}
    \label{fig:Convergence_hp_N}
\end{figure}

Значение Омега потенциала для метода простой итерации убывает значительно медленнее в сравнении с VF-DFT как видно на рис.~\ref{fig:Convergence_hp_N}. 
Также можно заметить, что сходимость стохастических алгоритмов не является строго монотонной. Это связано с принципом работы самих алгоритмов. Для стохастических алгоритмов нет точных критериев сходимости, а это значит, что решение, которое выдает алгоритм, не обязано быть лучшим. В таком случае, приходится либо продолжить расчет и надеяться, что решение улучшится, либо запустить алгоритм заново. От того, как настроен алгоритм, зависит насколько устойчив он будет и насколько быстро он будет находить решение.

Алгоритмы на основе ГА, МРЧ справились с задачей на порядок быстрее, но качество полученных решений немного уступает классическому DFT на основе метода простой итерации. Однако стоит отметить, что структурную особенность поведения флюида VF-DFT удалось воспроизвести хорошо. VF-DFT на основе МРЧ выдал значение $\Omega$~потенциала для посчитанной равновесной плотности $\Omega\left[\rho^*_\text{МРЧ}\right] = -0.2630$, VF-DFT на основе ГА $\Omega\left[\rho^*_\text{ГА}\right] = -0.2628$, оба оказались очень близки к значению, которое выдал классический DFT $\Omega\left[\rho^*_\text{МПИ}\right] = -0.2634$.

На рис.~\ref{fig:Density_hp_N_C} показан результат работы гибридных методов в сравнении с классическим DFT с методом простой итерации. Время работы немного отличается от VF-DFT, но конечное значение $\Omega$~потенциала у всех трех профилей плотности совпало $\Omega\left[\rho^*_\text{Г-МРЧ}\right] =\Omega\left[\rho^*_\text{Г-ГА}\right] =\Omega\left[\rho^*_\text{Г-МПИ}\right] =-0.2634$. 

Безвариационный подход позволяет получать решение в несколько раз быстрее, причем объединение VF-DFT с классическим DFT обеспечивает, кроме высокой скорости вычислений, приемлемое качество решения. 
\FloatBarrier
\subsubsection{Низкое давление}
\begin{figure}[htb!]
    \centering
    \includegraphics[page=1]{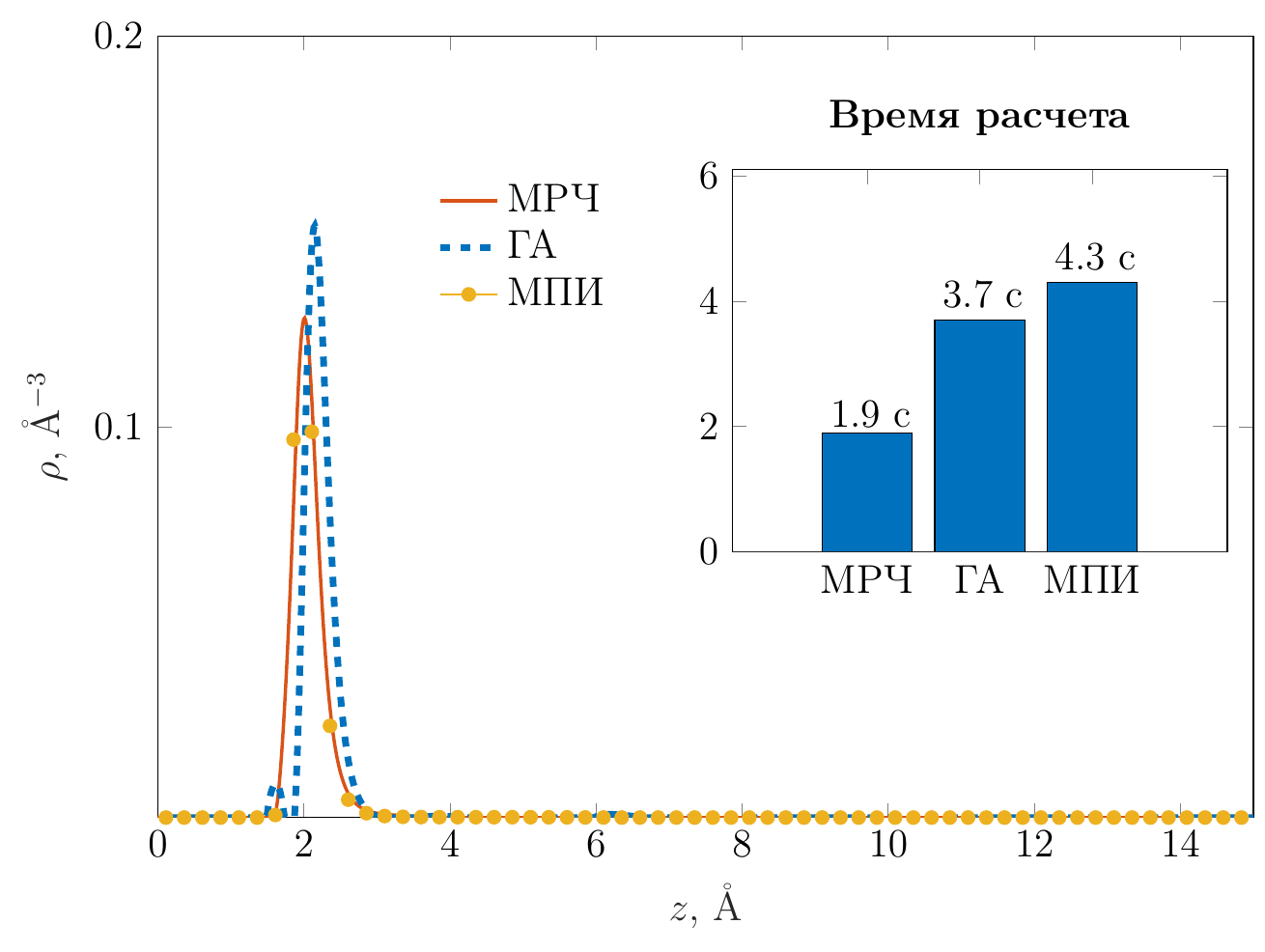}
    \caption{График равновесной плотности при $P/P_0 = 0.0044$ для азота. Красная сплошная линия --- VF-DFT с МРЧ $\Omega_\text{МРЧ} = -0.0305$, синяя прерывистая линия --- VF-DFT с ГА $\Omega_\text{ГА} = -0.0286$, желтыми кругами обозначен классический DFT с МПИ $\Omega_\text{МПИ} = -0.0308$}
    \label{fig:Density_lp_N}
\end{figure}

\addetoc{subsubsection}{Low pressure}
На рис.~\ref{fig:Density_lp_N} видно, что при низких относительных давлениях $P/P_0 = 0.0044$ выигрыш в скорости вычислений у VF-DFT не так значителен, как при высоких давлениях. При низких относительных давлениях классические алгоритмы работают быстрее, параметр $\gamma$ больше. Если посмотреть на полученный профиль плотности, качество решения кажется хуже, чем у классического DFT.
\begin{figure}[htb!]
    \centering
    \includegraphics[page=1]{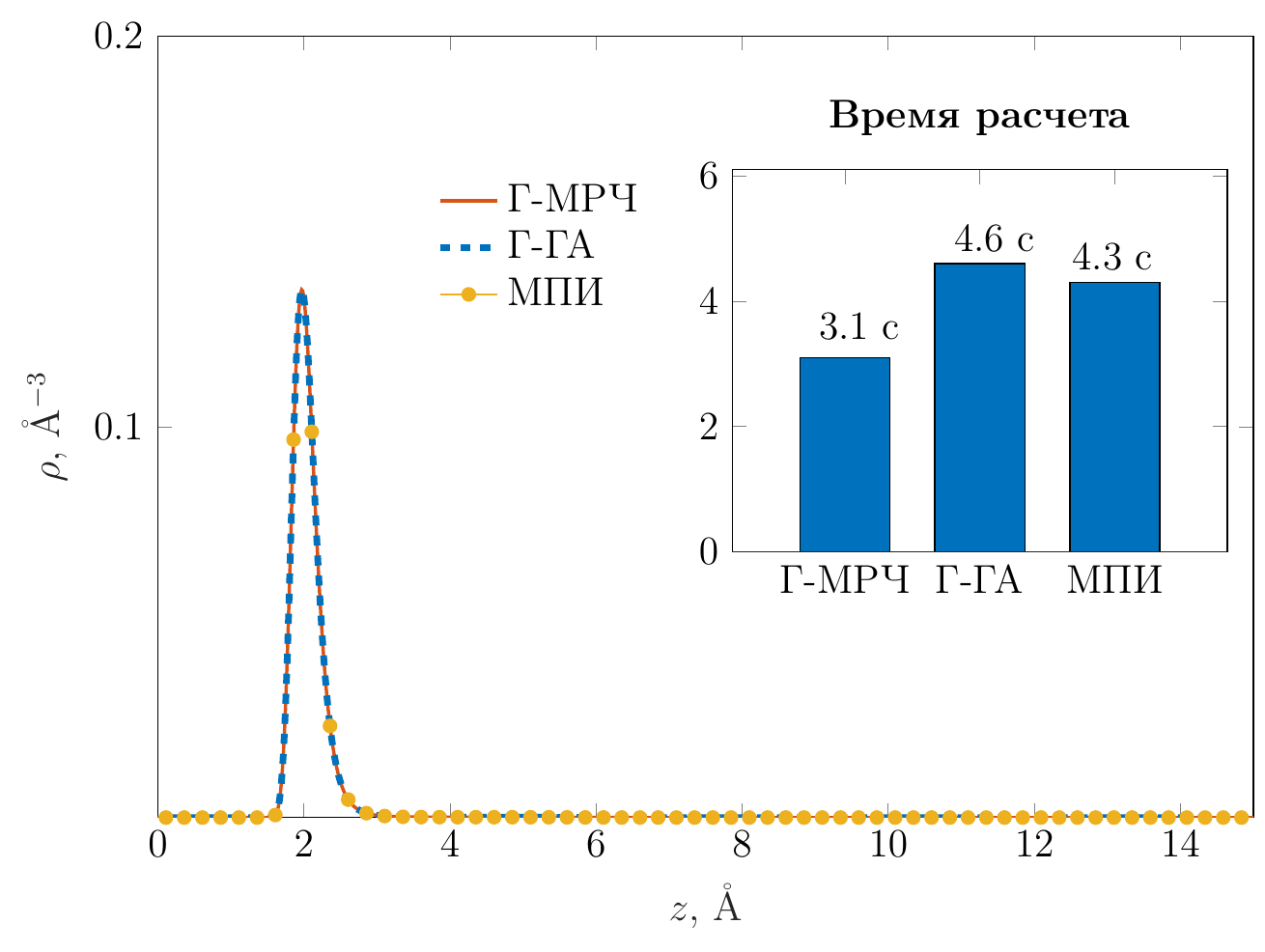}
    \caption{График равновесной плотности при $P/P_0 = 0.0044$ для азота. Красная сплошная линия --- H-DFT с МРЧ, синяя прерывистая линия --- H-DFT с ГА, желтыми кругами обозначен классический DFT с МПИ. Графики полностью совпадают, но время работы гибридных методов на порядок меньше $\Omega_\text{Г-МРЧ} =\Omega_\text{Г-ГА} =\Omega_\text{МПИ} =-0.0308$}
    \label{fig:Density_lp_N_C}
\end{figure}

Для гибридных алгоритмов рис.~\ref{fig:Density_lp_N_C} оказалось, что генетический алгоритм работает медленнее, чем метод простой итерации, МРЧ по прежнему быстрее обоих (однако стоит учитывать., что в гибридном подходе тратится время на имплементацию оптимизационных алгоритмов), а значение $\Omega$~потенциала для всех трех одинаково $\Omega\left[\rho^*_\text{Г-МРЧ}\right] =\Omega\left[\rho^*_\text{Г-ГА}\right] =\Omega\left[\rho^*_\text{МПИ}\right] =-0.0308$.






\FloatBarrier
\subsection{Аргон}
\addetoc{subsection}{Argon}
В этом разделе представлены наиболее ценные результаты работы нового алгоритма VF-DFT. Чтобы убедиться в том, что безвариационный подход можно применять не только для расчета равновесной плотности азота при температуре $77.4$~K (на основе азота считался базис), но и для других флюидов, было решено применить VF-DFT для аргона при температуре $87.3$~K. Результаты данного раздела подтверждают идею применимости безвариационного метода для расчета плотности флюида со сложной структурой свободной энергии Гельмгольца без вычисления вариаций. 
\subsubsection{Высокое давление}
\begin{figure}[htb!]
    \centering
    \includegraphics[page=1]{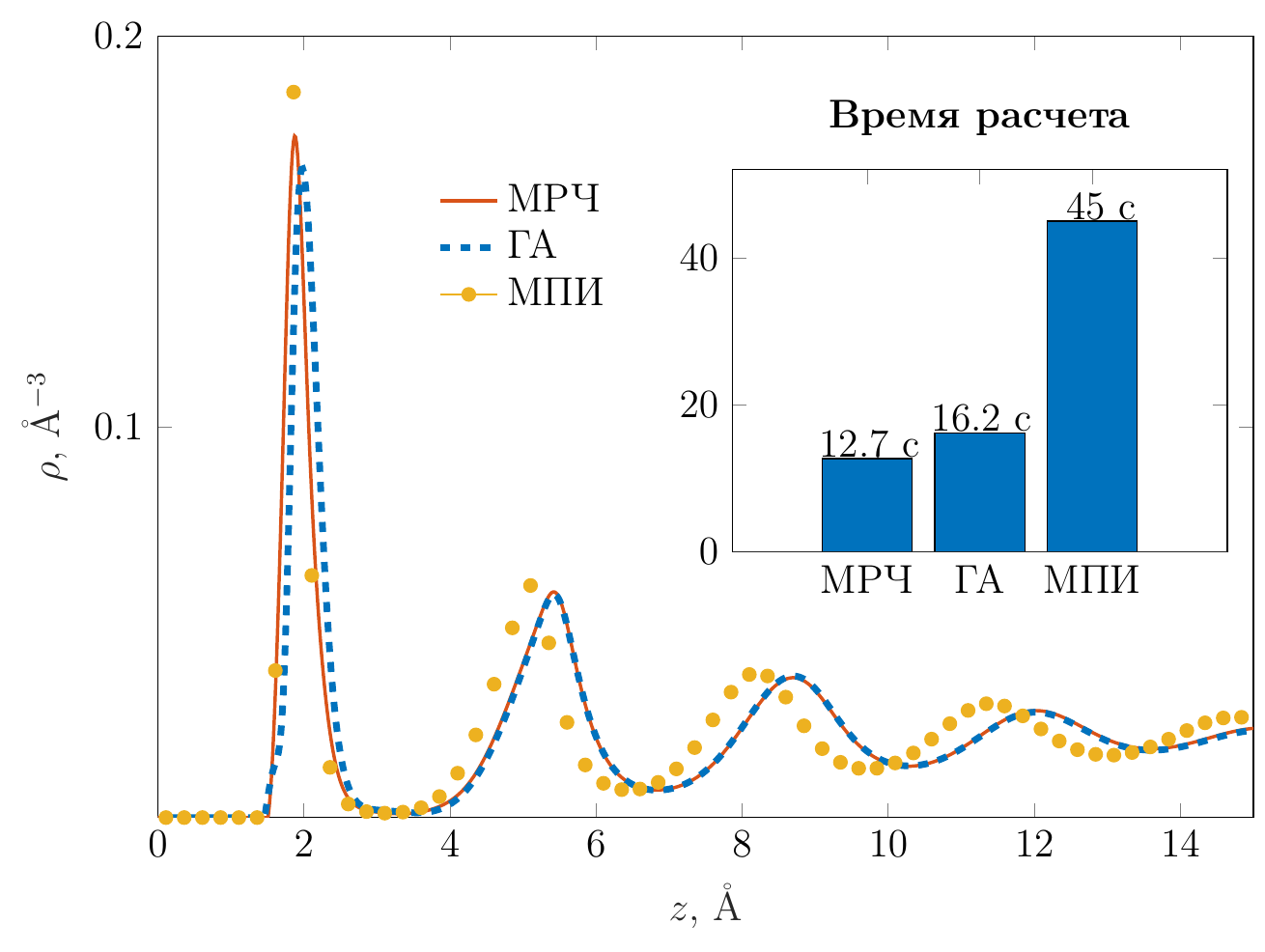}
    \caption{График равновесной плотности при $P/P_0 = 0.9932$ для аргона. Красная сплошная линия --- VF-DFT с МРЧ $\Omega_\text{МРЧ} = -0.1674$, синяя прерывистая линия --- VF-DFT с ГА $\Omega_\text{ГА} = -0.1626$, желтыми кругами обозначен классический DFT с МПИ $\Omega_\text{МПИ} = -0.1712$}
    \label{fig:Density_hp_A}
\end{figure}

\addetoc{subsubsection}{High pressure}
На рис.~\ref{fig:Density_hp_A} видно, что положение пиков совпало не точно, но стоит отметить, что количество и их амплитуда VF-DFT смог восстановить. Кроме того, VF-DFT по прежнему справляется с расчетами в разы быстрее, чем классический DFT с методом простой итерации.
\begin{figure}[htb!]
    \centering
    \includegraphics[page=1]{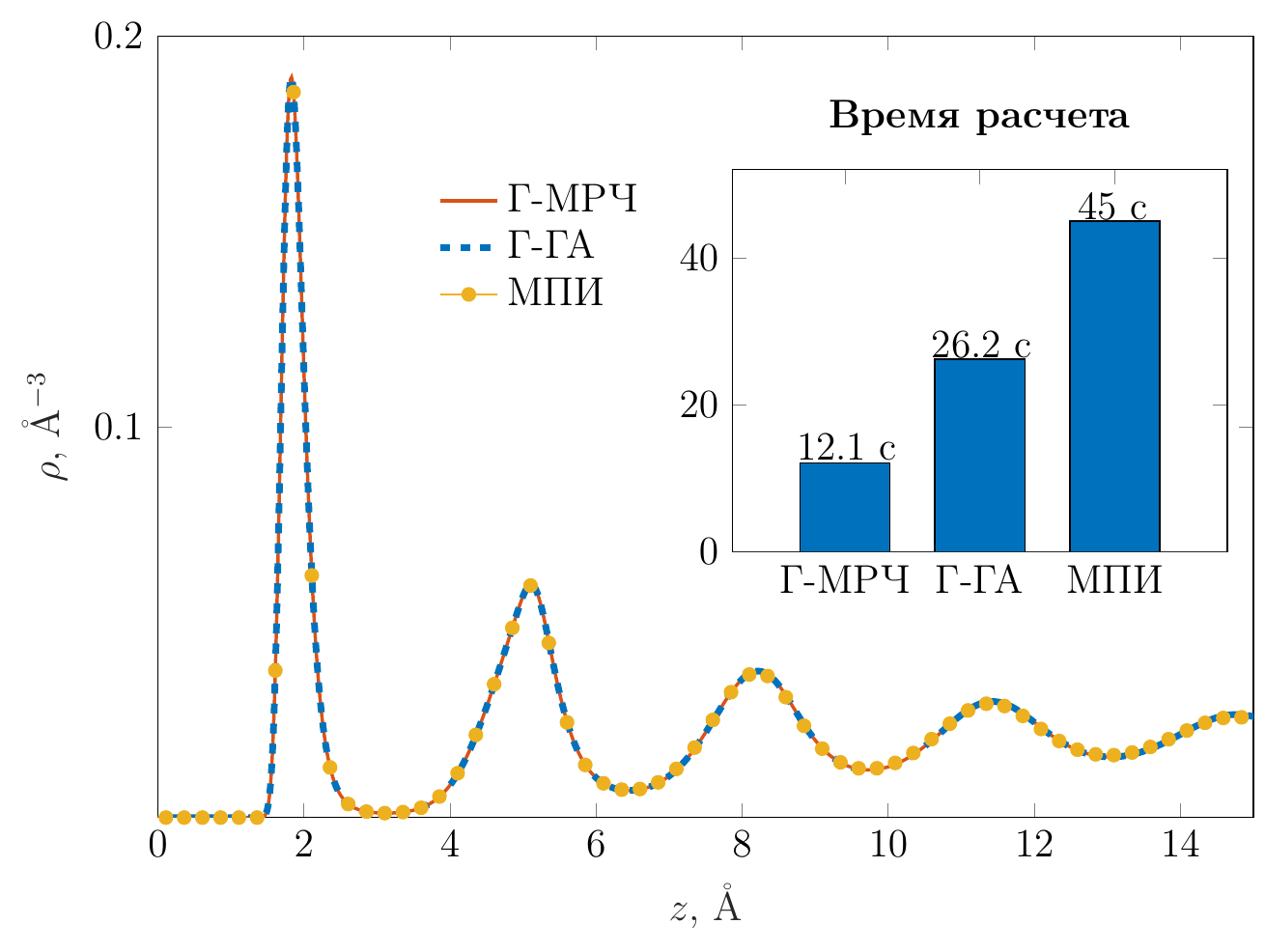}
    \caption{График равновесной плотности при $P/P_0 = 0.9932$ для аргона. Красная сплошная линия --- H-DFT с МРЧ, синяя прерывистая линия --- H-DFT с ГА, желтыми кругами обозначен классический DFT с МПИ. Графики полностью совпадают, но время работы гибридных методов на порядок меньше $\Omega_\text{Г-МРЧ} =\Omega_\text{Г-ГА} =\Omega_\text{МПИ} = -0.1712$}
    \label{fig:Density_hp_A_C}
\end{figure}

При использовании гибридного алгоритма рис.~\ref{fig:Density_hp_A_C} решения полностью совпали при меньшем времени расчета.
\FloatBarrier
\subsubsection{Среднее давление}
\begin{figure}[htb!]
    \centering
    \includegraphics[page=1]{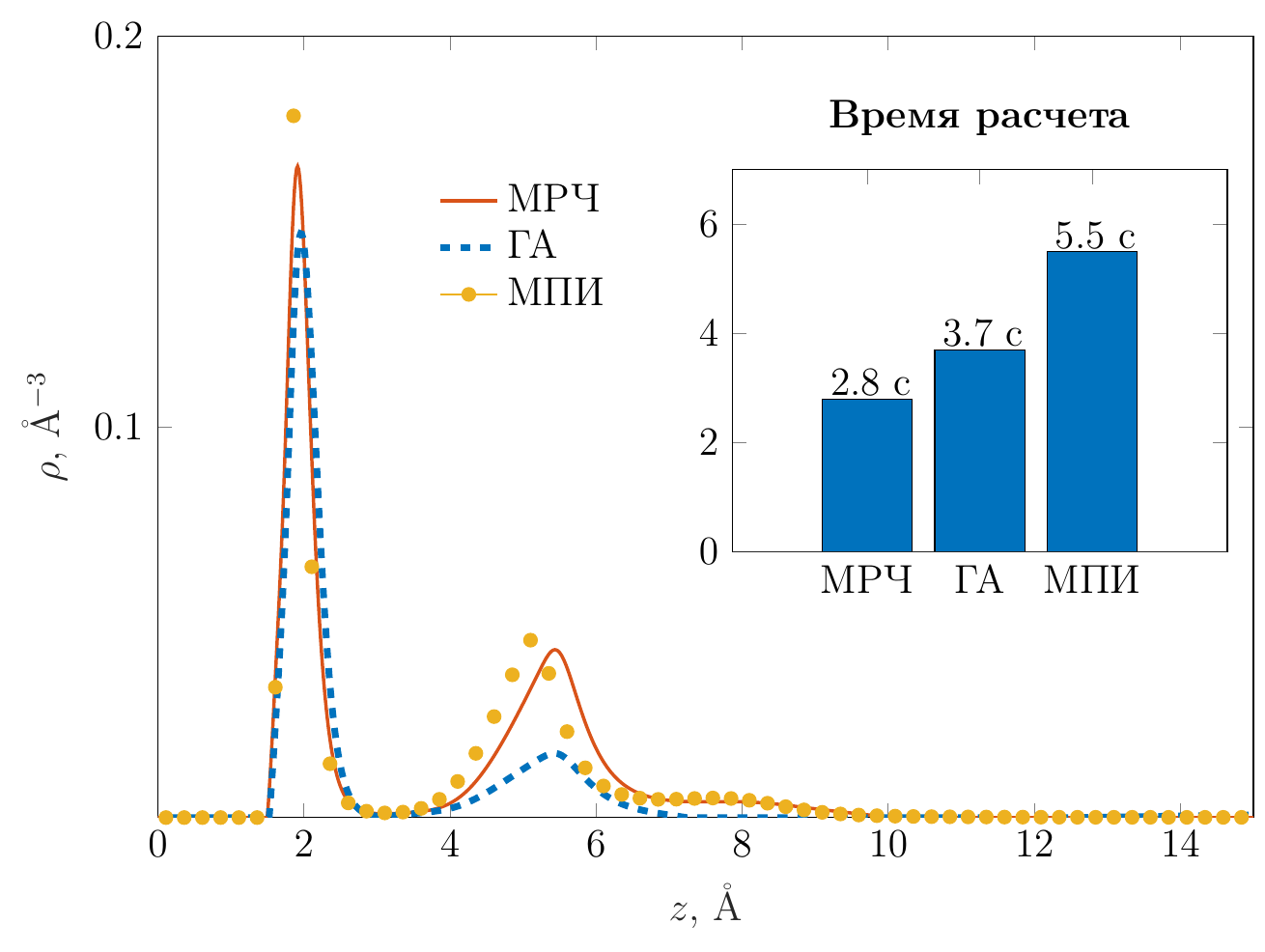}
    \caption{График равновесной плотности при $P/P_0 = 0.4739$ для аргона. Красная сплошная линия --- VF-DFT с МРЧ $\Omega_\text{МРЧ} = -0.1049$, синяя прерывистая линия --- VF-DFT с ГА $\Omega_\text{ГА} = -0.1019$, желтыми кругами обозначен классический DFT с МПИ $\Omega_\text{МПИ} = -0.1077$}
    \label{fig:Density_mp_A}
\end{figure}

\addetoc{subsubsection}{Middle pressure}
С уменьшением относительного давления безвариационный подход дает меньший выигрыш в скорости рис.~\ref{fig:Density_mp_A}. В этом кейсе алгоритмам удалось уловить положение пиков, но величину пиков смог восстановить только VF-DFT с МРЧ.
\begin{figure}[htb!]
    \centering
    \includegraphics[page=1]{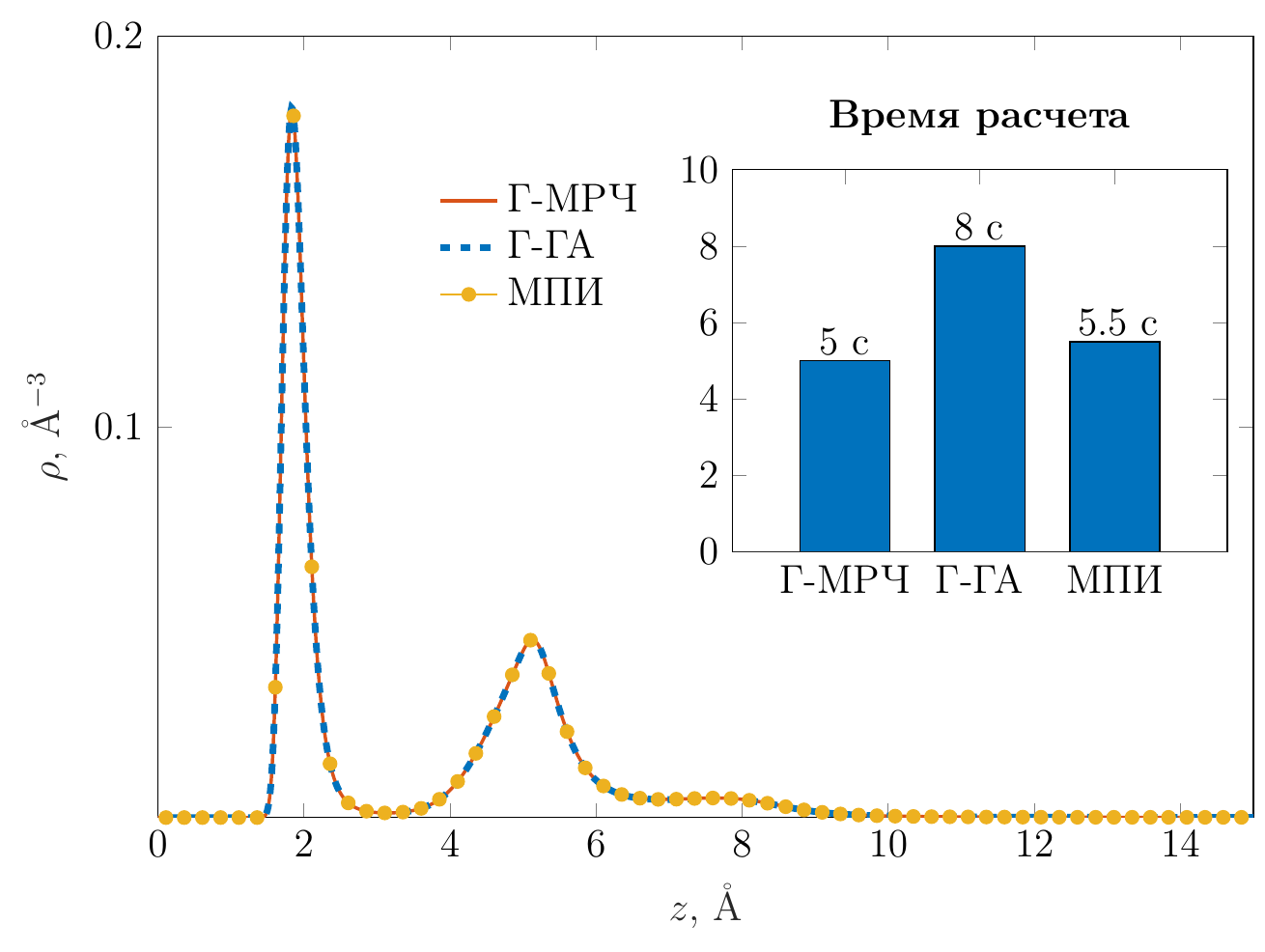}
    \caption{График равновесной плотности при $P/P_0 = 0.4739$ для аргона. Красная сплошная линия --- H-DFT с МРЧ, синяя прерывистая линия --- H-DFT с ГА, желтыми кругами обозначен классический DFT с МПИ. Графики полностью совпадают, но время работы гибридных методов на порядок меньше $\Omega_\text{Г-МРЧ} =\Omega_\text{Г-ГА} =\Omega_\text{МПИ} = -0.1077$}
    \label{fig:Density_mp_A_C}
\end{figure}

Гибридный подход в случае с аргоном при средних давлениях рис.~\ref{fig:Density_mp_A_C} не дал ожидаемого выигрыша в скорости расчета, H-DFT с генетическим алгоритмом оказался даже медленнее классического DFT. Для таких давлений в методе простой итерации для аргона можно использовать достаточно большое значение $\gamma$ и время, которое тратится в H-DFT на получение приближенного решение, оказалось очень близким ко времени работы классического DFT.
\FloatBarrier
\paragraph{Выводы}

Таким образом, наглядно видно, что безвариационный и гибридный подходы превосходят в скорости расчета классические методы. Также было продемонстрировано, что VF-DFT применим для описания поведения не только азота, но и других флюидов. Наибольший выигрыш в скорости у разработанных методов наблюдается для случаев с высоким относительным давлением.
\section{Восстановление распределения пор по размерам}\label{sec:psd}
\setcounter{equation}{0}
\addetoc{section}{Pore Size Distribution (PSD)}
\subsection{Введение в задачу восстановления распределения пор по размерам}
\addetoc{subsection}{Introduction to the problem}

\begin{figure}[htb!]
    \centering
    \includegraphics{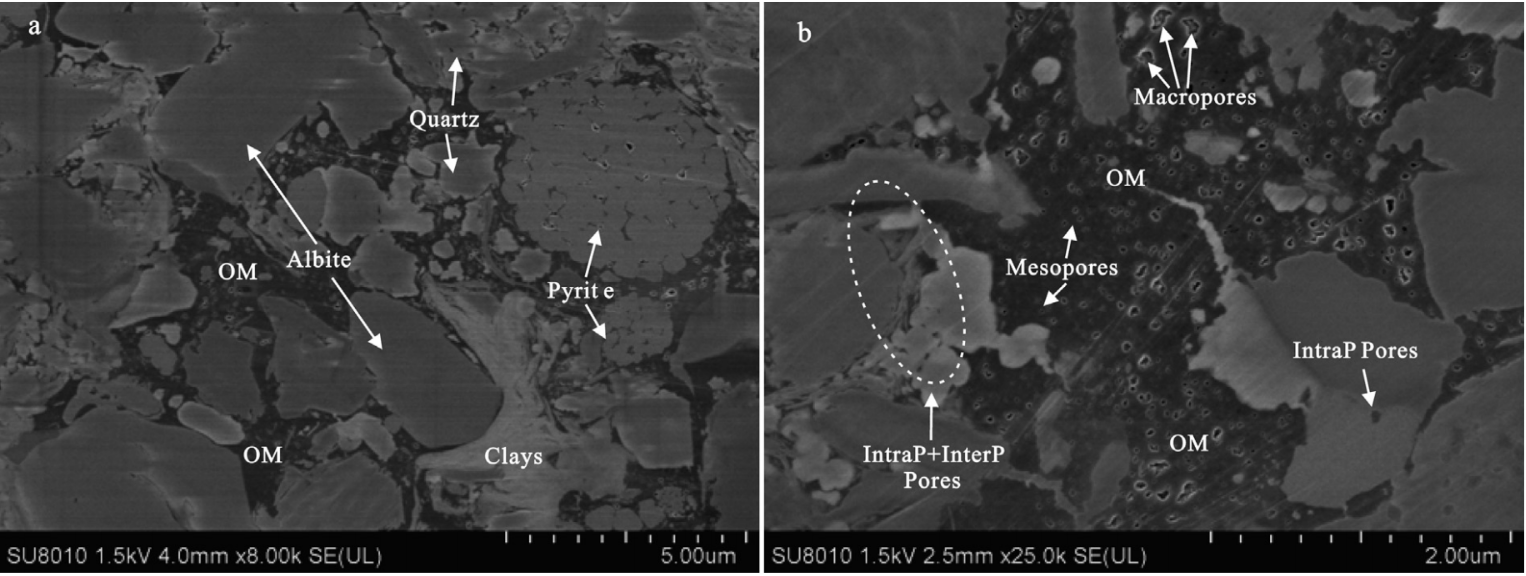}
    \caption{Изображение образца породы с месторождения сланцевого газа методом электронной микроскопии (FE-SEM) \cite{chalmers2012characterization}}
    \label{fig:FE-SEM}
\end{figure}

Сланцевый газ является перспективным нетрадиционным источником углеводородов, месторождения которого найдены по всему миру \cite{zhang2017dft}. В последнее время наблюдается значительный прогресс в оценке и разведке сланцевого газа в Китае \cite{zou2010geological}. Многочисленные исследования показали, что в поровой системе богатых органикой сланцев преобладают нанопоры \cite{loucks2009morphology}. Нанопоры обычно меньше 100 нм и включают микропоры (<2 нм), мезопоры (2-50 нм) и макропоры (50-100 нм) рис.~\ref{fig:FE-SEM} \cite{chalmers2012characterization}.
Богатые органикой сланцы имеют сложные и гетерогенные поровые системы с широкой вариабельностью типа пор, размера, формы и механизма течения \cite{clarkson2013pore}. Таким образом, исследование структуры нанопор может помочь лучше понять пути хранения и миграции газа в сланцах.

Для характеризации структуры порвого пространства в сланцах используют различные методы, например, адсорбция газов, ртутная порометрия, полевая эмиссионная электронная микроскопия или трансмиссионная электронная микроскопия (field emission scanning electron microscopy or transmission electron microscopy --- FE-SEM/TEM), ионная сканирующая электронная микроскопия (focused ion beam scanning electron
microscopy --- FIB-SEM) и малоугловое или ультра-малоуглового рассеяния нейтронов (small-angle or ultra-small-angle neutron scattering --- SANS/USANS). Каждый метод или техника применимы к определенному диапазону для характеристики размера пор. Низкотемпературная адсорбция газ стала наиболее популярным методом для характеристики нанопористой структуры сланцев, так как она позволяет оценить полный диапазон размеров нанопор \cite{zhang2017dft}.

В работе \cite{seaton1989new} представлена методика определения распределения пор по размерам для пористых углеродсодержащих материалов, основанная на применении классической теории функционала плотности к решению интегрального уравнения адсорбции. В этом подходе измеренное количество зондирующего газа, адсорбированного в пористом материале, вычисляется из свертки между распределением пор по размерам и ядром адсорбции. Эта процедура показала лучшие результаты для описания адсорбции газа в нанопористых материалах и восстановления распределения пор по размерам, по сравнению с методами, основанными на уравнении Кельвина, такими как метод Барретта-Джойнера-Халенды (Barrett-Joyner-Halend --- BJH).

\begin{figure}[htb!]
    \centering
    \includegraphics[page=1]{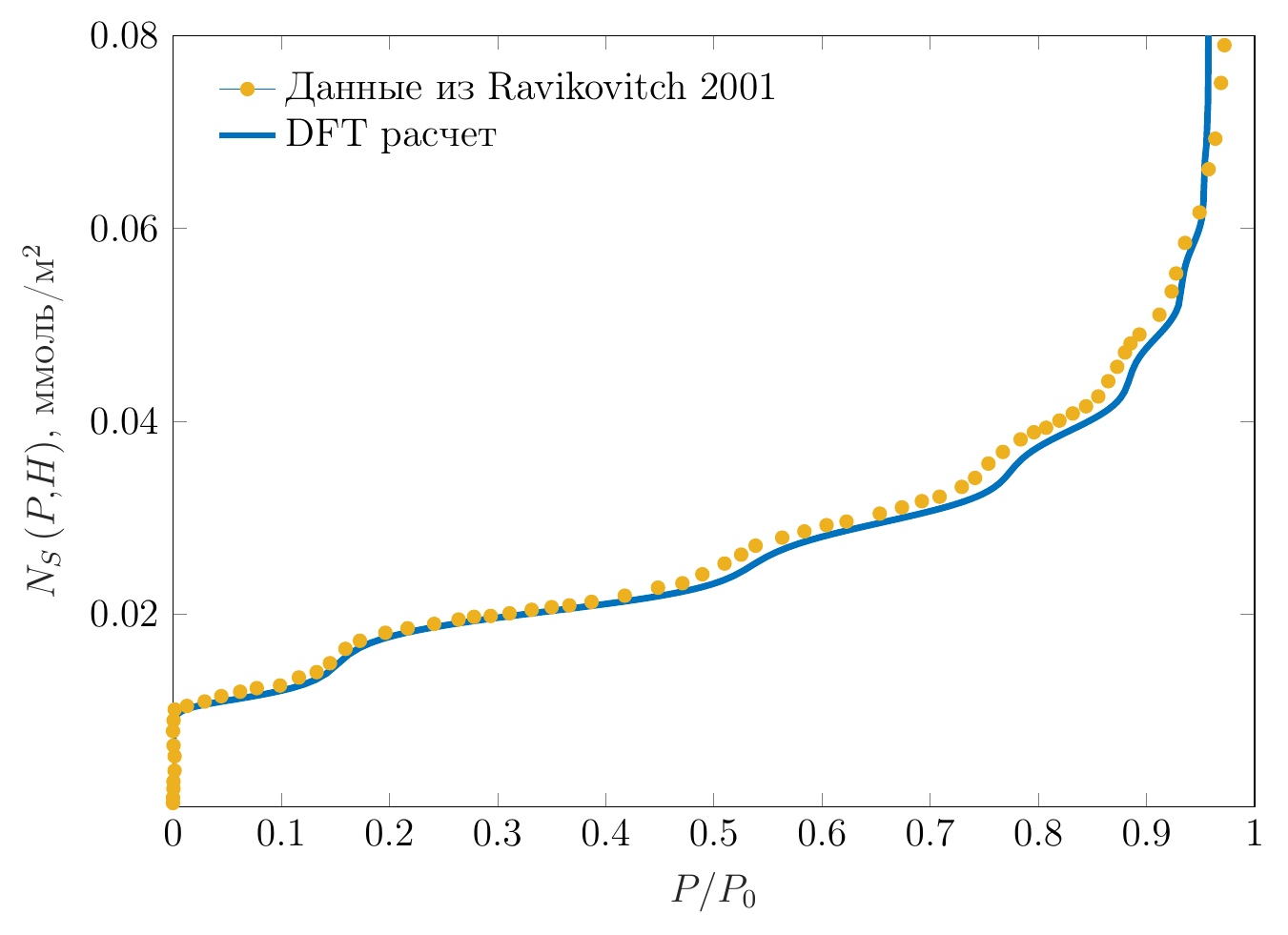}
    \caption{Изотерма адсорбции азота в поре $H_{cc} = 71.5$ \AA, рассчитанная с помощью классического DFT и свалидированная с \cite{Ravikovitch2001DensityNanopores}}
    \label{fig:isotherm71}
\end{figure}

Большинство моделей адсорбции газа в нанопорах представляют структуру пор как совокупность щелевидных пор со стенками из атомов углерода. Это представление аппроксимирует типичную пластинчатую геометрию пор, наблюдаемую в углеродистых материалах, геометрическая неоднородность моделируется изменением размера пор. Таким образом, экспериментальная изотерма описывается как комбинация изотерм в отдельных щелевидных порах с использованием интегрального уравнения адсорбции \cite{Ravikovitch2000UnifiedIsotherms}
\begin{equation}\label{eq:IAE}
    N_{exp}\left(P\right) = \int\limits_{H_{min}}^{H_{max}} N_{S}\left(P,H\right)\chi\left(H\right)dH,
\end{equation}
где $N_{exp}$ --- экспериментальная изотерма адсорбции, $H$ --- ширина поры, $P$ --- давление, $\chi\left(H\right)$ --- распределение пор по размерам в материале, $N_{S}\left(P,H\right)$ --- изотерма адсорбции в одиночной поре фиксированной ширины. Величина $N_{S}\left(P,H\right)$ может быть рассчитана с помощью теории функционала плотности следующим образом:
\begin{equation}
    N_{S}\left(P,H\right) = \dfrac{1}{2N_A}\int\limits_0^H \left(\rho\left(z\right) - \rho_{bulk}\left(P\right)\right)dz,
\end{equation}
где $N_A$ --- число Авогадро, $\rho\left(z\right)$ --- равновесный профиль плотности флюида в поре, $\rho_{bulk}$ --- плотность флюида в свободном объеме при давлении $P$. На рис.~\ref{fig:isotherm71} представлен характерный вид изотермы адсорбции в одиночной плоской поре, одна точка на такой кривой по сути является проинтегрированным профилем плотности типа рис.~\ref{fig:attraction} при фиксированном давлении. Рассчет изотермы производился классическим DFT с методом простой итерации, а результаты были свалидированы с \cite{Ravikovitch2001DensityNanopores}.

\begin{figure}[htb!]
    \centering
    \includegraphics[page=1]{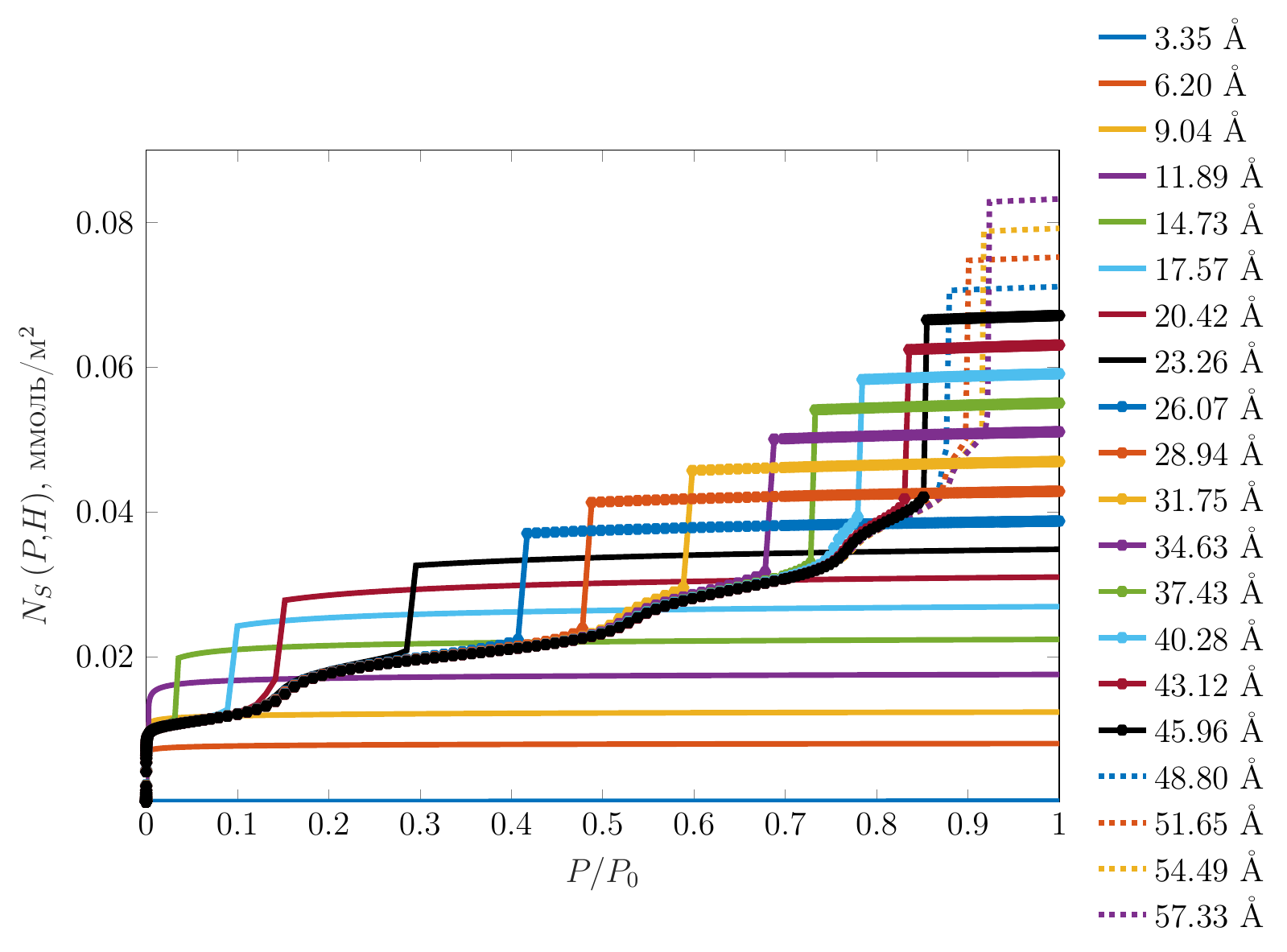}
    \caption{Набор двадцати изотерм адсорбции азота для пор шириной от $H_{in}=3.35$~\AA\, до $H_{in}=57.33$ \AA}
    \label{fig:isotherm_kernels}
\end{figure}

\begin{figure}[htb!]
    \centering
    \includegraphics[page=1]{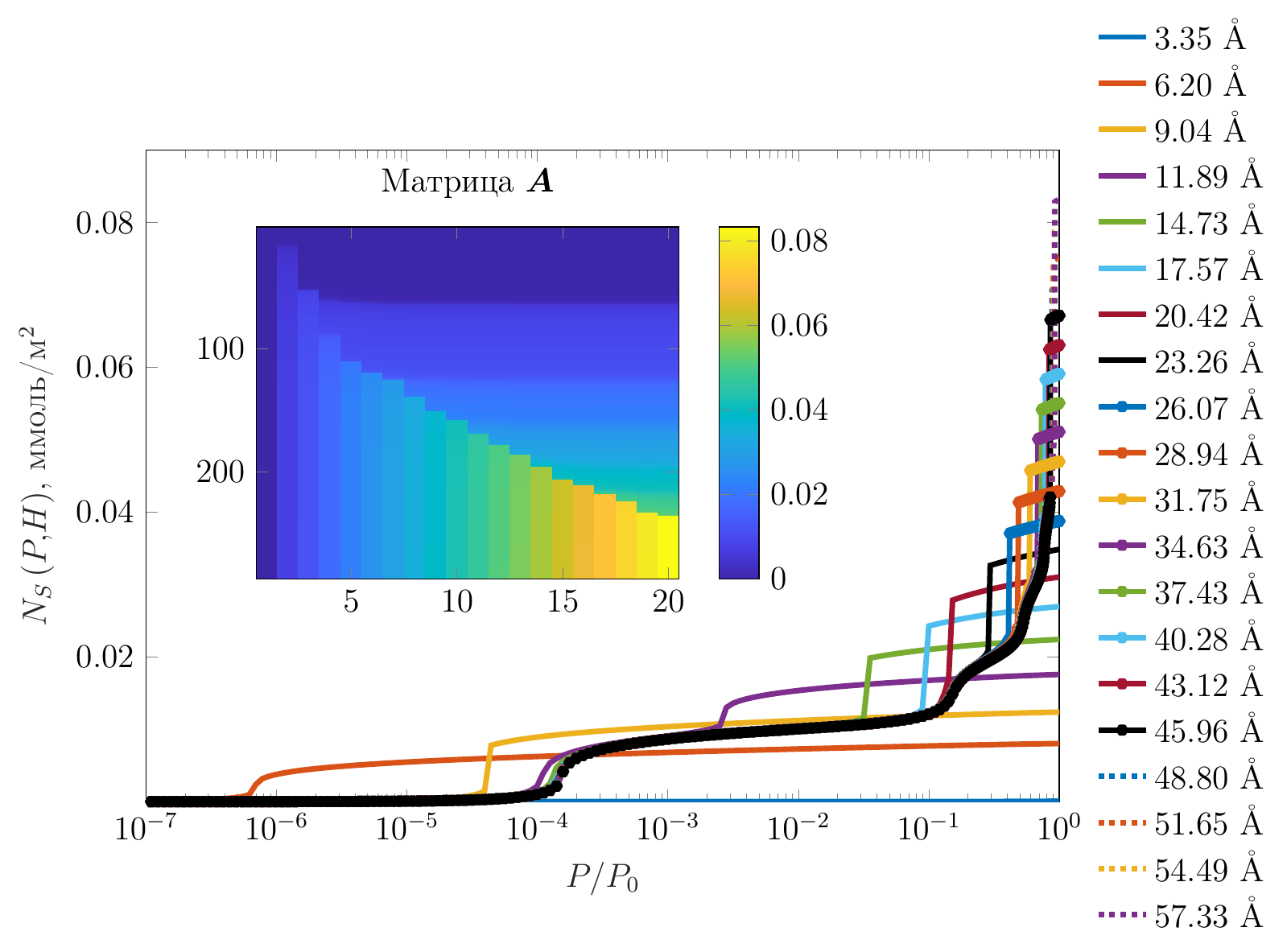}
    \caption{Набор двадцати изотерм адсорбции азота для пор шириной от $H_{in}=3.35$~\AA\, до $H_{in}=57.33$ \AA\, в логарифмическом масштабе. На вставленном изображении продемонстрирована матрица ядер адсорбции $\bm{A}$, где цветом отмечено значение изотермы для заданной поры и относительного давления}
    \label{fig:isotherm_kernels_log}
\end{figure}

Обращение уравнения \eqref{eq:IAE} для решения обратной задачи и нахождения распределения пор по размерам $\chi\left(H\right)$ как известно является плохо поставленной или некорректной задачей. Можно переписать интегральное уравнение \eqref{eq:IAE} в матричном виде \cite{Ravikovitch2000UnifiedIsotherms}
\begin{equation}
    \bm{A}\bm{x} = \bm{b}
\end{equation}
где $\bm{A}\left(m\times n\right)$ --- матрица изотерм для одиночных пор, которая считается с помощью теории функционала плотности ($m$ --- количество точек по давлению $P$, при которых считается изотерма для одной поры; $n$ --- количество одиночных пор, для которых были рассчитаны изотермы), $\bm{b}$ --- вектор с экспериментальными значениями изотермы адсорбции размера ($m\times 1$), $\bm{x}$ --- неизвестный вектор ($n\times 1$) распределения пор по размерам. На рис.~\ref{fig:isotherm_kernels} и рис.~\ref{fig:isotherm_kernels_log} изображены изотермы адсорбции азота в различных порах, всего в данной работе $n = 20$ изотерм и $m = 286$ точек по давлению для каждой изотермы. Набор таких изотерм адсорбции для одиночных пор называют ядрами адсорбции и каждый столбец матрицы $\bm{A}$ является изотермой для одиночной поры.

В данной работе используется метод регуляризации Тихонова, в котором решение вычисляется путем минимизации следующей квадратичной формы 
\begin{equation}\label{eq:inv_problem}
    \mathcal{L} = \| \bm{A}\bm{x} - \bm{b}\|^2_2 + \lambda^2 \|\bm{L}\bm{x}\|^2_2 \rightarrow \min
\end{equation}
где $\bm{L}$ --- линейный оператор (матрица $p\times n,\; p\le n
$), который накладывает ограничения на решение, например, на гладкость решения, а $\lambda$ --- параметр регуляризации. Выбор метода регуляризации не уникален и для разных задач может быть своя матрица $\bm{L}$. В данной работе полагается, что $\bm{L} = \bm{I}$, единичная матрица. Продифференцируем выражение \eqref{eq:inv_problem}, чтобы выразить искомое распределение пор по размерам $\bm{x}$
\begin{equation}
    \dfrac{\partial\mathcal{L}}{\partial \bm{x}} = 0 = \bm{A}^{\text{T}}\left(\bm{A}\bm{x} - \bm{b}\right) + \lambda^2\bm{x}
\end{equation}
\begin{equation}
    \bm{x} = \bm{A}^{\text{T}}\bm{b}\left(\bm{A}^{\text{T}}\bm{A} + \lambda^2 \bm{I}\right)^{-1}.
\end{equation}
C помощью сингулярного разложения матрицы $\bm{A}$ можно избежать обращения матриц.
\begin{equation}
    \bm{A} = \bm{U}\bm{S}\bm{V}^{\text{T}} = \sum\limits^n_{i=1} \bm{u}_i s_i \bm{v}_i^{\text{T}}
\end{equation}
где $\bm{U} = \left( \bm{u}_1 ,\dots, \bm{u}_n\right)$ и $\bm{V} = \left( \bm{v}_1 ,\dots, \bm{v}_n\right)$ ортогональные матрицы, а $\bm{S}$ --- диагональная матрица сингулярных чисел матрицы $\bm{A}$, которые расположены в порядке убывания. Таким образом, решение \eqref{eq:inv_problem} дается выражением \cite{Ravikovitch2000UnifiedIsotherms}
\begin{equation}\label{eq:PSD_answer}
    \bm{x} = \sum\limits_{i=1}^n \dfrac{s_i^2}{s_i^2 + \lambda^2} \dfrac{\bm{u}_i^{\text{T}}\bm{b}}{s_i} \bm{v}_i .
\end{equation}
\FloatBarrier
\subsection{Результаты восстановления распределения пор по размерам}\label{sec:results_psd}
\setcounter{equation}{0}
\addetoc{subsection}{Results of reconstruction the pore size distribution}

В качестве иллюстрации работы метода восстановления пор по размерам, рассматривается несколько модельных задач. В этих задачах будем задавать распределение пор по размерам, и с помощью \eqref{eq:IAE} считать <<экспериментальную>> изотерму адсорбции. Затем, на основании \eqref{eq:PSD_answer} будет рассчитываться искомое распределение, в такой модельной задаче есть несколько параметров, которые будут влиять на ответ, изменяя их можно увидеть сильные и слабые стороны данного подхода. По своей природе, распределение пор по размерам подчиняется гауссовскому распределению, поэтому именно такие распределения и будут исследоваться ниже
\begin{equation}
f\left(x\right) = \dfrac{1}{\sqrt{2\pi}\sigma}\exp{\left(-\dfrac{\left(x-\mu\right)^2}{2\sigma^2}\right)}.
\end{equation}
\subsubsection{Влияние параметров распределения на вид изотермы}
\begin{figure}[htb!]
    \centering
    \includegraphics[page=1]{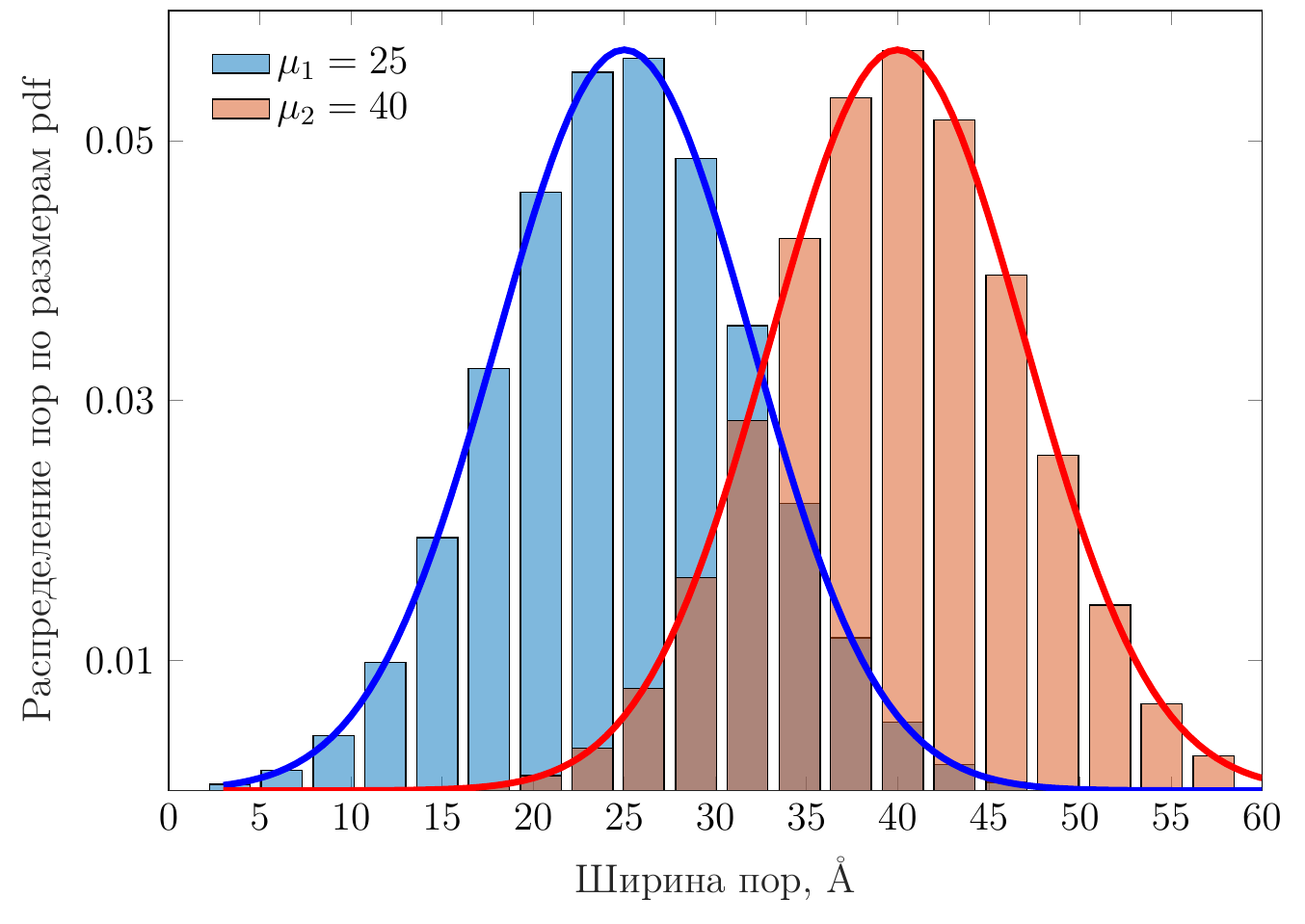}
    \caption{Два различны распределения пор по размерам с одинаковыми дисперсиями $\sigma_1 = \sigma_2 = 7$, но с различными математическим ожиданием $\mu_1 = 25$, $\mu_2 = 40$ }
    \label{fig:PSD_mu}
\end{figure}

\begin{figure}[htb!]
    \centering
    \includegraphics[page=1]{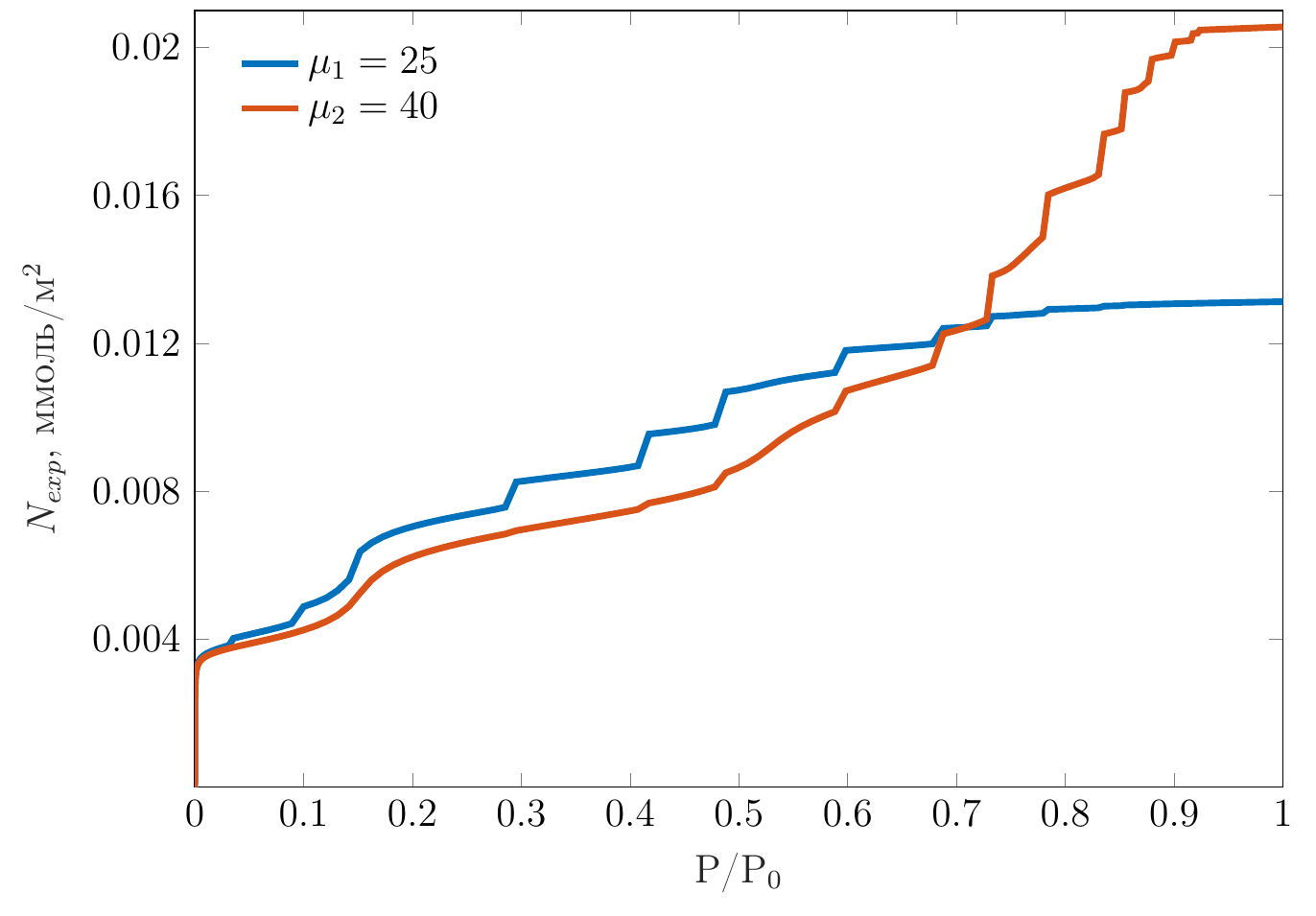}
    \caption{Расчетные изотермы адсорбции для распределений пор по размерам с одинаковой дисперсией, но различными математическими ожиданиями $\mu_1 = 25$, $\mu_2 = 40$}
    \label{fig:Ads_mu}
\end{figure}

\addetoc{subsubsection}{Influence of distribution parameters on the isotherm}
В данном разделе приводится иллюстрация того, насколько будет меняться вид рассчитанной изотермы адсорбции, при изменении параметров распределения пор по размерам (по сути это отражает чувствительность приведенного в работе метода восстановления распределения пор по размерам к начальным данным). На рис.~\ref{fig:PSD_mu} показаны два распределения с одинаковой дисперсией $\sigma_1 = \sigma_2 = 7$, но с различным математическим ожиданием $\mu_1 = 25$, $\mu_2 = 40$. При этом, после свертки данных распределений с ядрами адсорбции из матрицы $\bm{A}$ получаются две различные изотермы. Из рис.~\ref{fig:Ads_mu} видно, что расчетная изотерма адсорбции чувствительна к значению математического ожидания распределения пор по размерам. 

\FloatBarrier

\begin{figure}[htb!]
    \centering
    \includegraphics[page=1]{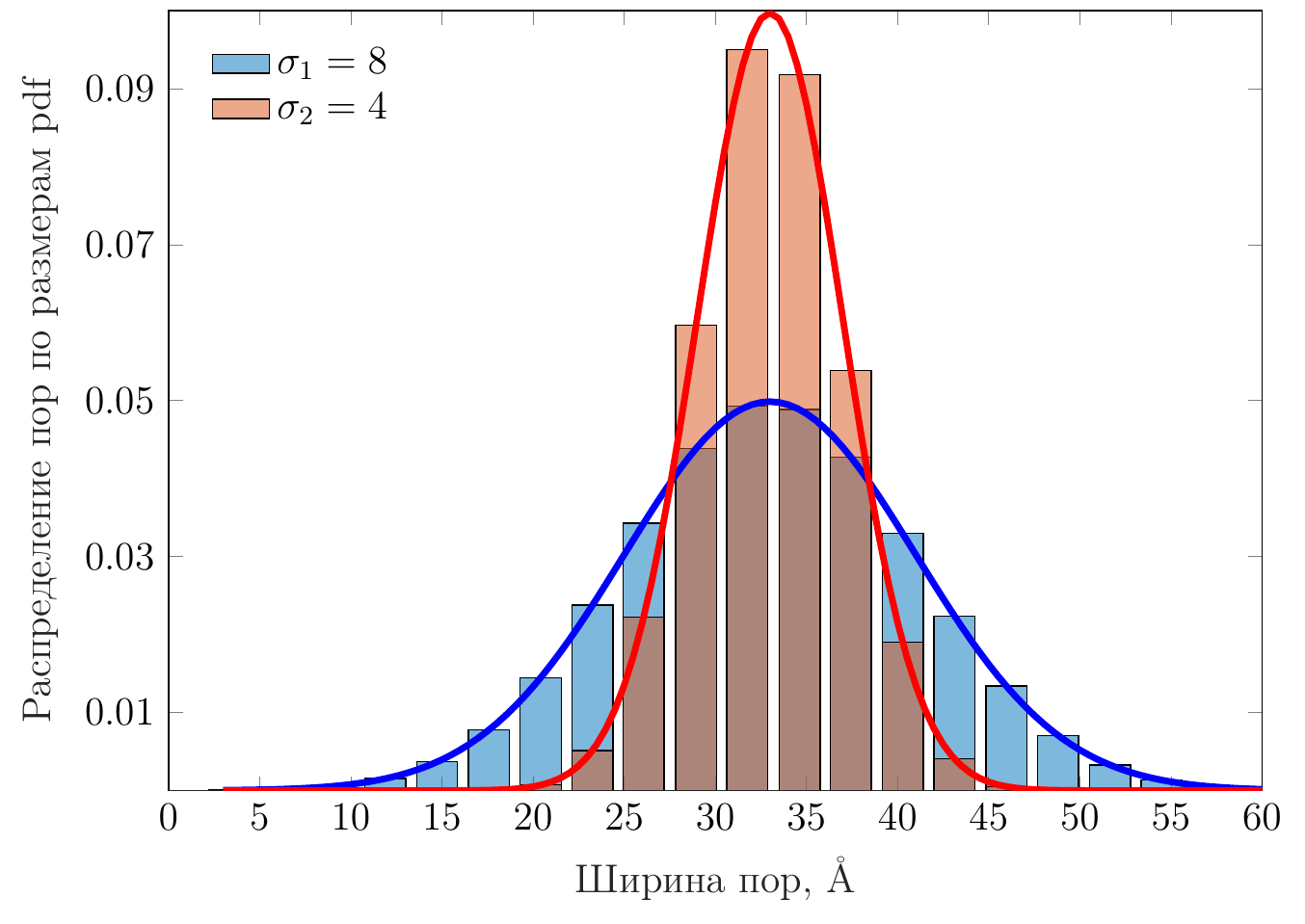}
    \caption{Два различны распределения пор по размерам с одинаковым математическим ожиданием $\mu_1 = \mu_2 = 33$, но с различными дисперсиями $\sigma_1 = 8$, $\sigma_2 = 4$}
    \label{fig:PSD_sigma}
\end{figure}

\begin{figure}[htb!]
    \centering
    \includegraphics[page=1]{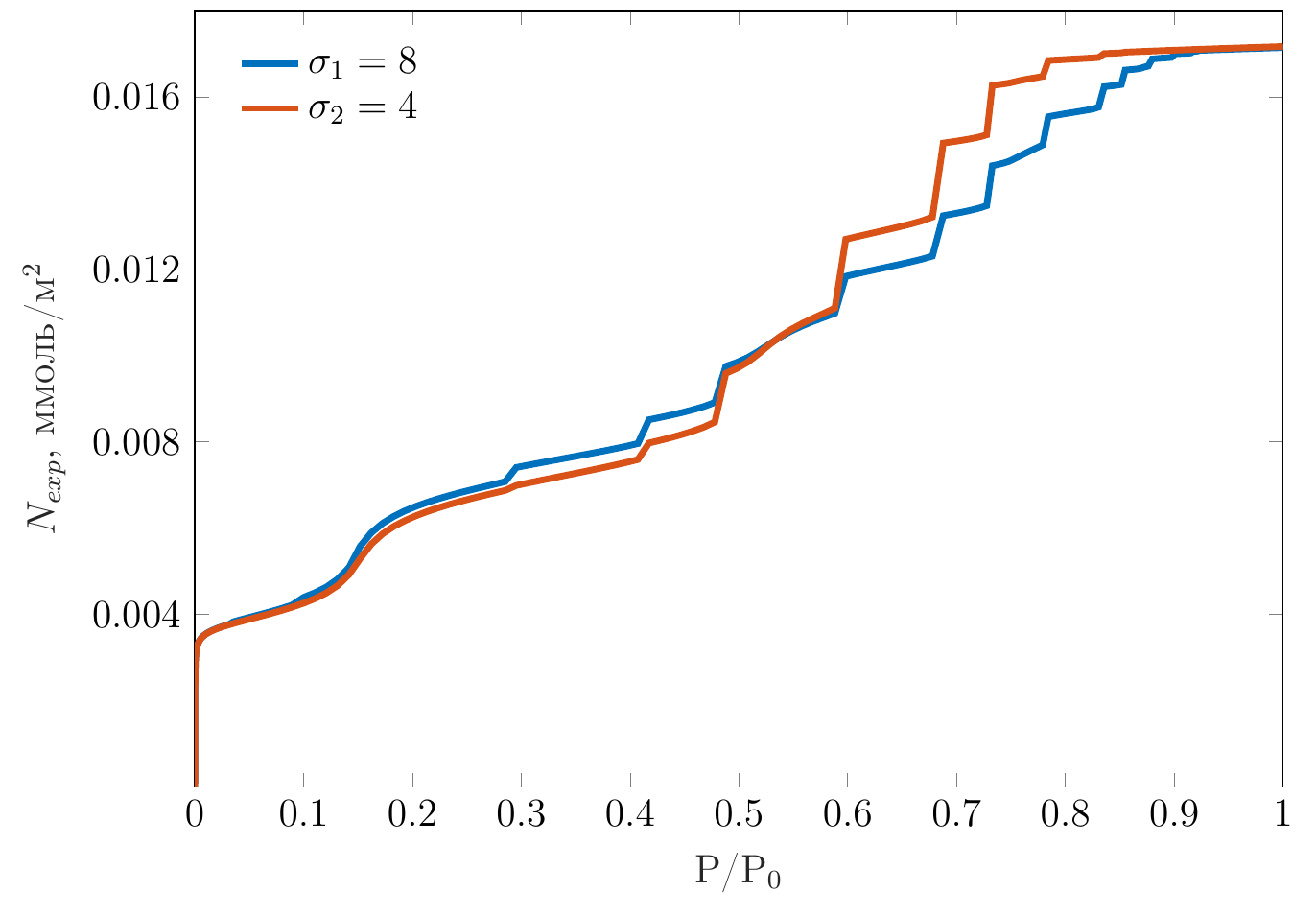}
    \caption{Расчетные изотермы адсорбции для распределений пор по размерам с одинаковой математическими ожиданиями, но различными дисперсиями $\sigma_1 = 8$, $\sigma_2 = 4$}
    \label{fig:Ads_sigma}
\end{figure}

Дисперсия распределения также влияет на вид расчетной изотермы. На рис.~\ref{fig:PSD_sigma} показаны два распределения с одинаковыми математическими ожиданиями $\mu_1 = \mu_2$, но с различными дисперсиями $\sigma_1 = 8$, $\sigma_2 = 4$. После свертки таких распределений с ядрами адсорбции рис.~\ref{fig:isotherm_kernels} получились две различные изотермы рис.~\ref{fig:Ads_sigma}.
\FloatBarrier

\subsubsection{Зависимость решения от параметра регуляризации}
\addetoc{subsubsection}{Effect of the regularization parameter on PSD reconstruction}

\paragraph{Отсутствие регуляризации $\lambda = 0$.}

\begin{figure}[htb!]
    \centering
    \includegraphics[page=1]{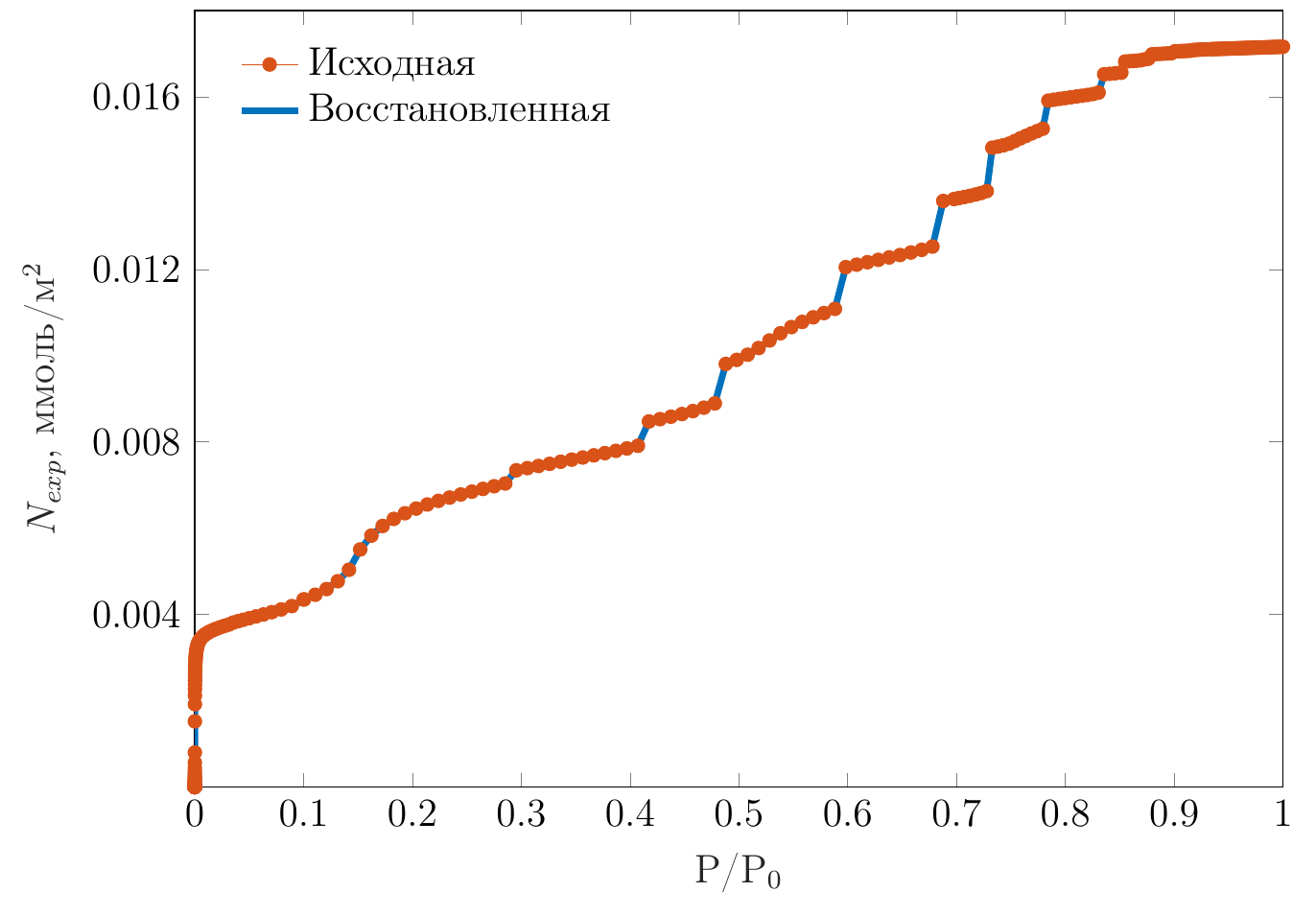}
    \caption{Сравнение между восстановленной изотермой адсорбцией в образце (синяя линия) и исходной <<экспериментальной>> изотермой для случая $\lambda=0$}
    \label{fig:Ads_lambda0}
\end{figure}

\begin{figure}[htb!]
    \centering
    \includegraphics[page=1]{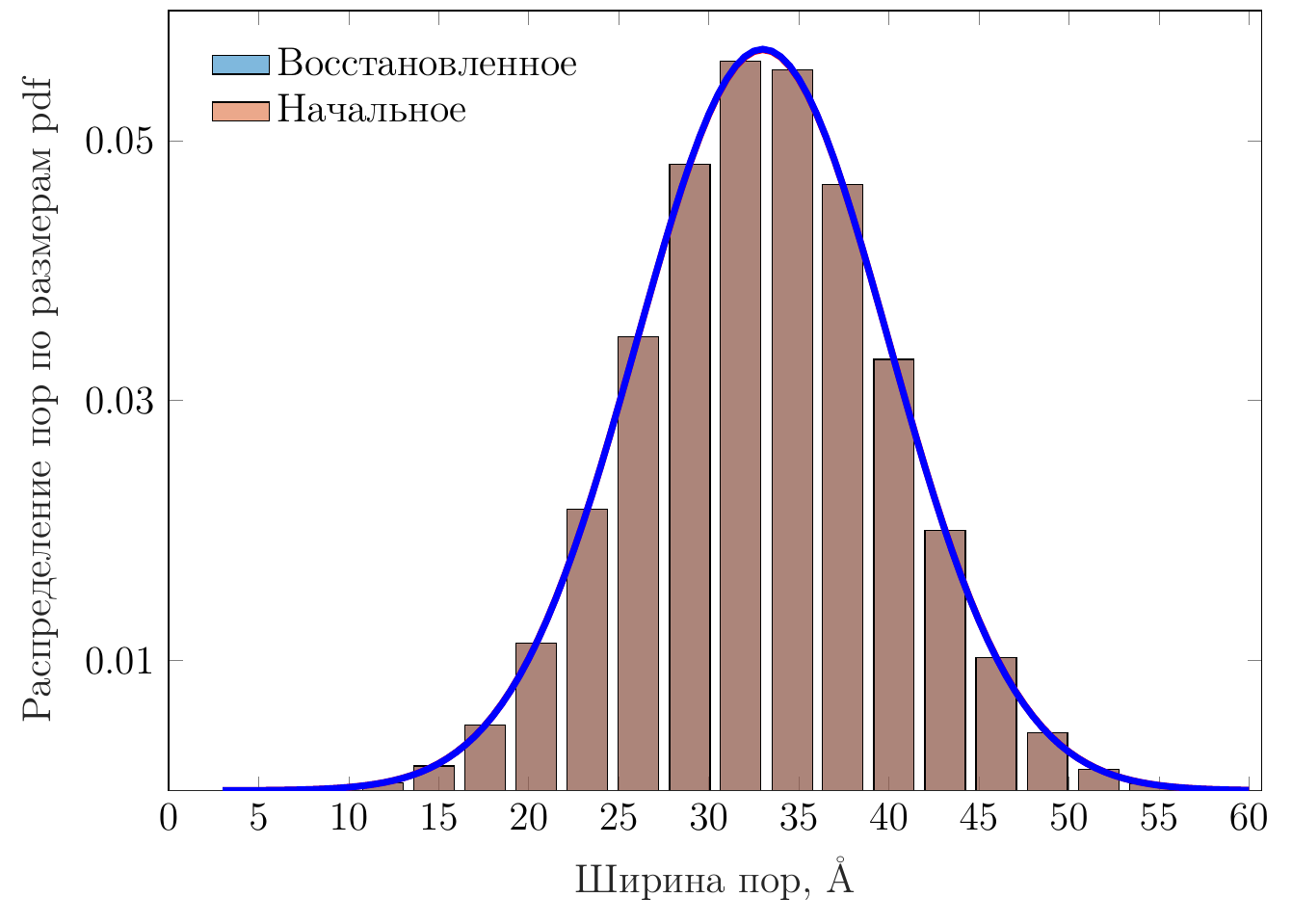}
    \caption{Сравнение восстановленного (синий цвет) и исходного (красный цвет) распределения пор по размером при $\lambda = 0$}
    \label{fig:PSD_lambda0}
\end{figure}

В самом простом приближении стоит рассмотреть задачу, когда регуляризационный параметр отсутствует и $\lambda = 0$. В таком случае восстановленное распределение должно полностью совпасть с изначально заданным распределением  рис.~\ref{fig:PSD_lambda0}. Исходное распределение было смоделировано по гауссовскому закону распределения с параметрами $\mu_{init} = 33$, $\sigma_{init} = 7$, у восстановленного распределения получились параметры $\mu_{rec}=32.99$, $\sigma_{rec}=6.99$. На рис.~\ref{fig:Ads_lambda0} представлена <<экспериментальная>> изотерма адсорбции в образце и восстановленная, которая рассчитывалась как свертка восстановленного распределения с ядрами адсорбции, видно хорошее соответствие между двумя изотермами.

\begin{figure}[htb!]
    \centering
    \includegraphics[page=1]{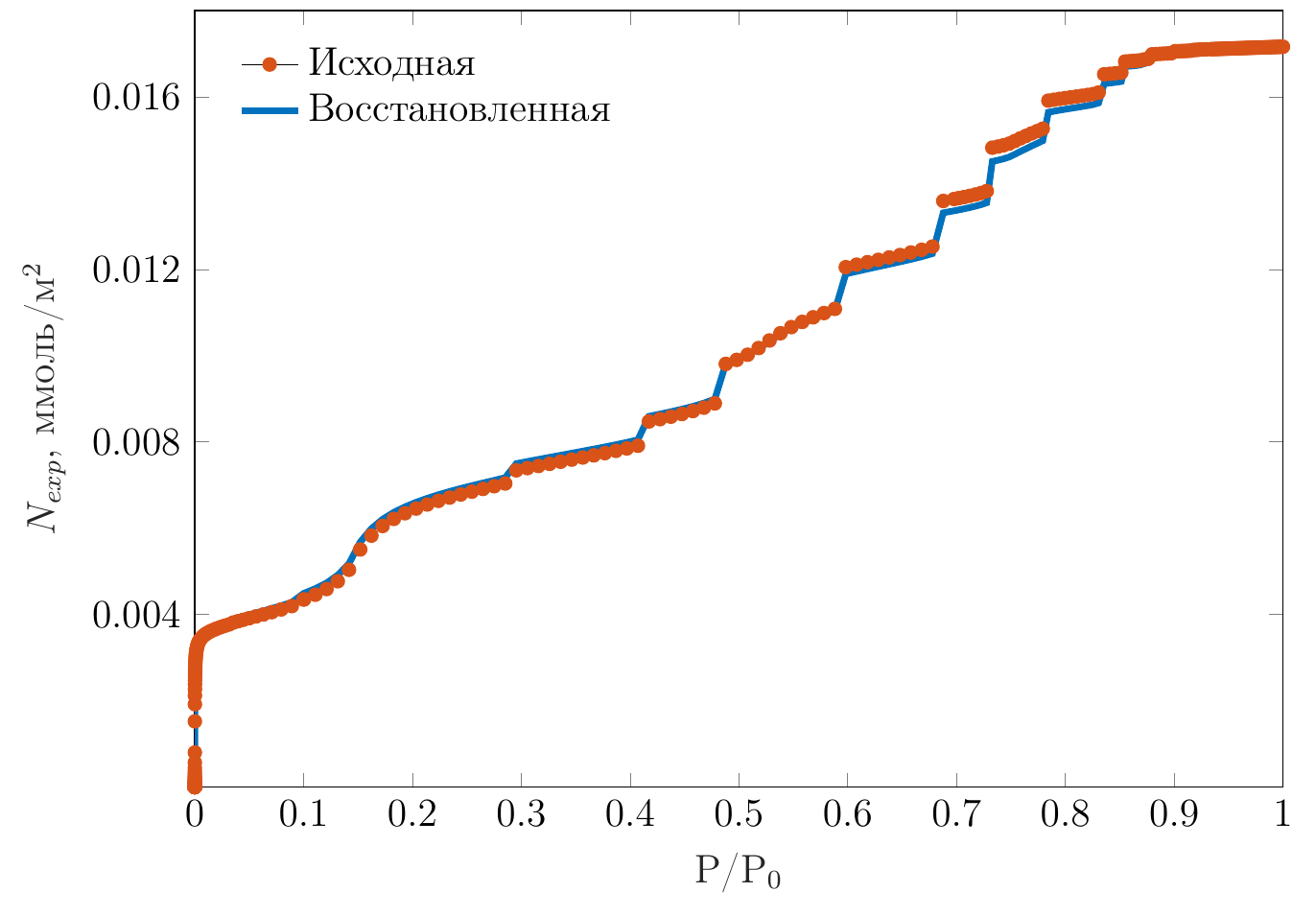}
    \caption{Сравнение между восстановленной изотермой адсорбцией в образце (синяя линия) и исходной <<экспериментальной>> изотермой для случая $\lambda=0.07$}
    \label{fig:Ads_lambda007}
\end{figure}

\begin{figure}[htb!]
    \centering
    \includegraphics[page=1]{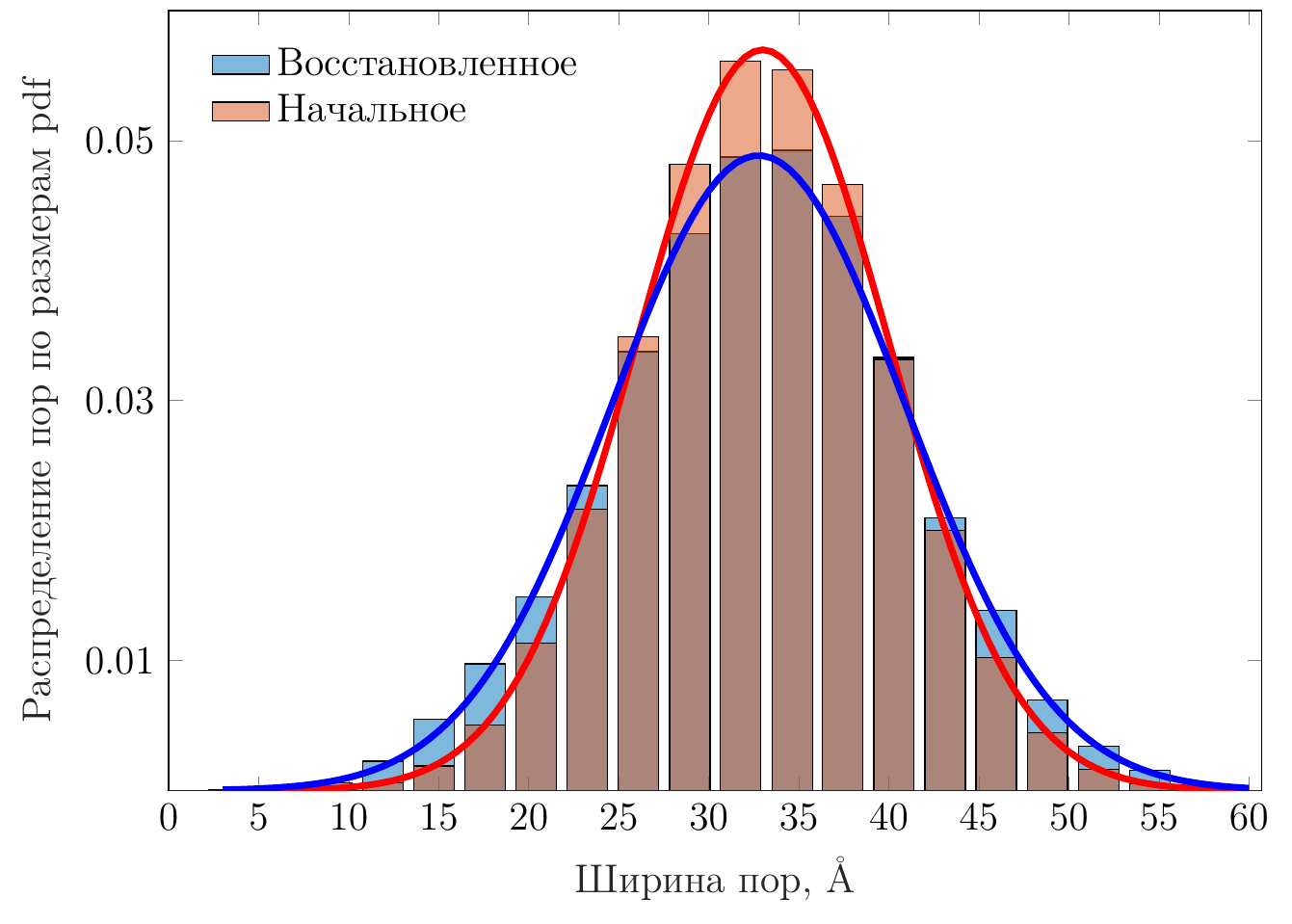}
    \caption{Сравнение восстановленного (синий цвет) и исходного (красный цвет) распределения пор по размером при $\lambda = 0.07$}
    \label{fig:PSD_lambda007}
\end{figure}

\FloatBarrier
\paragraph{Решение при регуляризации $\lambda = 0.07$.}
При изменении параметра регуляризации исходное и восстановленное распределение получаются отличными друг от друга, причем при увеличении $\lambda$ у восстановленного распределения увеличивается дисперсия, а восстановленная изотерма адсорбции становится более гладкой. У исходного распределения были параметры $\mu_{init} = 33$, $\sigma_{init} = 7$, а у восстановленного получились $\mu_{rec}=32.76$, $\sigma_{rec}=8.17$. На рис.~\ref{fig:Ads_lambda007} продемонстрирована восстановленная изотерма адсорбции и первоначальная изотерма, видно, что соответствие не такое хорошее как на рис.~\ref{fig:Ads_lambda0} для случая $\lambda=0$. На рис.~\ref{fig:PSD_lambda007} показаны распределения пор по размерам для случая $\lambda=0.07$.
\FloatBarrier
\subsubsection{Восстановление сдвинутого и бимодального распределения}
\begin{figure}[htb!]
    \centering
    \includegraphics[page=1]{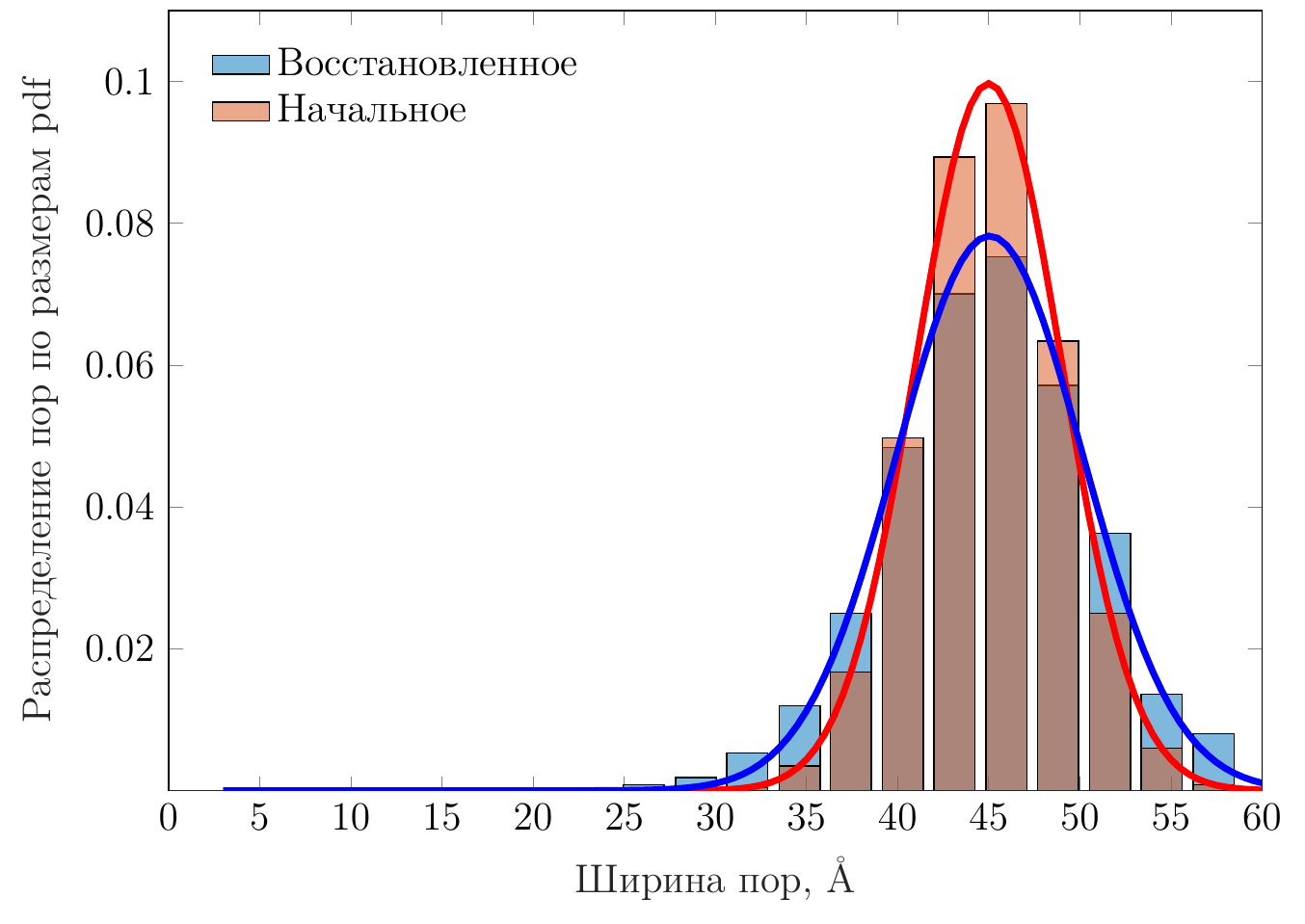}
    \caption{Сравнение восстановленного (синий цвет) и исходного (красный цвет) распределения пор по размером при $\lambda = 0.07$, $\mu_{init}= 45$, $\sigma_{init} = 4$, $\mu_{rec}= 45.05$, $\sigma_{rec} = 5.1$}
    \label{fig:PSD_skew}
\end{figure}

\begin{figure}[htb!]
    \centering
    \includegraphics[page=1]{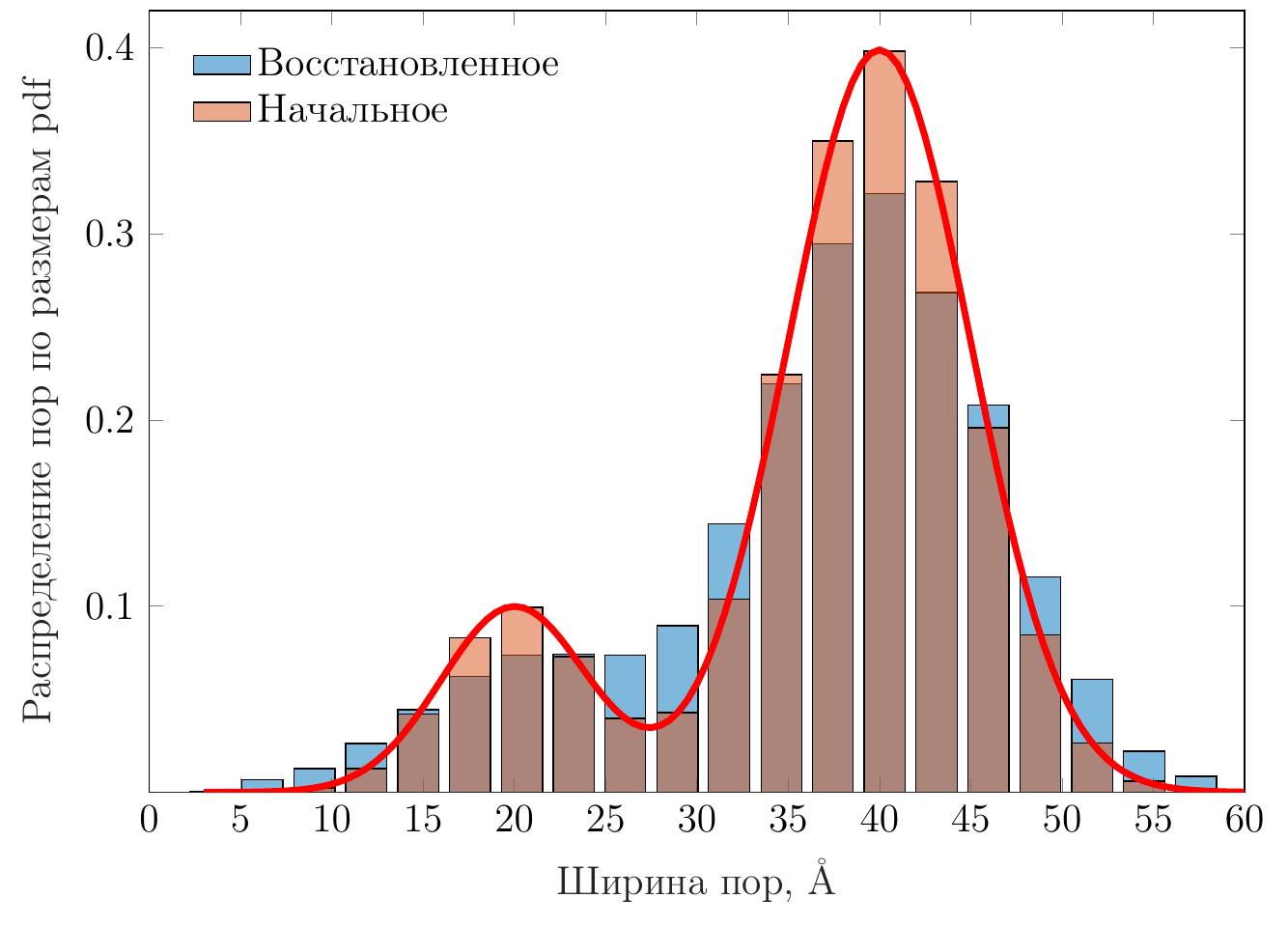}
    \caption{Сравнение восстановленного (синий цвет) и исходного (красный цвет) распределения пор по размером для случая с бимодальным распределением при $\lambda = 0.07$, $\mu^1_{init}=20$, $\sigma^1_{init}=4$, $\mu^2_{init}=40$, $\sigma^2_{init}=5$}
    \label{fig:PSD_bimodal}
\end{figure}

\addetoc{subsubsection}{Reconstruction shifted and bimodal distributions}
Для случая, когда распределение не центрировано, метод из раздела выше достаточно хорошо справляется с восстановлением распределения пор по размерам. На рис.~\ref{fig:PSD_skew} приведены результаты работы метода при параметрах $\lambda=0.07$, $\mu_{init}= 45$, $\sigma_{init} = 4$, В результате получилось распределение с параметрами $\mu_{rec}= 45.05$, $\sigma_{rec} = 5.1$. При этом стоит отметить, что как и для самого первого случая с $\lambda=0$, при выборе такого значения параметра регуляризации начальное и восстановленное распределения полностью совпадают.
\FloatBarrier
Чаще всего, реальное распределение пор по размерам является нормальным бимодальным. Для того, чтобы проверить работу алгоритма восстановления распределения пор по размерам было задано исходное распределение следующим образом:
\begin{equation}
    f\left(x\right) = \dfrac{1}{\sqrt{2\pi}\sigma_1}\exp{\left(-\dfrac{\left(x-\mu_1\right)^2}{2\sigma_1^2}\right)} + \dfrac{5}{\sqrt{2\pi}\sigma_2}\exp{\left(-\dfrac{\left(x-\mu_2\right)^2}{2\sigma_2^2}\right)}, 
\end{equation}
где параметры распределения $\mu_{init}^1 = 20$, $\sigma_{init}^1 = 4$, $\mu_{init}^2 = 40$, $\sigma_{init}^2 = 5$, а параметр регуляризации в алгоритме был выбран $\lambda = 0.07$. На рис.~\ref{fig:PSD_bimodal} представлен результат работы алгоритма, в сравнении с начальным распределением.
\FloatBarrier
\subsubsection{Зависимость решения от качества экспериментальных данных}
\begin{figure}[htb!]
    \centering
    \includegraphics[page=1]{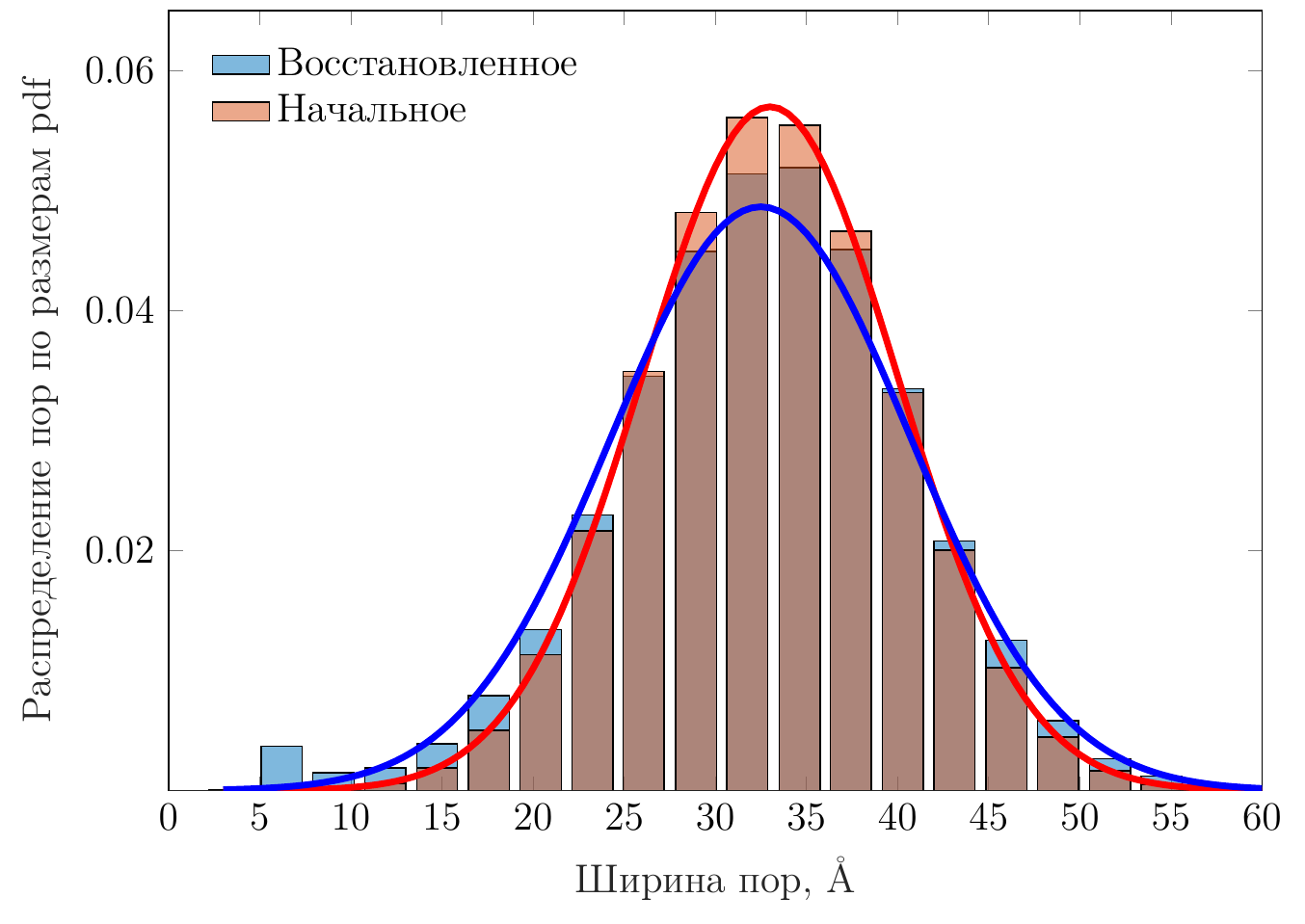}
    \caption{Сравнение восстановленного (синий цвет) и исходного (красный цвет) распределения пор по размером для случая с зашумленной изотермой при $\lambda = 0.05$, $\mu_{init}=33$, $\sigma_{init}=7$, $\mu_{rec}=32.5$, $\sigma_{rec}=8.2$}
    \label{fig:PSD_noise}
\end{figure}

\begin{figure}[htb!]
    \centering
    \includegraphics[page=1]{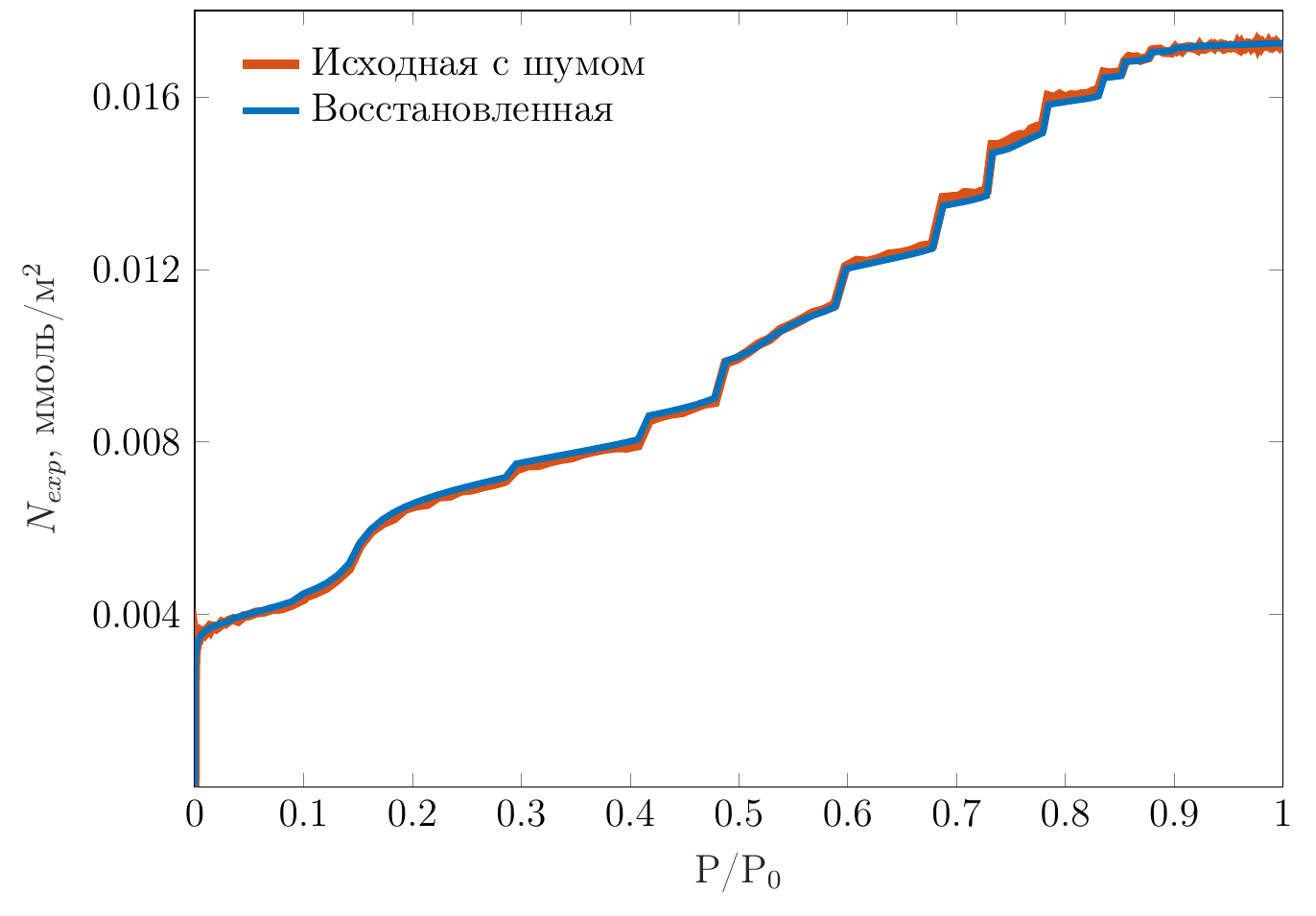}
    \caption{Сравнение между восстановленной изотермой адсорбцией в образце (синяя линия) и исходной зашумленной изотермой для случая $\lambda=0.05$}
    \label{fig:Ads_noise}
\end{figure}

\addetoc{subsubsection}{Influence of experimental data quality}
Данная модельная задача демонстрирует как важна точность экспериментальных данных. С помощью заданного нормального распределения с параметрами $\mu_{init} = 33$, $\sigma_{init}=7$ была рассчитана <<экспериментальная>> изотерма адсорбции. Затем, к изотерме была добавлена случайна величина, равномерно распределенная на отрезке $\left[10^{-5},\, 10^{-4}\right]$. В результате получилась изотерма, которая показана на рис.~\ref{fig:Ads_noise} красным цветом. С помощью зашумленной изотермы было восстановлено распределение пор по размерам рис.~\ref{fig:PSD_noise}. Для алгоритма восстановления распределения пор по размерам был выбран параметр $\lambda=0.05$. При таком значении регуляризационного параметра посчитанная изотерма адсорбции лучше всего совпадает с исходной изотермой. В такой задаче, с зашумленными данными у восстановленного распределения появляются артефакты на хвостах распределения и регуляризационный параметр $\lambda$ требует более тонкой настройки.

\FloatBarrier
\paragraph{Выводы}

Данный раздел не содержит в себе научной новизны, но наглядно демонстрирует применение теории функционала плотности для исследования структуры нанопористых материалов. Методы восстановление распределения пор по размерам опираются на результаты DFT по адсорбции флюида в изолированной поре фиксированной ширины. Усовершенствование моделей DFT, принципов расчета равновесной плотности в поре и скорость этих расчетов становятся существенными для быстрого и качественного описания порового пространства. Разработанные в данной работе методы VF-DFT и H-DFT позволят исследовать сложные флюиды, учитывать различную геометрию порового пространства и значительно сократить время расчета изотермы адсорбции флюида в поре заданной ширины. 

\section{Заключение}
\addetoc{section}{Conclusion}
Являясь универсальным методом статистической механики, классическая теория функционала плотности предлагает мощную альтернативу множеству традиционных теоретических методов и молекулярному моделированию для связывания микроскопических свойств физико-химических систем со структурными и термодинамическими свойствами. Практическая ценность DFT проявляется не только в ее универсальности для описания термодинамических свойств классических систем, но и в универсальности для решения задач, которые не могут быть достигнуты традиционными теориями. Разработанный в данной работе безвариационный подход для поиска равновесной плотности может позволить в дальнейшем описывать сложные взаимодействия флюидов, для которых представляется трудоемким вычисления вариации свободной энергии Гельмгольца. Комбинирование безвариационного подхода с методом простой итерации позволило в разы ускорить расчет равновесной плотности при высоких относительных давлениях, но с незначительной потерей качества решения ($\sim 5\%$). Наибольший эффект от ускорения расчетов наблюдается для случаев с высоким относительным давлением. В дальнейшем, комбинированный алгоритм можно использовать для ускорения расчета изотерм адсорбции. При этом стоит отметить, что скорость и качество решения, которые показывает безвариационный подход VF-DFT, напрямую зависет от базиса. Размер базиса и вид базисных функций можно постоянно совершенствовать и дорабатывать, чтобы получать решения более высокого качества. Кроме того, на скорость работы алгоритма влияет настройка параметров оптимизационных алгоритмов. Было проведено исследование по зависимости скорости работы стохастических алгоритмов в данных задачах и предложены рекомендации по первоначальной настройке. В дальнейшем предлагается исследовать безвариационный подход в комбинации с другими стохастическими методами оптимизации или с любыми другими методами оптимизации нулевого порядка.

Как одно из наиболее важных применений теории функционала плотности, в данной работе была рассмотрена задача восстановления распределения пор по размерам на основании экспериментальных данных по низкотемпературной адсорбции. Рассмотренный подход для решению данной проблемы показал себя весьма действенным и наглядным для описания структуры порового пространства.

\newpage
\addcontentsline{toc}{section}{Список Литературы}
\addcontentsline{tec}{section}{References}%
\printbibliography

\end{document}